%% file: main.tex
\documentclass[pdflatex,sn-mathphys-num]{sn-jnl}

\usepackage{graphicx}
\usepackage{multirow}
\usepackage{amsmath,amssymb,amsfonts}
\usepackage{amsthm}
\usepackage{mathrsfs}
\usepackage[title]{appendix}
\usepackage{xcolor}
\usepackage{textcomp}
\usepackage{manyfoot}
\usepackage{booktabs}
\usepackage{algorithm}
\usepackage{algorithmicx}
\usepackage{algpseudocode}
\usepackage{listings}
\usepackage{bm}
\usepackage{float} 
\usepackage{subfigure}
\usepackage{makecell}
\usepackage{natbib}
\usepackage{color}
\usepackage{ulem}
\usepackage{soul}
\usepackage{wasysym}
\usepackage[utf8]{inputenc}

\newcommand{\ignore}[1]{}

\theoremstyle{thmstyleone}

\theoremstyle{thmstyletwo}

\theoremstyle{thmstylethree}

\providecommand{\ModelName}{\textsc{SCALE-Charge}}

\begin{document}
\renewcommand{\figurename}{Fig}

\title[Article Title]{From local learning to global electronic properties in large-scale systems}

\author[1]{\fnm{Yu} \sur{Cheng}}
\equalcont{These authors contributed equally to this work.}

\author[1]{\fnm{Zhichang} \sur{Fu}}
\equalcont{These authors contributed equally to this work.}

\author[1]{\fnm{Yechen} \sur{Chen}}

\author[1]{\fnm{Yonggang} \sur{Li}}

\author*[1,2]{\fnm{Weiqing} \sur{Zhou}}
\email{weiqingzhou@whu.edu.cn}

\author*[1,2,3]{\fnm{Shengjun} \sur{Yuan}}
\email{s.yuan@whu.edu.cn}

\affil[1]{Key Laboratory of Artificial Micro- and Nano-structures of Ministry of Education and School of Physics, Wuhan University, Wuhan 430072, P. R. China}
\affil[2]{Wuhan Institute of Quantum Technology, Wuhan 430206, P. R. China}
\affil[3]{School of Artificial Intelligence, Wuhan University, 430072, Wuhan, P. R. China}

\abstract{Machine learning from local atomic environments offers an efficient route to electronic structure, but evaluating global electronic properties in large systems remains computationally demanding. Here we combine real space charge density learning with stochastic propagation using Kohn-Sham Hamiltonians constructed from the predicted densities. The framework avoids target system self-consistent iterations and explicit diagonalization, while hierarchical blocking enables dense grid inference within limited GPU memory. We demonstrate density of states calculations for one million Si atoms and optical and transport calculations for over one hundred thousand atoms, without training separate models for each property. This framework extends the practical scale of electronic structure calculations within Kohn-Sham DFT, opening opportunities to investigate how atomic structure governs electronic, optical and transport properties at length scales relevant to real materials and devices.}

\maketitle

\section{Introduction}
\input{context/Introduction}

\section{Scalable real-space density inference}
\label{sec:scalable_framework}
\input{context/model}

\section{Results}


\subsection{Real-space charge-density accuracy}
\label{sec:density_accuracy}
\input{context/result_crystal}

\subsection{Density prediction across chemical space}
\label{sec:chemical_space}
\input{context/result_MP}

\subsection{Density prediction across system sizes and structural environments}
\label{sec:structural_transferability}
\input{context/result_defect}

\subsection{Electronic-structure properties from predicted densities}
\label{sec:properties}
\input{context/result_properties}

\subsection{Large-scale spectral and transport calculations from predicted densities}
\label{sec:large_scale_properties}
\input{context/result_ON}

\section{Discussion and conclusion}
\label{sec:discussion}
\input{context/conclusion}

\section{Methods}
\label{sec:methods}

\input{context/method_dataset}

\input{context/method_DFT}

\input{context/model_detail}

\input{context/method_sPM}

\section*{Data availability}

\section*{Author contributions}

\section*{Competing interests}
The authors declare no competing financial or non-financial interests.

\section*{Acknowledgements}
This work was supported by the National Natural Science Foundation of China (Grants No. 12425407 and 12547165). W.Z. gratefully acknowledges support from the Hubei Provincial Young Science and Technology Talent “Morning Light” Support Program and the Hongyi postdoctoral fellowship of Wuhan University. We thank the Core Facility of Wuhan University for providing the computational resources.

\bibliography{references}

\end{document}


\pdfbookmark[0]{Supplementary Materials}{sm:title}
\maketitle

\thispagestyle{empty}
\clearpage
\setcounter{page}{1}

\pdfbookmark[1]{Supplementary figures}{sm:figures}
\section*{Supplementary figures}


\pdfbookmark[2]{Fig. S1 -- Angular-order accuracy and inference cost}{sm:figure1}
\begin{figure}[H]
    \centering
    \includegraphics[keepaspectratio,width=0.98\linewidth]{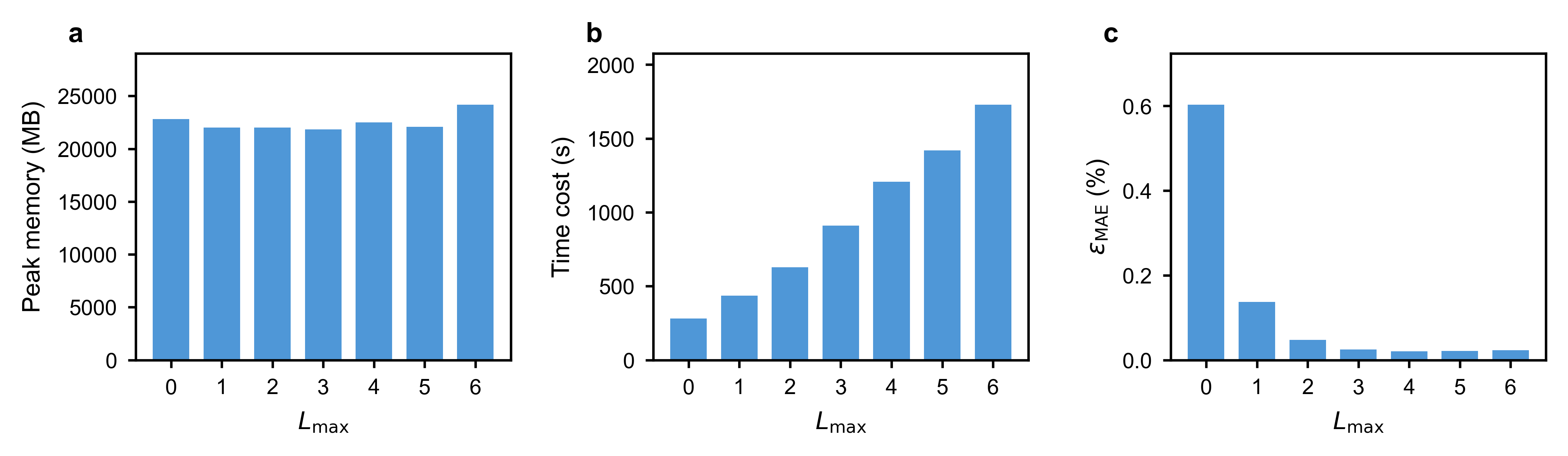}
    \caption{\textbf{Angular-order accuracy and inference cost.}
    \textbf{a--c}, Peak GPU memory, total network execution time, and electron-number-normalized absolute density error $\epsilon_{\mathrm{MAE}}$ for crystalline C as the maximum angular momentum $L_{\max}$ varies from $0$ to $6$. Memory and timing measurements use million-atom systems. Peak memory is the maximum recorded GPU allocation across the atom-network and probe-decoding stages, while total network execution time is the sum of their runtimes across all blocks. Density errors are evaluated on 64-atom validation configurations and reported as percentages. Most of the accuracy gain is obtained by $L_{\max}=3$--$4$, whereas higher orders continue to increase computational cost without a systematic further reduction in error.}
    \label{fig:model_L}
\end{figure}

\pdfbookmark[2]{Fig. S2 -- Bond-length distributions}{sm:figure2}
\begin{figure}[H]
    \centering
    \includegraphics[keepaspectratio,width=0.98\linewidth]{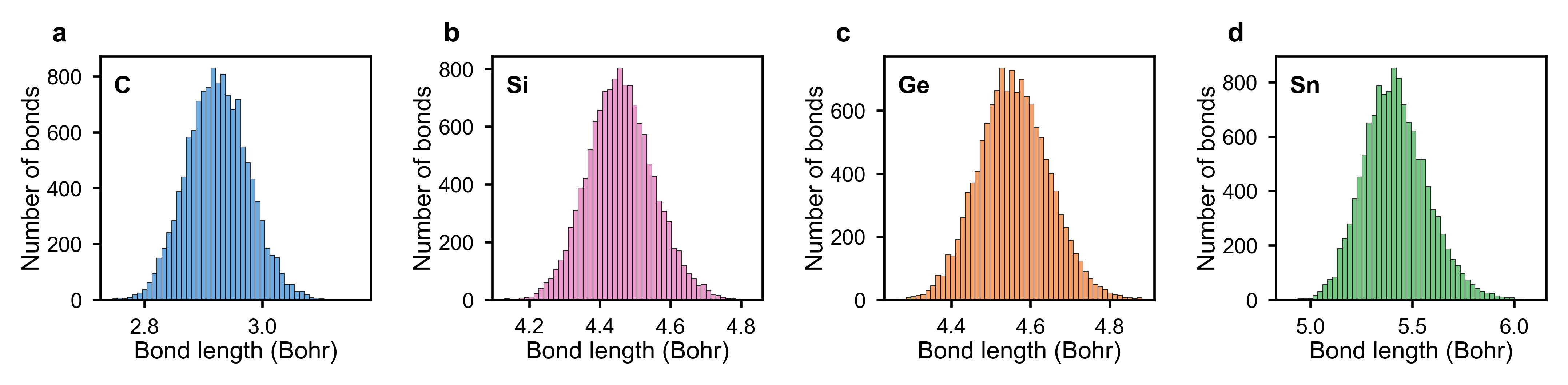}
    \caption{\textbf{Bond-length distributions of 64-atom group-IV training structures.}
    \textbf{a--d}, Histograms of C--C, Si--Si, Ge--Ge, and Sn--Sn bond lengths, respectively, pooled over the corresponding training structures. Bonds are identified when the interatomic distance does not exceed the sum of the covalent radii plus 0.45~\AA. Periodic distances follow the minimum-image convention, and each atom pair is counted once. The vertical axes show bond counts; each panel uses 50 bins spanning that material's observed bond-length range.}
    \label{fig:bonds}
\end{figure}


\clearpage
\pdfbookmark[2]{Fig. S3 -- Training losses}{sm:figure3}
\begin{figure}[H]
    \centering
    \includegraphics[keepaspectratio,width=0.98\linewidth]{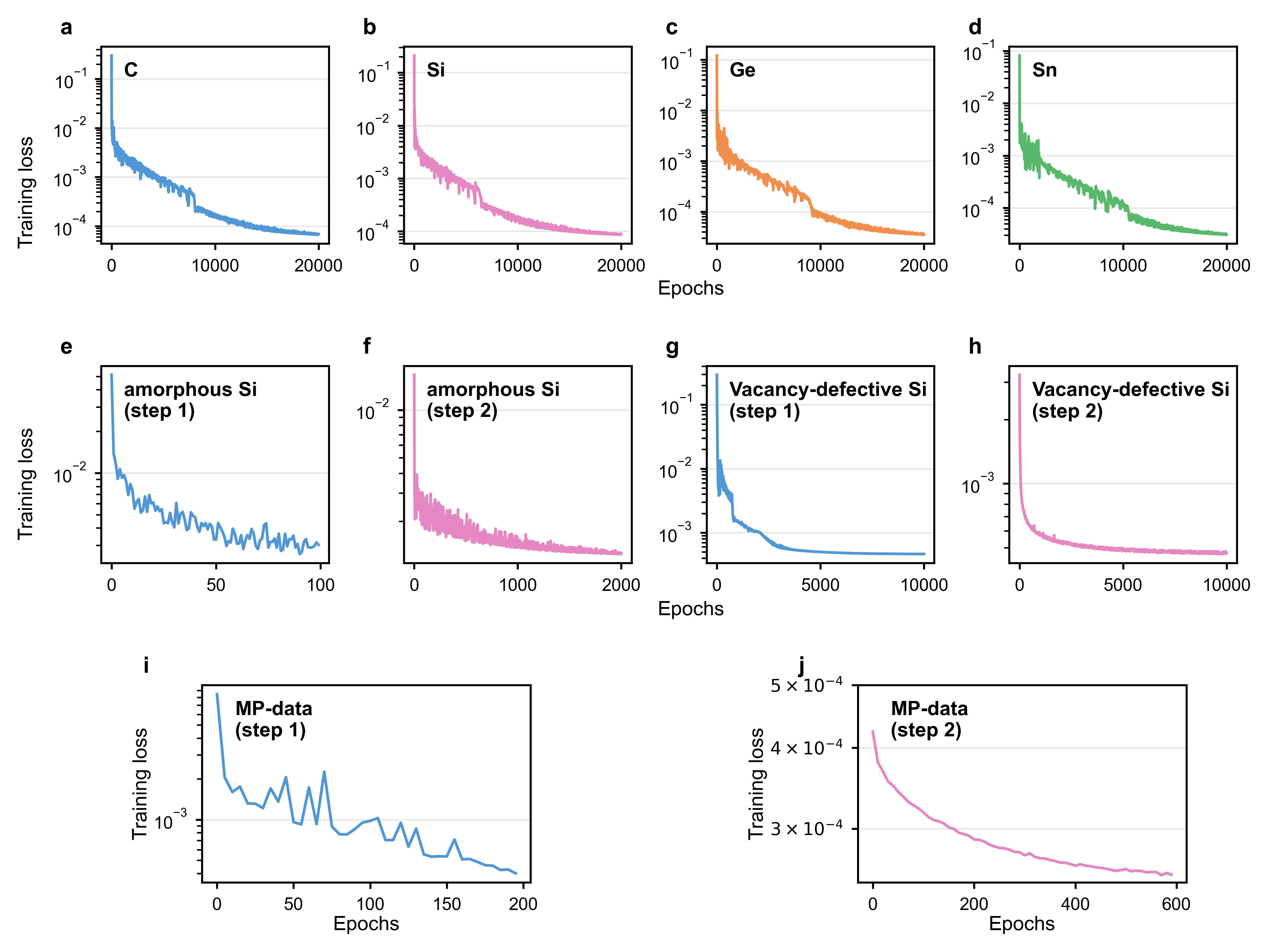}
    \caption{\textbf{Training-loss evolution across the datasets.}
    \textbf{a--d}, Crystalline C, Si, Ge, and Sn, respectively.
    \textbf{e,f}, Step 1 and step 2 of training for amorphous Si.
    \textbf{g,h}, Corresponding training stages for vacancy-defective Si.
    \textbf{i,j}, Corresponding training stages for the Materials Project (MP) dataset.
    Each value is the arithmetic mean of the recorded batch training losses within an epoch. The vertical axes are logarithmic. The C and Si charge densities are rescaled for training by factors of $3.3333$ and $10.0$, respectively.}
    \label{fig:training_loss}
\end{figure}


\clearpage
\pdfbookmark[2]{Fig. S4 -- Element-wise structure counts}{sm:figure4}
\begin{figure}[H]
    \centering
    \includegraphics[keepaspectratio,width=0.98\linewidth]{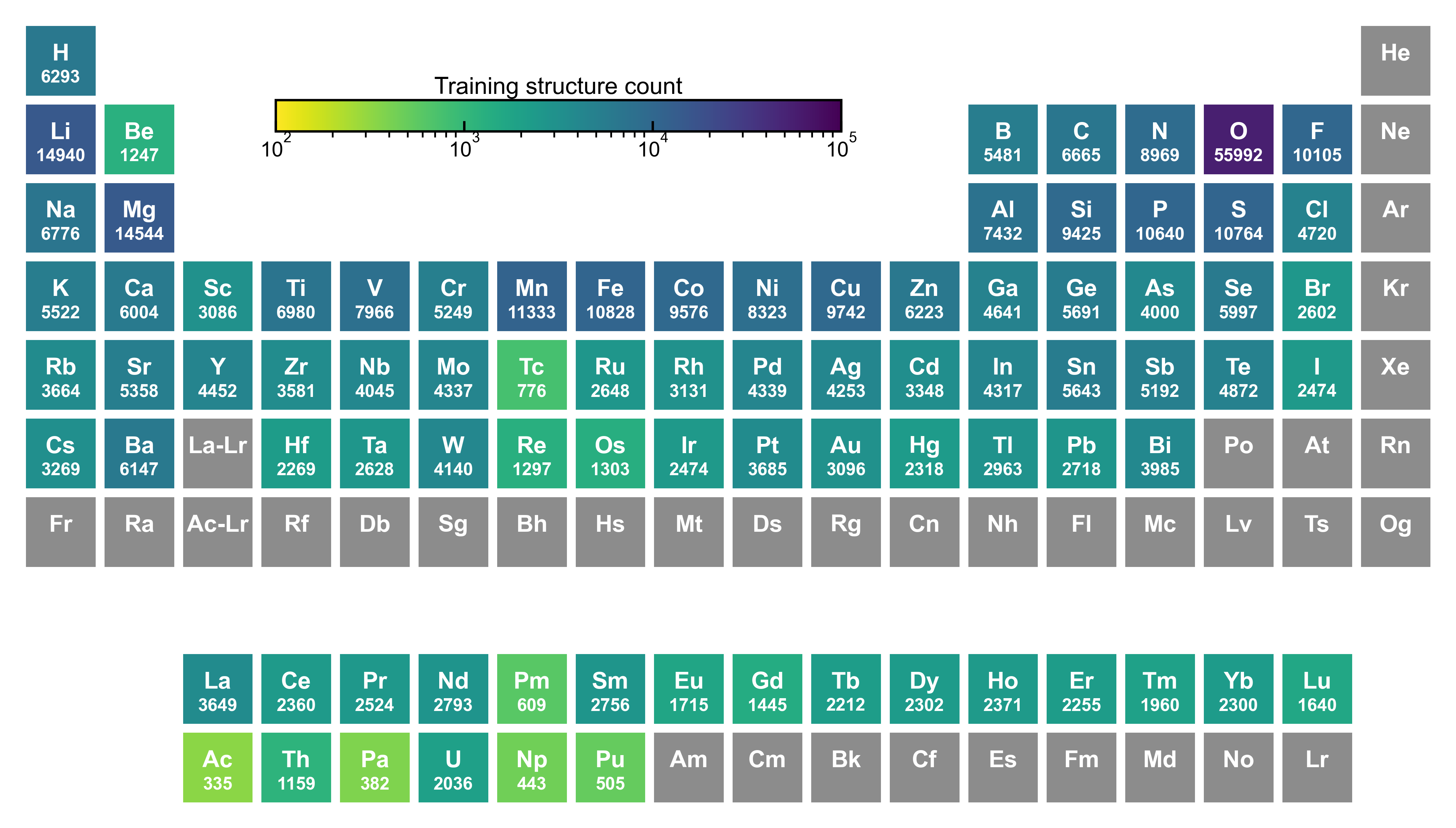}
    \caption{\textbf{Element-wise structure counts in the MP training set.}
    The periodic table reports the number of training structures containing each element in the MP dataset used to train \ModelName{}. Each element is counted once per structure, irrespective of the number of atoms of that species. Numerical labels and cell colors encode these counts over the complete training set, with colors shown on a logarithmic scale. Only elements appearing in at least $10$ validation structures are assigned values; gray cells indicate elements for which no value is reported.}
    \label{fig:mp_frequency}
    \SMPreviewNote{The image shown here is recovered from the earlier SM PDF and reports the older atom-count statistics. Replace it with the current structure-count figure corresponding to the caption above.}
\end{figure}

\pdfbookmark[2]{Fig. S5 -- Temperature-dependent Si DOS}{sm:figure5}
\begin{figure}[H]
    \centering
    \includegraphics[keepaspectratio,width=0.98\linewidth]{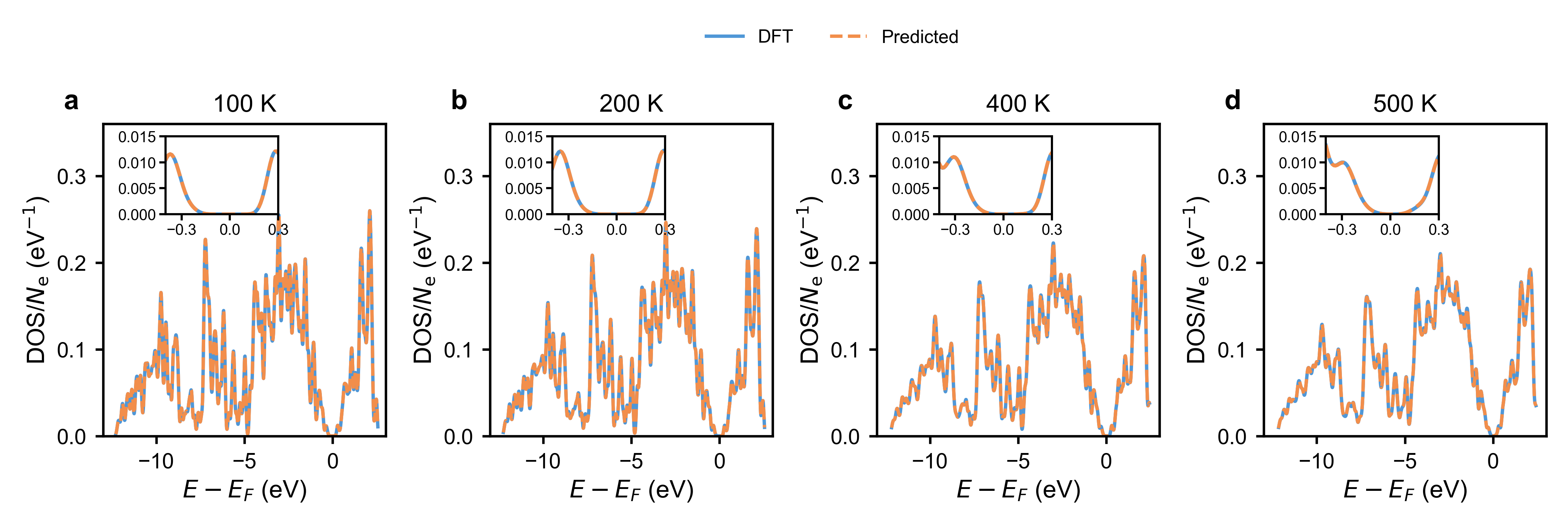}
    \caption{\textbf{Temperature-dependent DOS of crystalline Si.}
    \textbf{a--d}, Density of states averaged over 100 Si configurations at 100, 200, 400, and 500~K, respectively, using the model trained at 300~K without retraining. Solid blue and dashed orange curves denote results obtained from the reference and predicted charge densities. Before averaging, each spectrum is aligned to its Fermi energy, divided by the corresponding electron number to give an approximately unit integral below the Fermi level, and interpolated onto a common energy grid. Insets enlarge the region near the Fermi level. Energies are shown as $E-E_F$ in eV.}
    \label{fig:si_dos_temp}
\end{figure}

\clearpage
\pdfbookmark[2]{Fig. S6 -- Properties of C, Ge, and Sn}{sm:figure6}
\begin{figure}[H]
    \centering
    \includegraphics[keepaspectratio,width=0.98\linewidth]{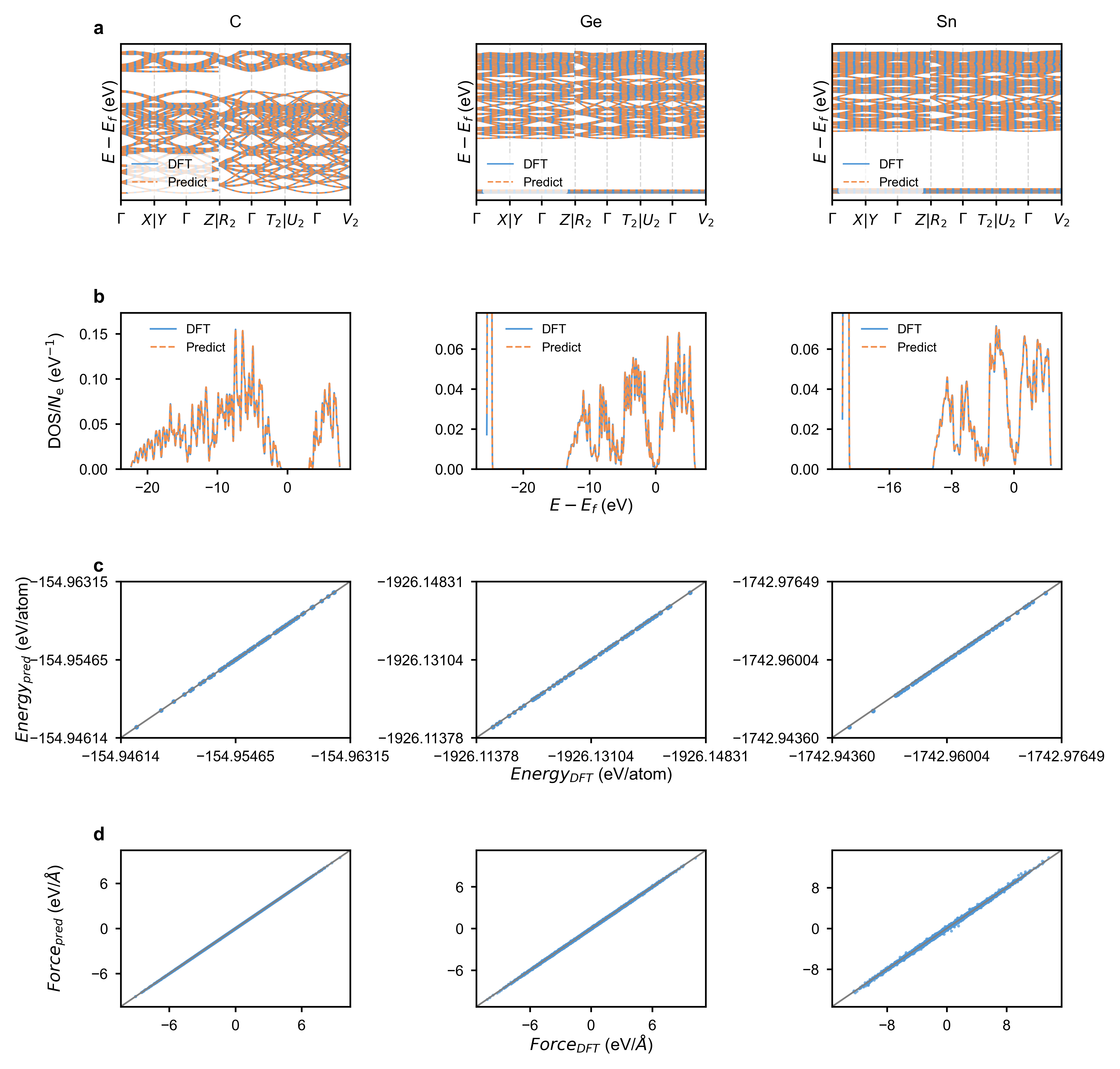}
    \caption{\textbf{Electronic-structure properties from predicted charge densities for C, Ge, and Sn.}
    Columns correspond to 64-atom C, Ge, and Sn systems, respectively. Fixed-density calculations using predicted charge densities are compared with those using DFT reference densities.
    \textbf{a}, Kohn--Sham band structures for one representative configuration of each material, with energies referenced to the corresponding Fermi level.
    \textbf{b}, DOS averaged over 100 configurations at 300~K for each material. Before averaging, each spectrum is aligned to its Fermi energy, divided by its corresponding electron number, and interpolated onto a common energy grid. Solid blue and dashed orange curves in \textbf{a,b} denote reference- and predicted-density results, respectively.
    \textbf{c}, Total energies per atom for 100 configurations per material.
    \textbf{d}, Cartesian force components for the same configurations. Gray diagonal lines in \textbf{c,d} indicate equality. Band energies are in eV, normalized DOS in eV$^{-1}$, total energies in eV/atom, and forces in eV/\AA.}
    \label{fig:nscf_CGeSn}
\end{figure}

\begin{figure}[H]
    \centering
    \includegraphics[keepaspectratio,width=0.98\linewidth]{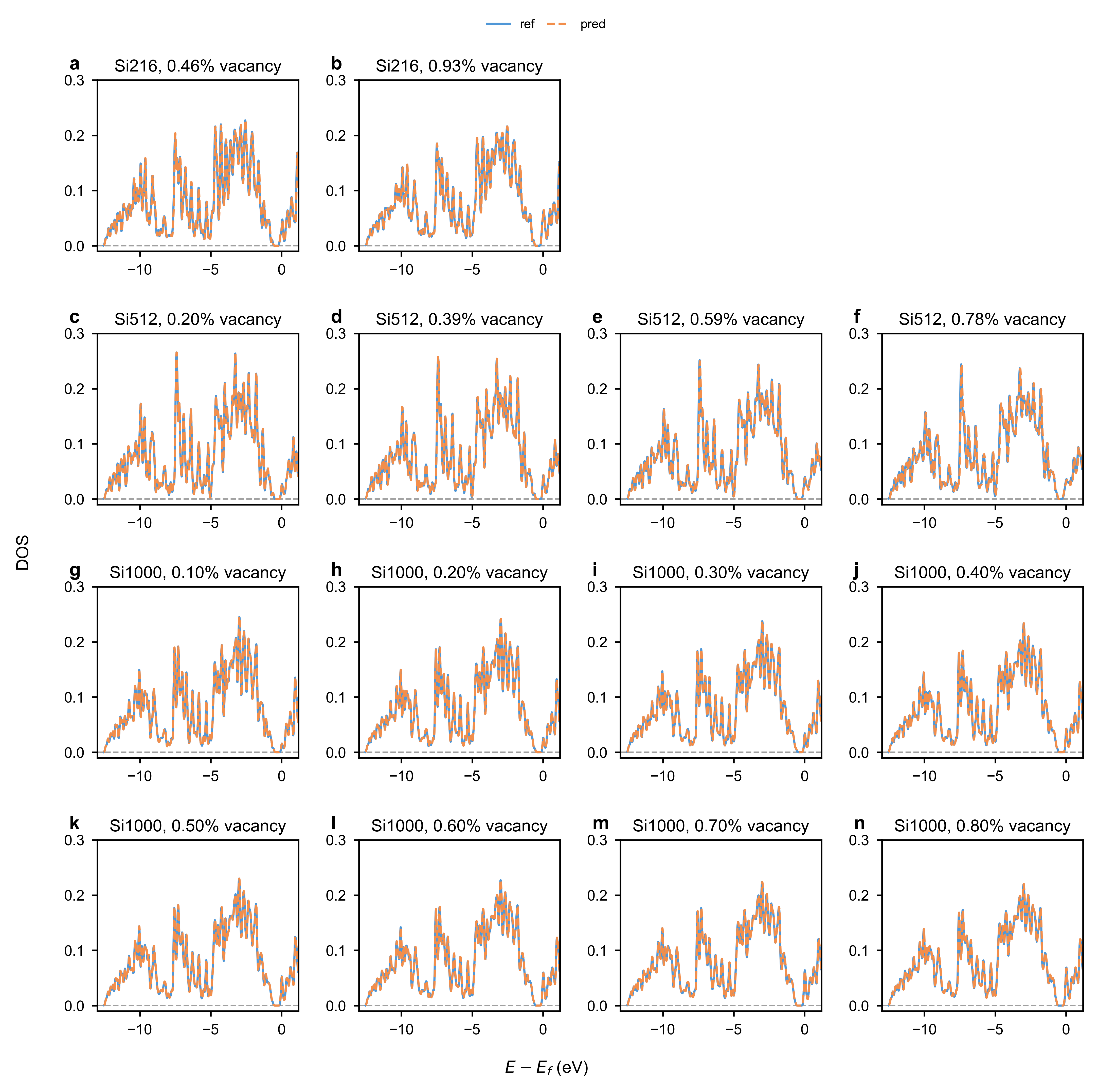}
    \caption{\textbf{ABACUS DOS from reference and \ModelName{}-predicted charge densities.}
    \textbf{a--n}, DOS of vacancy-containing Si validation structures with different system sizes and vacancy concentrations. \textbf{a,b}, Si216 structures containing one and two vacancies, respectively. \textbf{c--f}, Si512 structures containing one to four vacancies. \textbf{g--n}, Si1000 structures containing one to eight vacancies. Solid and dashed curves denote the DOS obtained from the reference and \ModelName{}-predicted charge densities, respectively, using ABACUS non-self-consistent calculations. For each system size and vacancy number, the DOS is averaged over multiple structural configurations, with the DOS broadening parameter set to $0.045$~eV. Energies are aligned to the Fermi level of each structure, and the DOS is normalized by the corresponding electron number. Vacancy concentrations are defined as the number of removed atoms divided by the number of sites in the pristine parent supercell.}
    \label{fig:vac_dos_abacus}
\end{figure}

\begin{figure}[H]
    \centering
    \includegraphics[keepaspectratio,width=0.98\linewidth]{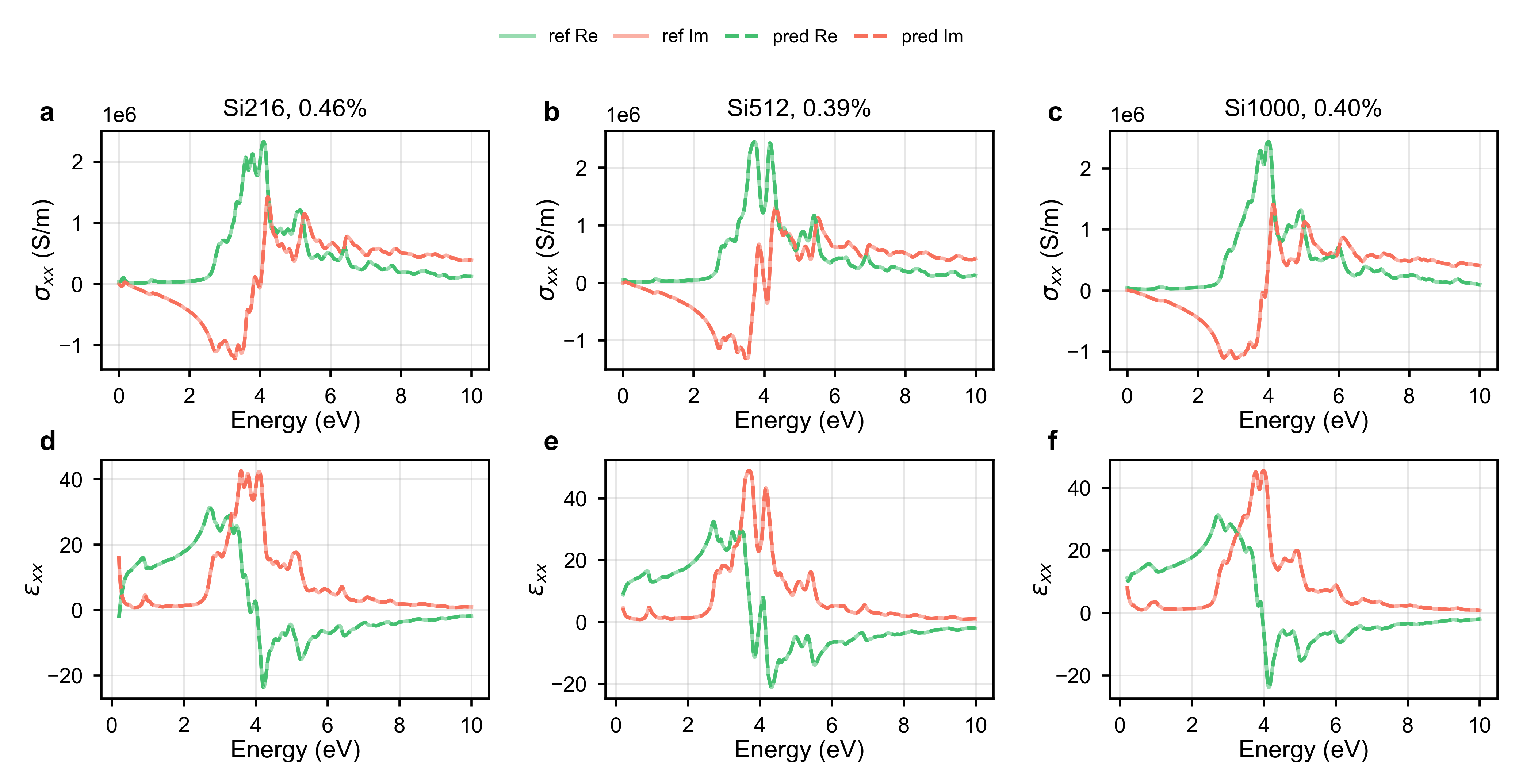}
    \caption{\textbf{ABACUS--PYATB optical response from reference and \ModelName{}-predicted charge densities.}
    \textbf{a--c}, Real and imaginary parts of the $xx$ optical conductivity for vacancy-containing Si216, Si512, and Si1000 structures with $1$, $2$, and $4$ vacancies, corresponding to vacancy concentrations of $0.46\%$, $0.39\%$, and $0.40\%$, respectively. Optical responses are calculated using ABACUS and PYATB from the reference and predicted charge densities. Solid and dashed curves denote the reference- and prediction-based results, respectively, while green and orange curves represent the real and imaginary components. \textbf{d--f}, Corresponding real and imaginary parts of the $xx$ dielectric function for the same structures. Horizontal axes show photon energy $\hbar\omega$ in eV. Optical conductivity is expressed in $\mathrm{S\,m^{-1}}$, whereas the dielectric function is dimensionless.}
    \label{fig:vac_ac_abacus}
\end{figure}

\begin{figure}[H]
    \centering
    \includegraphics[keepaspectratio,width=0.98\linewidth]{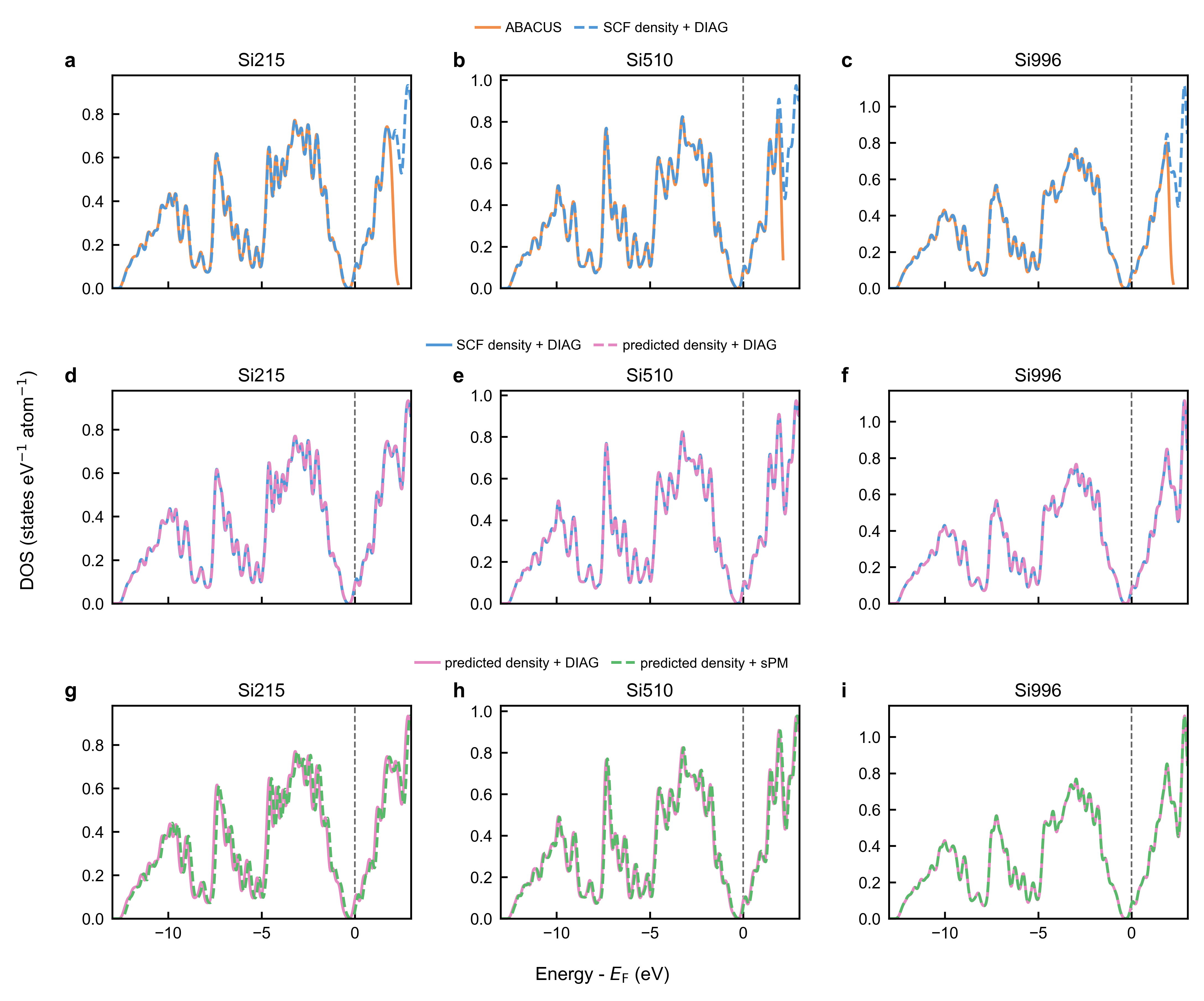}
    \caption{\textbf{Validation of ABPLaS--DIAG and sPM DOS from \ModelName{}-predicted charge densities.}
    \textbf{a--i}, DOS obtained using different charge densities and electronic-structure methods for vacancy-containing Si. The three columns correspond to Si215, Si510, and Si996, generated by removing $1$, $2$, and $4$ atoms from pristine Si216, Si512, and Si1000 supercells, respectively. \textbf{a--c}, Comparison between the ABACUS DOS obtained from its self-consistent density and the ABPLaS--DIAG DOS obtained from the ABPLaS self-consistent density. The ABACUS DOS broadening parameter is set to $0.1$~eV. \textbf{d--f}, DIAG DOS obtained from the ABPLaS self-consistent and \ModelName{}-predicted densities. \textbf{g--i}, DIAG and sPM DOS obtained from the \ModelName{}-predicted density, with the sPM curves averaged over $256$ stochastic samples. All DOS curves are normalized by the atom number. Each energy zero is the Fermi energy obtained by integrating the corresponding DOS to the target electron count. In each row, solid and dashed curves denote the two calculations listed in the legend.}
    \label{fig:vac_dos_abplas}
\end{figure}

\begin{figure}[H]
    \centering
    \includegraphics[keepaspectratio,width=0.98\linewidth]{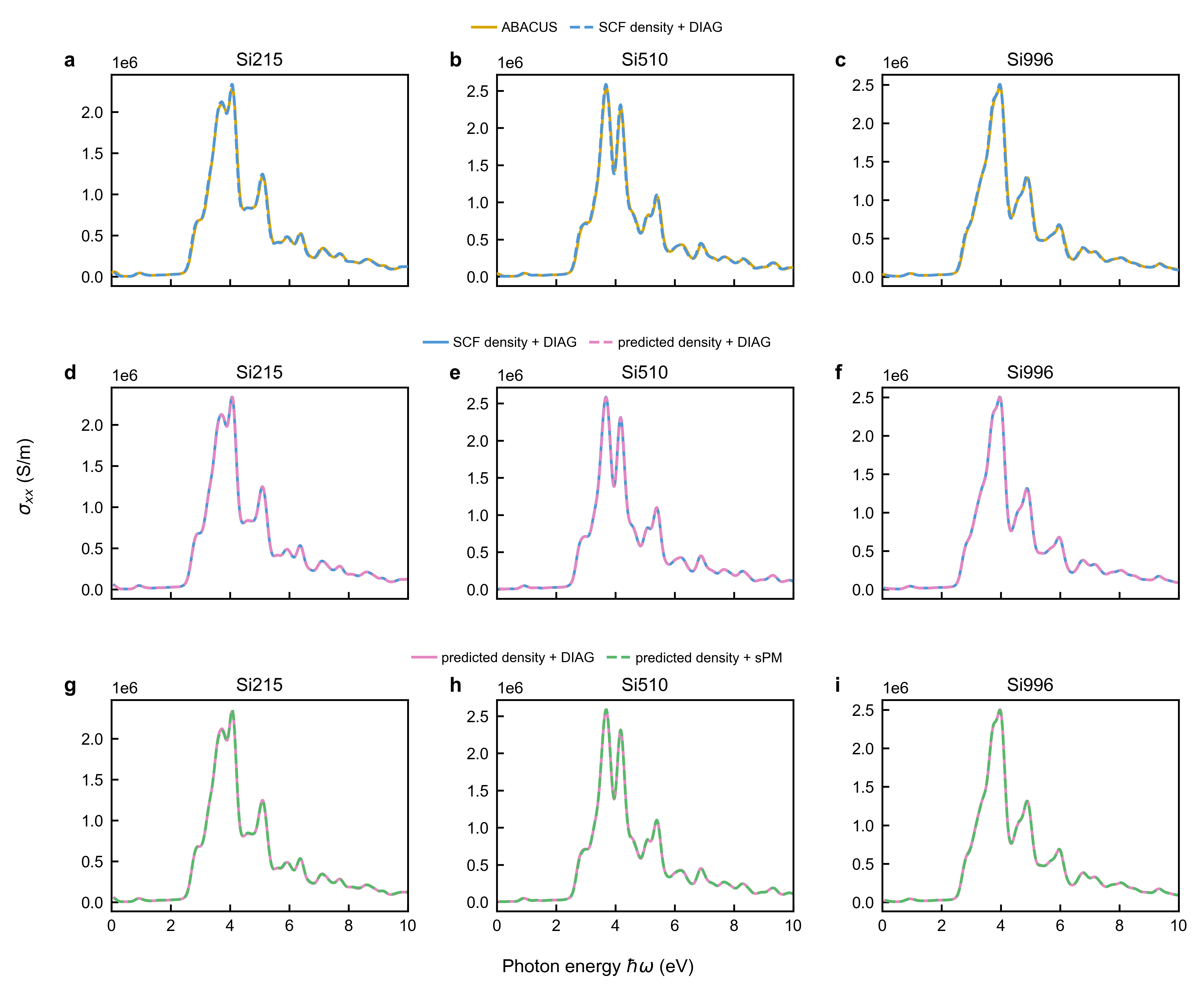}
    \caption{\textbf{Validation of ABPLaS--DIAG and sPM optical conductivity from \ModelName{}-predicted charge densities.}
    \textbf{a--i}, Real $xx$ optical conductivity, $\sigma_{xx}$, obtained using different charge densities and electronic-structure methods for vacancy-containing Si. The columns correspond to Si215, Si510, and Si996, generated by removing $1$, $2$, and $4$ atoms from pristine Si216, Si512, and Si1000 supercells, respectively. \textbf{a--c}, Comparison between the ABACUS--PYATB result and the ABPLaS--DIAG result obtained from its self-consistent density. The ABACUS--PYATB input spectra use Lorentzian broadening of $\eta=0.005$~eV, followed by convolution with a Gaussian kernel of standard deviation $0.10$~eV truncated at five standard deviations. Convolution uses the full energy grid before selecting the plotted range. ABPLaS conductivities are converted from internal units to $\mathrm{S\,m^{-1}}$. \textbf{d--f}, DIAG $\sigma_{xx}$ obtained from the ABPLaS self-consistent and \ModelName{}-predicted densities. \textbf{g--i}, DIAG and sPM $\sigma_{xx}$ obtained from the predicted density, with the sPM curves averaged over $128$ stochastic samples. Horizontal axes show photon energy $\hbar\omega$ from $0$ to $10$~eV. Conductivity is expressed in $\mathrm{S\,m^{-1}}$. In each row, solid and dashed curves denote the calculations listed in the legend.}
    \label{fig:vac_ac_abplas}
\end{figure}

The calculations in the following 4 supplementary figures (Figs.~\ref{fig:vac_dos_abacus}, \ref{fig:vac_ac_abacus}, \ref{fig:vac_dos_abplas}, and \ref{fig:vac_ac_abplas}) use a $k$-point spacing of approximately $0.1\,\mathrm{Bohr}^{-1}$. This corresponds to $3\times3\times3$, $2\times2\times2$, and $2\times2\times2$ $k$-point meshes for structures derived from Si216, Si512, and Si1000 parent supercells, respectively.

\clearpage
\pdfbookmark[2]{Fig. S7 -- ABACUS DOS comparisons}{sm:figure7}

\clearpage
\pdfbookmark[2]{Fig. S8 -- ABACUS optical-response comparisons}{sm:figure8}

\clearpage
\pdfbookmark[2]{Fig. S9 -- ABPLaS DOS validation}{sm:figure9}

\clearpage
\pdfbookmark[2]{Fig. S10 -- ABPLaS optical-conductivity validation}{sm:figure10}

\clearpage
\pdfbookmark[2]{Fig. S11 -- Stochastic DOS convergence}{sm:figure11}
\begin{figure}[H]
    \centering
    \includegraphics[keepaspectratio,width=0.98\linewidth]{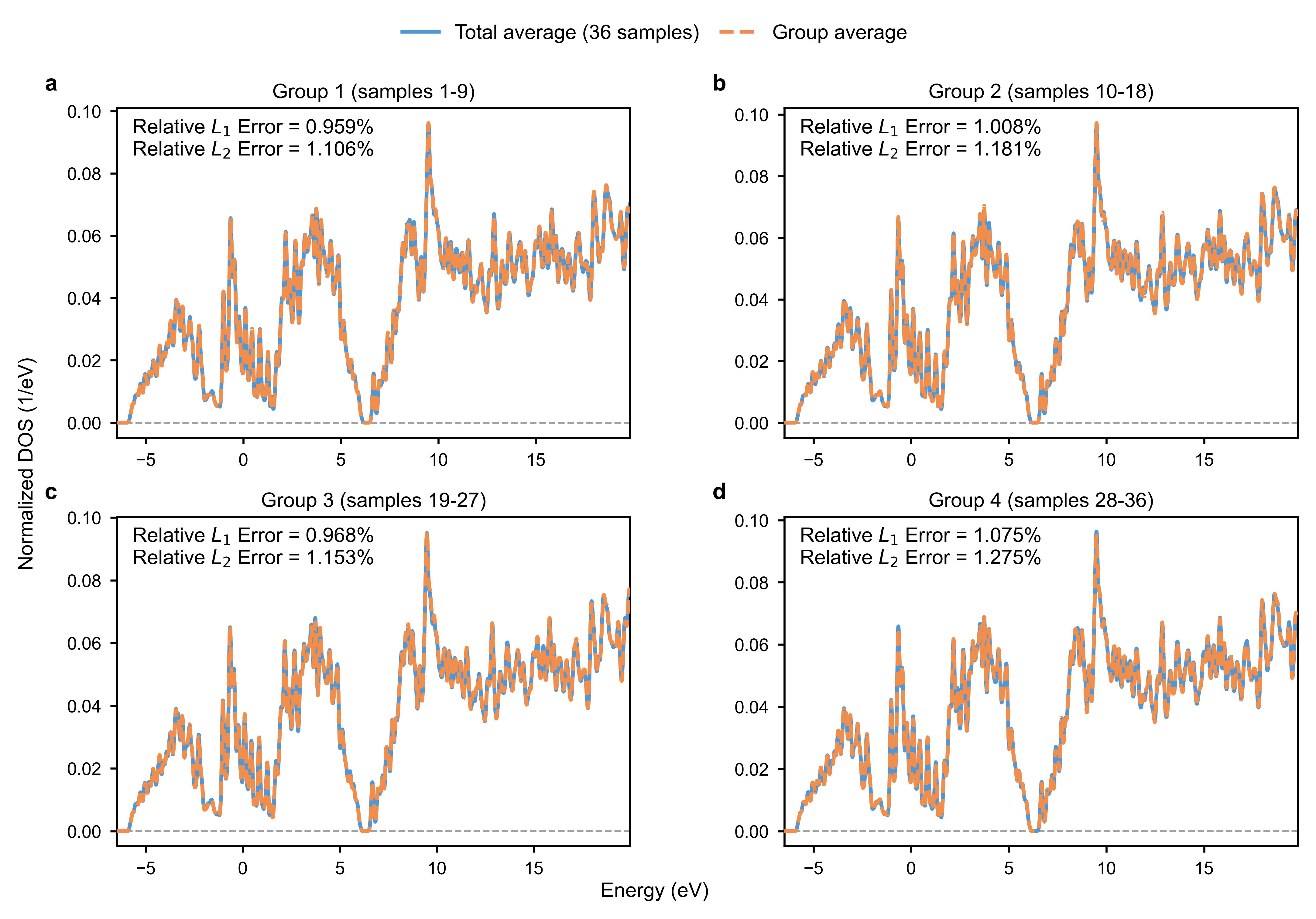}
    \caption{\textbf{Convergence of the DOS averaged over nine stochastic samples.}
    For a Si structure obtained by removing $64$ atoms from a $13{,}824$-site supercell, leaving $13{,}760$ atoms, $36$ stochastic states are generated using different random seeds, and a DOS is calculated from each state using sPM, with a single $k$ point used for each calculation. \textbf{a--d}, Four groups contain samples $1$--$9$, $10$--$18$, $19$--$27$, and $28$--$36$, respectively. Dashed curves show the group-averaged DOS, while solid curves show the average over all $36$ samples as an internal convergence reference. Within each panel's comparison energy window $W_k$, which is also its plotted range, the two curves are separately normalized to unit integral. The close agreement indicates that nine samples achieve the DOS statistical accuracy required here relative to this reference. The error measures are defined below.}
    \label{fig:abplas_dos_sample}
\end{figure}

\begingroup\small
\noindent\textbf{Error measures for Fig.~\ref{fig:abplas_dos_sample}.}
Denoting the normalized nine-sample and 36-sample averages by
$\widetilde D_{9}^{(k)}(E_i)$ and $\widetilde D_{36}^{(k)}(E_i)$,
respectively, the errors in window $W_k$ are
\begin{align*}
L_1^{(k)}
&=\sum_{E_i\in W_k}
\left|\widetilde D_{9}^{(k)}(E_i)-\widetilde D_{36}^{(k)}(E_i)\right|\Delta E,
\\[0.3em]
L_2^{(k)}
&=\left[
\frac{\displaystyle\sum_{E_i\in W_k}
\left(\widetilde D_{9}^{(k)}(E_i)-\widetilde D_{36}^{(k)}(E_i)\right)^2}
{\displaystyle\sum_{E_i\in W_k}
\left(\widetilde D_{36}^{(k)}(E_i)\right)^2}
\right]^{1/2}.
\end{align*}
\endgroup

\clearpage
\pdfbookmark[1]{Supplementary tables}{sm:tables}
\section*{Supplementary tables}

\pdfbookmark[2]{Table S1 -- Reported model comparisons}{sm:table1}
\begin{table}[H]
\centering
\begin{threeparttable}
\caption{Density-prediction errors reported for different models.}
\label{tab:comparing}
\small
\setlength{\tabcolsep}{4pt}
\renewcommand{\arraystretch}{1.25}
\begin{tabularx}{\linewidth}{@{}p{0.20\linewidth}*{4}{>{\centering\arraybackslash}X}@{}}
\toprule
Dataset & \ModelName{} & ChargE3Net & EAC-Net & DeepCharge \\
\midrule
Materials Project
& $0.21\%$--$1.89\%$
& \makecell{$0.523\pm0.010\%$\tnote{1}}
& $0.39\%$--$1.83\%$
& --- \\
\addlinespace[0.5em]
Bulk Si
& \makecell{$0.032\%$\\$(0.000064\,e/\text{\AA}^{3})$}
& --- & $\approx0.2\%$ & $0.000489\,e/\text{\AA}^{3}$ \\
\addlinespace[0.5em]
Amorphous Si
& \makecell{$0.448\%$\\$(0.000877\,e/\text{\AA}^{3})$}
& --- & $\approx0.8\%$ & $0.004017\,e/\text{\AA}^{3}$ \\
\bottomrule
\end{tabularx}
\begin{tablenotes}[flushleft]
\footnotesize
\item[1] The Materials Project result for ChargE3Net is the mean normalized density error, with one standard error.
\item[] Reported results can differ because of dataset versions, training/validation splits, and preprocessing. For bulk and amorphous Si, structure-generation procedures and DFT settings also differ between studies. These values therefore provide contextual comparisons rather than a benchmark under matched conditions.
\end{tablenotes}
\end{threeparttable}
\end{table}

\pdfbookmark[2]{Table S2 -- Crystalline model settings}{sm:table2}
\begin{table}[H]
\centering
\begin{threeparttable}
\caption{Model configurations and training settings for crystalline C, Si, Ge, and Sn.}
\label{tab:Training setups 1}
\small
\setlength{\tabcolsep}{4pt}
\renewcommand{\arraystretch}{1.14}
\begin{tabularx}{\linewidth}{@{}p{0.27\linewidth}*{4}{>{\centering\arraybackslash}X}@{}}
\toprule
Parameter & C & Si & Ge & Sn \\
\midrule
\multicolumn{5}{@{}l}{\textit{Model and representation}}\\[0.25em]
Atom--atom cutoff (Bohr) & 5.2 & 8.1 & 8.2 & 10.1 \\
Atom--probe cutoff (Bohr) & 5.2 & 8.1 & 8.2 & 10.1 \\
Strictly local & False & False & False & False \\
$N_{\mathrm{norm}}$ & By type & By type & By type & By type \\
Layers & 3 & 3 & 3 & 3 \\
\makecell[l]{Irreducible\\representations}
& \makecell{$26\times0\mathrm{e}+9\times1\mathrm{o}$\\$+5\times2\mathrm{e}+4\times3\mathrm{o}$\\$+3\times4\mathrm{e}$}
& \makecell{$26\times0\mathrm{e}+9\times1\mathrm{o}$\\$+5\times2\mathrm{e}+4\times3\mathrm{o}$\\$+3\times4\mathrm{e}$}
& \makecell{$26\times0\mathrm{e}+9\times1\mathrm{o}$\\$+5\times2\mathrm{e}+4\times3\mathrm{o}$\\$+3\times4\mathrm{e}$}
& \makecell{$26\times0\mathrm{e}+9\times1\mathrm{o}$\\$+5\times2\mathrm{e}+4\times3\mathrm{o}$\\$+3\times4\mathrm{e}$} \\
Residual update & True & True & True & True \\
Number of parameters & 523,534 & 525,582 & 530,190 & 534,798 \\
\addlinespace[0.65em]
\multicolumn{5}{@{}l}{\textit{Density targets and probe sampling}}\\[0.25em]
Target scale factor & 0.3 & 0.1 & 1.0 & 1.0 \\
Uniform sampling fraction & 0.8 & 0.8 & 0.9 & 0.9 \\
Gaussian sampling fraction & 0.2 & 0.2 & 0.1 & 0.1 \\
Gaussian sampling $\sigma$ & 10.0 & 10.0 & 10.0 & 10.0 \\
Gaussian sampling $\epsilon$ & 0.05 & 0.05 & 0.05 & 0.05 \\
Probes per structure & 5000 & 5000 & 5000 & 5000 \\
\addlinespace[0.65em]
\multicolumn{5}{@{}l}{\textit{Optimization}}\\[0.25em]
Epochs & 20,000 & 20,000 & 20,000 & 20,000 \\
Batch size & 32 & 32 & 32 & 32 \\
Learning rate & 0.005 & 0.005 & 0.005 & 0.005 \\
Learning-rate scheduler
& \texttt{ExponentialLR} & \texttt{ExponentialLR}
& \texttt{ExponentialLR} & \texttt{ExponentialLR} \\
Decay $\beta$ / factor & 0.99977 & 0.99977 & 0.99977 & 0.99977 \\
Patience & None & None & None & None \\
Loss function & $L_1$ loss & $L_1$ loss & $L_1$ loss & $L_1$ loss \\
\bottomrule
\end{tabularx}
\begin{tablenotes}[flushleft]
\footnotesize
\item[] Density targets are divided by the target scale factor to improve training stability. The same definition applies to Tables~\ref{tab:Training setups 2} and \ref{tab:Training setups 3}.
\end{tablenotes}
\end{threeparttable}
\end{table}

\clearpage

\pdfbookmark[2]{Table S3 -- Amorphous and vacancy model settings}{sm:table3}
\begin{table}[H]
\centering
\begin{threeparttable}
\caption{Model configurations and staged training settings for amorphous and vacancy-containing Si.}
\label{tab:Training setups 2}
\small
\setlength{\tabcolsep}{4pt}
\renewcommand{\arraystretch}{1.20}
\begin{tabularx}{\linewidth}{@{}p{0.27\linewidth}*{4}{>{\centering\arraybackslash}X}@{}}
\toprule
& \multicolumn{2}{c}{Amorphous Si} & \multicolumn{2}{c}{Vacancy-containing Si} \\
\cmidrule(lr){2-3}\cmidrule(l){4-5}
Parameter & Stage 1 & Stage 2 & Stage 1 & Stage 2 \\
\midrule
\multicolumn{5}{@{}l}{\textit{Model and representation}}\\[0.25em]
Atom--atom cutoff (Bohr) & 10.6 & 10.6 & 7.6 & 7.6 \\
Atom--probe cutoff (Bohr) & 10.6 & 10.6 & 7.6 & 7.6 \\
Strictly local & False & False & True & True \\
$N_{\mathrm{norm}}$ & By $N_{\mathrm{edges}}$ & By $N_{\mathrm{edges}}$ & By $N_{\mathrm{edges}}$ & By $N_{\mathrm{edges}}$ \\
Layers & 3 & 3 & 5 & 5 \\
\makecell[l]{Irreducible\\representations}
& \makecell{$26\times0\mathrm{e}+9\times1\mathrm{o}$\\$+5\times2\mathrm{e}+4\times3\mathrm{o}$\\$+3\times4\mathrm{e}$}
& \makecell{$26\times0\mathrm{e}+9\times1\mathrm{o}$\\$+5\times2\mathrm{e}+4\times3\mathrm{o}$\\$+3\times4\mathrm{e}$}
& \makecell{$13\times0\mathrm{e}+4\times1\mathrm{o}$\\$+3\times2\mathrm{e}+2\times3\mathrm{o}$\\$+1\times4\mathrm{e}$}
& \makecell{$13\times0\mathrm{e}+4\times1\mathrm{o}$\\$+3\times2\mathrm{e}+2\times3\mathrm{o}$\\$+1\times4\mathrm{e}$} \\
Residual update & True & True & True & True \\
Number of parameters & 525,582 & 525,582 & 608,301 & 608,301 \\
\addlinespace[0.65em]
\multicolumn{5}{@{}l}{\textit{Density targets and probe sampling}}\\[0.25em]
Target scale factor & 0.1 & 0.1 & 0.1 & 0.1 \\
Uniform sampling fraction & 0.8 & 0.8 & 0.8 & 0.6 \\
Gaussian sampling fraction & 0.2 & 0.2 & 0.2 & --- \\
Gaussian sampling $\sigma$ & 10.0 & 10.0 & 10.0 & --- \\
Gaussian sampling $\epsilon$ & 0.05 & 0.05 & 0.05 & --- \\
Probes per structure & 10,240 & 5120 & 10,240 & 10,240 \\
\addlinespace[0.65em]
\multicolumn{5}{@{}l}{\textit{Optimization}}\\[0.25em]
Epochs & 100 & 2000 & 10,000 & 10,000 \\
Batch size & 2 & 10 & 20 & 20 \\
Learning rate & 0.01 & 0.005 & 0.01 & 0.0001 \\
Learning-rate scheduler
& \makecell{\texttt{ReduceLR}\\\texttt{OnPlateau}}
& \makecell{\texttt{ReduceLR}\\\texttt{OnPlateau}}
& \texttt{ExponentialLR} & \texttt{ExponentialLR} \\
Decay $\beta$ / factor & 0.950 & 0.990 & 0.99915 & 0.99961 \\
Patience & 5 & 5 & None & None \\
Loss function & $L_1$ loss & $L_1$ loss & $L_1$ loss & $L_1$ loss \\
\bottomrule
\end{tabularx}
\end{threeparttable}
\end{table}

\clearpage

\pdfbookmark[2]{Table S4 -- Materials Project model settings}{sm:table4}
\begin{table}[H]
\centering
\begin{threeparttable}
\caption{Model configuration and staged training settings for the Materials Project dataset.}
\label{tab:Training setups 3}
\small
\setlength{\tabcolsep}{6pt}
\renewcommand{\arraystretch}{1.23}
\begin{tabularx}{\linewidth}{@{}p{0.36\linewidth}*{2}{>{\centering\arraybackslash}X}@{}}
\toprule
Parameter & Stage 1 & Stage 2 \\
\midrule
\multicolumn{3}{@{}l}{\textit{Model and representation}}\\[0.25em]
Atom--atom cutoff (Bohr) & 7.6 & 7.6 \\
Atom--probe cutoff (Bohr) & 7.6 & 7.6 \\
Strictly local & False & False \\
$N_{\mathrm{norm}}$ & By $N_{\mathrm{edges}}$ & By $N_{\mathrm{edges}}$ \\
Layers & 3 & 3 \\
Irreducible representations
& \makecell{$102\times0\mathrm{e}+34\times1\mathrm{o}+20\times2\mathrm{e}$\\$+15\times3\mathrm{o}+11\times4\mathrm{e}$}
& \makecell{$102\times0\mathrm{e}+34\times1\mathrm{o}+20\times2\mathrm{e}$\\$+15\times3\mathrm{o}+11\times4\mathrm{e}$} \\
Residual update & True & True \\
Number of parameters & 8,139,592 & 8,139,592 \\
\addlinespace[0.65em]
\multicolumn{3}{@{}l}{\textit{Density targets and probe sampling}}\\[0.25em]
Target scale factor & 1.0 & 1.0 \\
Uniform sampling fraction & 1.0 & 1.0 \\
Gaussian sampling fraction & 0.0 & 0.0 \\
Gaussian sampling $\sigma$ & --- & --- \\
Gaussian sampling $\epsilon$ & --- & --- \\
Probes per structure & 32 & 32 \\
\addlinespace[0.65em]
\multicolumn{3}{@{}l}{\textit{Optimization}}\\[0.25em]
Epochs & 200 & 600 \\
Batch size & 64 & 256 \\
Learning rate & 0.005 & 0.001 \\
Learning-rate scheduler & \texttt{ReduceLROnPlateau} & \texttt{ReduceLROnPlateau} \\
Decay $\beta$ / factor & 0.950 & 0.980 \\
Patience & 5 & 3 \\
Loss function & $L_1$ loss & $L_1$ loss \\
\bottomrule
\end{tabularx}
\end{threeparttable}
\end{table}

%% file: context/Introduction.tex
Local atomic environments have been shown to be an effective basis of machine learning in atomistic simulations~\cite{2019-Carleo-RMP,2022-roadmap}. Models encoding chemical species and geometry within a finite spatial range have been developed for interatomic potentials~\cite{2007-BPNN,18PRL-Zhang,2022-batzner}, electron densities~\cite{2021-lewis,2024-koker-higher}, and electronic Hamiltonians~\cite{2022-deeph,2023-zhong-transferable,2026-deephr}. However, efficient local predictions do not provide directly the spectra and transport properties of large and structurally complex systems.

The density of states (DOS), optical response, and charge transport depend on the electronic states and dynamics of the complete system. In materials with thermal disorder or dilute defects, these properties can depend on both local perturbations and their spatial arrangement, motivating large supercells that capture the relevant structural and propagation length scales. Traditional Kohn--Sham density functional theory (KS-DFT) becomes prohibitively expensive at this scale, as the solution to the orbital eigenvalue problem and the orthogonalization of the orbitals typically have asymptotically cubic cost in the system size~\cite{1999-Goedecker,2012-bowler-review,2019-lin-numerical}. An efficient route must therefore address both the generation of electronic input and the subsequent evaluation of global observables. Previous work has demonstrated large-scale electronic-property calculations by combining machine-learned tight-binding Hamiltonians with stochastic propagation~\cite{2024-gu-deeptb}.

Learning the electron density in real space provides an alternate route to the self-consistent-field (SCF) approach. The predicted density, together with a given ionic structure, pseudopotentials, and a density-dependent exchange-correlation approximation, determine the Hartree and exchange-correlation potentials in the Kohn-Sham Hamiltonian. Density as a real-space field bridges the gap between reference-data generation and property evaluation in different numerical representations. Existing density-learning methods have shown improved SCF initialization and property calculations from a fixed predicted density~\cite{2023-lv-deep,2024-koker-higher,2025-li-resnet}. In the latter case, density inference replaces SCF iterations of the target system and the resulting fixed Hamiltonian can support multiple observables without property-specific retraining.

Linear-scaling methods offer ways to obtain electronic properties without resorting to full diagonalization, and there are several families of such methods~\cite{1999-Goedecker,2012-bowler-review,2026-zhou-review}. Stochastic propagation methods (sPM) evaluate spectral and dynamical response through correlation functions of propagated random states~\cite{2010-yuan-graphene,2023-tbplas,2026-fu-large,2026-zhou-review}. Fourier transformation of the corresponding time correlations yields spectra across the sampled energy or frequency range without repeating the propagation for each spectral point. Propagation under the full-system Hamiltonian allows electronic motion beyond the spatial range of the learned descriptors. For sparse Hamiltonians with bounded connectivity, the leading Hamiltonian--vector operations scale linearly with system size at fixed spectral resolution and number of random states.

Generating full real-space density fields at this scale also poses an additional inference challenge. Density-learning approaches include models that predict expansion coefficients in an auxiliary atom-centered basis~\cite{2021-lewis} and models that directly evaluate the density at spatial query points~\cite{2022-jorgensen-equivariant,2024-koker-higher}. Within the latter class, heterogeneous atom--grid architectures such as EAC-Net separate atomic representation learning from spatial decoding, allowing equivariant atomic descriptors to be reused across many grid points~\cite{2025-qin-eac}. When the atomic and probe working sets exceed GPU memory, spatial blocking and query batching limit the size of each inference task. In a single-level blocking scheme, one block size controls both the atomic context and the probe batch: smaller blocks reduce decoding memory but increase repeated atomic encoding, whereas larger blocks reduce this redundancy at a higher memory cost. Decoupling these two sizes can therefore improve descriptor reuse while controlling the memory required for probe decoding.

Here, we link local charge-density learning to global electronic properties by using memory-efficient density inference and stochastic propagation at the Kohn-Sham DFT level. We combine the equivariant atom--grid density model SCALE-Charge with the real-space DFT code ABPLaS~\cite{2023-zhou-time} and its stochastic propagation solver~\cite{2026-fu-large}. We propose hierarchical atom-probe blocking that allows independent control of atomic context and probe batch sizes, avoiding repetitive atomic encoding and improving the time-memory trade-off of dense-grid inference. Predicted densities correspond to fixed Kohn-Sham Hamiltonians, which are used without target-system self-consistent field iterations, and their spectral and transport properties are computed via stochastic propagation without explicit diagonalization. We present calculations of DOS on million atom Si supercells and optical, quasi-eigenstate and transport calculations on the hundred-thousand atom scale. In Si with randomly distributed vacancies, pronounced defect-related spectral weight coexists with strongly inhomogeneous state distributions and suppressed diffusion, showing that spectral prominence need not translate into efficient charge transport.

%% file: context/model.tex
\begin{figure*}[t]
    \centering
    \includegraphics[width=\textwidth]{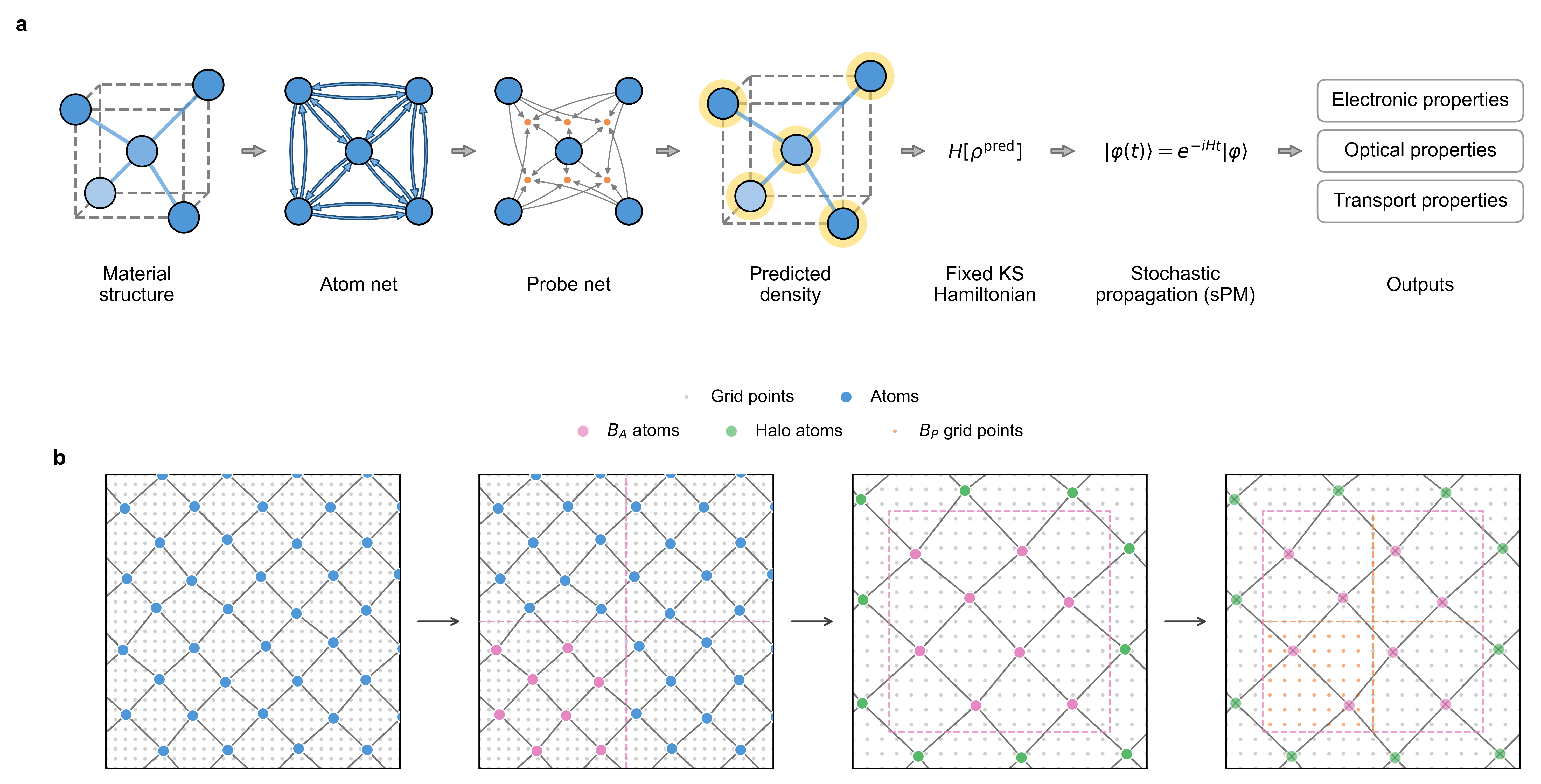}
    \caption{%
\textbf{From local charge-density learning to global electronic properties in large-scale systems.}
\textbf{a}, Overall workflow. An equivariant atom network encodes local chemical environments into reusable atomic descriptors. An atom--probe decoder combines these descriptors with spatial query positions to assemble the real-space charge-density field. The complete predicted density is used to construct a fixed Kohn--Sham Hamiltonian. Stochastic propagation evaluates the global properties over the complete supercell.
\textbf{b}, Hierarchical atom--probe blocking and descriptor reuse. The target grid is partitioned into outer blocks with nominal edge length $B_A$. For each outer block, the atoms contributing to its probes and the complete surrounding neighborhoods required for their descriptors are gathered. Atomic encoding is performed once, and the required final descriptors are cached. Inner probe blocks with nominal edge length $B_P$ are then decoded sequentially using the same cache. Gray points denote grid points, blue points denote atoms, pink points identify atoms whose descriptors are required for decoding, green points denote additional halo atoms, and orange points mark the active probe subblock.
}
    \label{fig:scale_framework}
\end{figure*}

Figure~\ref{fig:scale_framework}a illustrates the procedure how SCALE-Charge provides the real space density to build a fixed Kohn--Sham Hamiltonian in ABPLaS~\cite{2023-zhou-time}, the stochastic propagation of which is used to calculate global electronic properties~\cite{2026-fu-large}. This portion details how reusable atomic representations and hierarchical blocking build the complete density field with bounded GPU working memory and minimal repeated atomic computation.

\subsection{Heterogeneous atom--grid representation}

SCALE-Charge uses a heterogeneous graph with atom and probe nodes. Atom--atom connections supply the chemical environments used to construct equivariant descriptors, while atom--probe connections couple these descriptors to spatial query points. Atom--grid architectures such as EAC-Net exploit this separation to reuse atomic representations during spatial decoding~\cite{2025-qin-eac}. For atomic species $\{Z_i\}$, coordinates $\{\mathbf{R}_i\}$, and probe positions $\{\mathbf{r}_p\}$, the mapping is
\begin{equation}
\begin{aligned}
\{\mathbf{n}_i^{(T)}\}
&=
\mathcal{A}\!\left(\{Z_i,\mathbf{R}_i\}\right),\\
\hat{\rho}_p
&=
\mathcal{D}\!\left(
\left\{
\left(\mathbf{n}_i^{(T)},\mathbf{r}_p-\mathbf{R}_i\right)
\right\}_{i\in\mathcal{N}_{\mathrm{pa}}(p)}
\right).
\end{aligned}
\end{equation}
Here, $\mathcal{A}$ is the atom encoder, $\mathbf{n}_i^{(T)}$ is the descriptor after $T$ atom layers, and $\mathcal{D}$ is the probe decoder. Atomic encoding uses neighborhoods defined by the atom--atom cutoff $R_{\mathrm{aa}}$, while $\mathcal{N}_{\mathrm{pa}}(p)$ contains atoms within the probe--atom cutoff $R_{\mathrm{pa}}$. Periodic images are treated consistently in the neighbor lists and relative displacement vectors.

The atomic descriptors depend on the atomic environment but are independent of the selected probe batch. Once constructed for a given atomic context, they can therefore be reused across multiple batches without repeating the encoding. The decoder combines these equivariant features with probe--atom distances and directions to produce a scalar density invariant under a joint rotation of the structure and query point. Atomic encoding and probe decoding can thus be scheduled with different working-set sizes, providing the basis for hierarchical inference. Here, $\hat{\rho}_p$ denotes the network prediction before normalization of the complete density field to the target electron number.

Unless otherwise stated, the models use equivariant features up to $L_{\max}=4$, with the angular-order accuracy--cost assessment provided in the Supplementary Materials (SM). Detailed feature updates, scalar readout, and training settings are given in Methods and SM. The implementation supports recursive message passing and a separately trained strictly local variant.
The models used in Sections 3.1–3.4 employ recursive message passing, except for the vacancy-containing Si model introduced in Section 3.3, which uses the strictly local variant of SCALE-Charge. This model is reused for the corresponding property benchmarks in Section 3.4 and the large-scale calculations in Section 3.5.

\subsection{Hierarchical atom--probe blocking}

For sufficiently large systems, processing all atomic features and probe points simultaneously exceeds the available GPU memory. Spatial blocking limits the working set, but a single-level scheme couples atomic reuse to probe-batch size. Smaller blocks reduce decoding memory while increasing repeated encoding of overlapping atomic environments and their halos. Larger blocks reduce this redundancy but require more memory for probe decoding.

Hierarchical atom--probe blocking separates these choices (Fig.~\ref{fig:scale_framework}b). We denote the nominal edge lengths of the outer target-grid blocks and inner probe blocks by $B_A$ and $B_P$, respectively, both measured in grid points. In particular, $B_A$ specifies the outer target region rather than the full extent of its gathered atomic context. For each outer block, we collect the atoms contributing to its probes together with the complete dependency domains required to construct their descriptors. These domains include the neighborhoods needed by initialization and subsequent encoder layers. The atom encoder is evaluated once, and the final descriptors required for decoding are cached. The target probes are then partitioned into disjoint inner blocks and decoded sequentially using the same cache. Probe intermediates are released between inner blocks, while the atomic cache is retained until the outer block is complete. Single-level blocking is recovered when each outer block is decoded as one probe batch, with $B_A=B_P=B$.

A schematic decomposition of the dominant feature-memory contributions is
\begin{equation}
M_{\mathrm{peak}}^{\mathrm{feat}}(B_A,B_P)
\simeq
\max\!\left\{
M_{\mathrm{atom}}(B_A),
M_{\mathrm{cache}}(B_A)+M_{\mathrm{probe}}(B_P)
\right\}.
\end{equation}
Here, $M_{\mathrm{atom}}$ is the peak feature-memory requirement during atomic encoding, $M_{\mathrm{cache}}$ is the storage occupied by the retained atomic descriptors, and $M_{\mathrm{probe}}$ is the additional feature workspace required to decode one inner block. The measured GPU allocations reported below also include model parameters, graph data, and other runtime allocations. In particular, the measured probe-stage peak already includes the atomic cache and should not be identified with $M_{\mathrm{probe}}$ alone.

Reducing $B_P$ lowers the decoding workspace and can make room for a larger $B_A$, allowing greater descriptor reuse. The two sizes must nevertheless be chosen jointly: very small probe batches can reduce GPU utilization and increase batching overhead, while larger atomic contexts require more encoding and cache memory.

Hierarchical blocking reorganizes atomic encoding and probe decoding while retaining the complete atomic dependencies required by every decoded probe. The computational savings arise from descriptor reuse and independent control of the two working sets, without truncating the model's required atomic environments.

After all blocks have been processed, the raw predictions are assembled into a complete density field and normalized once to the target electron number. The downstream Hamiltonian is constructed over the complete supercell and retains couplings across inference-block boundaries. This scheduling principle applies to heterogeneous architectures with bounded atomic dependency domains, query-independent atomic descriptors, and independently decodable probes.

\begin{figure}
    \centering
    \includegraphics[keepaspectratio,width=1.0\textwidth]{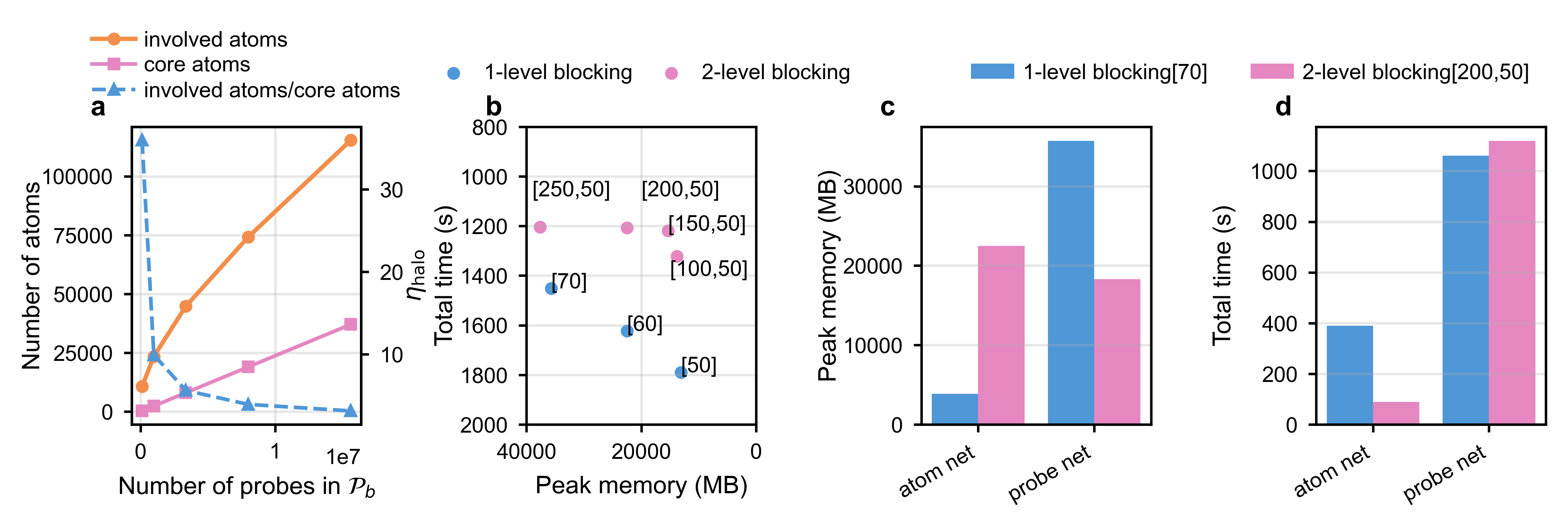}
    \caption{%
\textbf{Hierarchical atom--probe blocking improves the time--memory trade-off of real-space density inference.}
\textbf{a}, Estimated core-atom count $N_{\mathrm{core}}$ and total involved-atom count $N_{\mathrm{involved}}$ versus the number of probes in the outer block. The core count is estimated by multiplying the total atom count by the volume fraction occupied by the outer target block. The involved count is averaged over full-sized outer blocks and includes the complete atomic context required for descriptor construction. Atom counts and their ratio $\eta_{\mathrm{halo}}=N_{\mathrm{involved}}/N_{\mathrm{core}}$ are shown at the same time.
\textbf{b}, Total inference time, which includes atomic encoding and probe decoding, versus peak allocated GPU memory for single-level blocking (blue) and hierarchical blocking (pink). Labels $[B]$ specify the single-level block size, while $[B_A,B_P]$ specify the outer target-grid and inner probe block sizes.
\textbf{c,d}, Peak allocated GPU memory (\textbf{c}) and cumulative runtime (\textbf{d}) of atomic encoding and probe decoding for single-level blocking with $B=70$ and hierarchical blocking with $(B_A,B_P)=(200,50)$. Stage runtimes are summed over all calls covering the complete target grid, whereas stage memory values are the maximum allocations recorded during those calls. The measured allocations include model parameters, graph data, and other runtime allocations; the probe-stage peak also includes the retained atomic descriptors. The two stage peaks are not additive.
All measurements concern density inference and exclude Hamiltonian construction and subsequent property calculations. All benchmarks use a diamond-C supercell containing $10^6$ atoms and a uniform $750\times750\times750$ target grid, comprising 421,875,000 grid points.
}
    \label{fig:model_valid}
\end{figure}

\subsection{Time--memory trade-offs and inference scaling}

Fig.~\ref{fig:model_valid} compares the inference time and peak allocated GPU memory of single-level and hierarchical blocking. Stage runtimes are accumulated over all calls covering the complete target grid, while stage memory values are the largest recorded allocations. The measurement scope and benchmark settings are specified in the caption and Methods.

Fig.~\ref{fig:model_valid}a quantifies the atomic-context overhead associated with the outer block. Both the estimated geometric-core atom count, $N_{\mathrm{core}}$, and the total involved-atom count, $N_{\mathrm{involved}}$, increase with block size, while their ratio, $\eta_{\mathrm{halo}}$, decreases. Larger outer blocks thus reduce the relative boundary overhead and the repeated atomic encoding required to cover the complete grid.

For the single-level configurations tested in Fig.~\ref{fig:model_valid}b, increasing the block size reduces inference time but raises peak memory. Varying $B_A$ and $B_P$ independently improves the measured time--memory Pareto frontier among the tested configurations. In particular, hierarchical blocking with $(B_A,B_P)=(200,50)$ reduces total inference time by approximately 16.8\% and peak allocated GPU memory by approximately 37.0\%. These gains are obtained for the same trained model and target grid while retaining the atomic dependencies required for prediction.

The stage decomposition in Fig.~\ref{fig:model_valid}c,d explains the simultaneous reduction in time and memory. Enlarging the outer block increases the atomic-stage memory peak, but the smaller inner blocks lower the probe-stage peak sufficiently to reduce the overall maximum. Atomic encoding consequently sets the memory peak for the hierarchical configuration. Meanwhile, its cumulative runtime falls to approximately one-quarter of the single-level value, while probe-decoding time increases only slightly. Probe decoding remains the dominant network stage, but the reduction in repeated atomic encoding outweighs the additional batching overhead.

At fixed atomic and grid densities, cutoff radii, and network configuration, the number of probes grows in proportion to the number of atoms, while the mean neighbor counts remain bounded. With fixed block sizes, the leading network-evaluation cost therefore scales linearly with system size. Provided that local atomic populations remain bounded, increasing the system size primarily increases the number of blocks rather than their working-set sizes, allowing approximately size-independent peak GPU working memory during blockwise network inference. This working-memory estimate excludes storage of global input data, the complete density field, and the downstream electronic-structure calculation. The large-system density-inference measurements in Fig.~\ref{fig:dos_large}b,c provide an empirical assessment of this behavior.
The following benchmarks assess the accuracy and transferability of the predicted real-space densities.

%% file: context/result_crystal.tex
The predicted density supplies the electronic input for subsequent property calculations. We first test the real-space accuracy in thermally perturbed crystals. All density and electronic-property benchmarks presented in Secs.~\ref{sec:density_accuracy}--\ref{sec:properties} are computed using recursive message-passing models, except for the vacancy-containing Si model introduced in Sec.~3.3, which is computed using a separately trained strictly local variant of SCALE-Charge. 
SCALE-Charge models for C, Si, Ge and Sn are separately trained from molecular-dynamics configurations of 64-atom supercells at 300~K. For each material we sample 100 training and 100 validation snapshots from disjoint parts of the same trajectory.
Reference densities are calculated using self-consistent DFT calculations with ABACUS~\cite{2025-zhou-abacus} with sampling and electronic-structure parameters described in Methods.
For each structure, we evaluate the integrated absolute density error normalized by the reference electron number,
\begin{equation}
\epsilon_{\mathrm{MAE}}
=
\frac{
\sum_{g\in\mathcal{G}}
\left|\rho_g^{\mathrm{pred}}-\rho_g^{\mathrm{DFT}}\right|
\Delta v_g
}{
\sum_{g\in\mathcal{G}}\rho_g^{\mathrm{DFT}}\Delta v_g
},
\label{eq:density_mae}
\end{equation}
where $\mathcal{G}$ is the complete real-space grid and $\Delta v_g$ is the volume element at grid point $g$. The denominator is the electron number represented by the reference density; on a uniform grid, the common volume element cancels. We report $\epsilon_{\mathrm{MAE}}$ as a percentage.

\begin{figure*}[t]
    \centering
    \includegraphics[width=\textwidth]{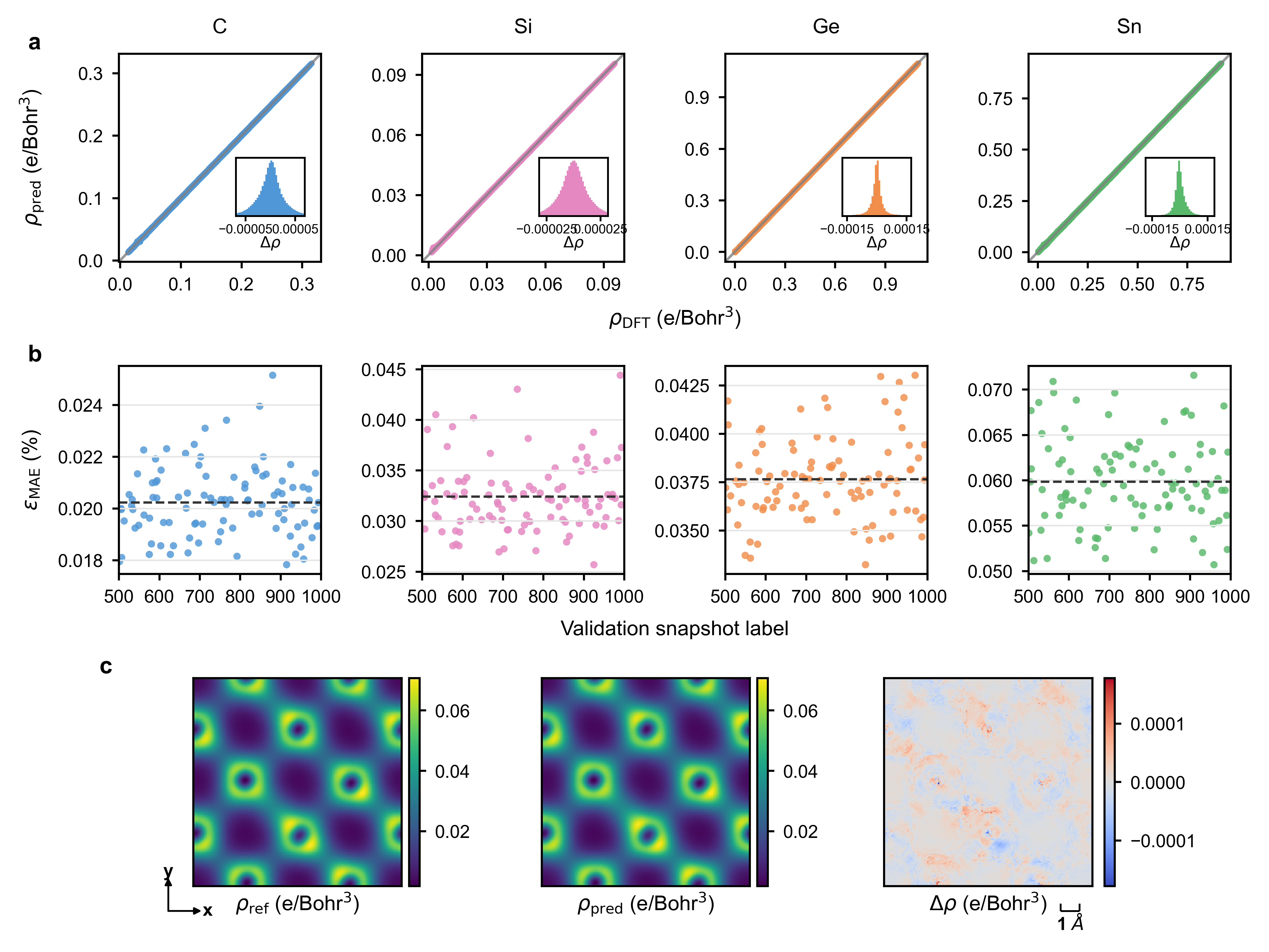}
    \caption{%
\textbf{Real-space charge-density accuracy in thermally perturbed crystals.}
Separate models are trained for C, Si, Ge, and Sn using 64-atom configurations at 300~K. Columns in \textbf{a--b} follow this material order.
\textbf{a}, Grid-point parity plots for the last snapshot in each validation set. Insets show signed residuals $\Delta\rho=\rho^{\mathrm{pred}}-\rho^{\mathrm{DFT}}$. Only the central residual ranges are displayed, while all grid points enter the complete-grid error.
\textbf{b}, Electron-number-normalized absolute density error $\epsilon_{\mathrm{MAE}}$ for 100 validation snapshots per material. The horizontal axis gives the original trajectory indices, and dashed lines indicate the means.
\textbf{c}, Reference density, predicted density, and signed difference for a representative Si slice. The reference and prediction share a color scale, while the difference has a separate scale centered at zero. Density values and residuals are in $e\,\mathrm{Bohr}^{-3}$, and the scale bar is 1~\AA.
}
    \label{fig:accuracy_crystal}
\end{figure*}

Across the 100 validation configurations, the mean errors are approximately $0.020\%$, $0.032\%$, $0.038\%$, and $0.060\%$ for C, Si, Ge, and Sn, respectively (Fig.~\ref{fig:accuracy_crystal}b). The representative grid-point comparisons in Fig.~\ref{fig:accuracy_crystal}a closely follow the reference densities, with signed residuals concentrated near zero. These results establish the accuracy of the separately trained models within the sampled thermal distributions and demonstrate the applicability of the same architecture to the four material-specific datasets.

In the representative Si slice shown in Fig.~\ref{fig:accuracy_crystal}c, the model reproduces the atomic and bonding features, with residual amplitudes approximately two to three orders of magnitude smaller than the displayed density variation.
Together, these tests establish a crystalline accuracy baseline for the material-specific models.

%% file: context/result_MP.tex
\begin{figure*}[t]
    \centering
    \includegraphics[width=\textwidth]{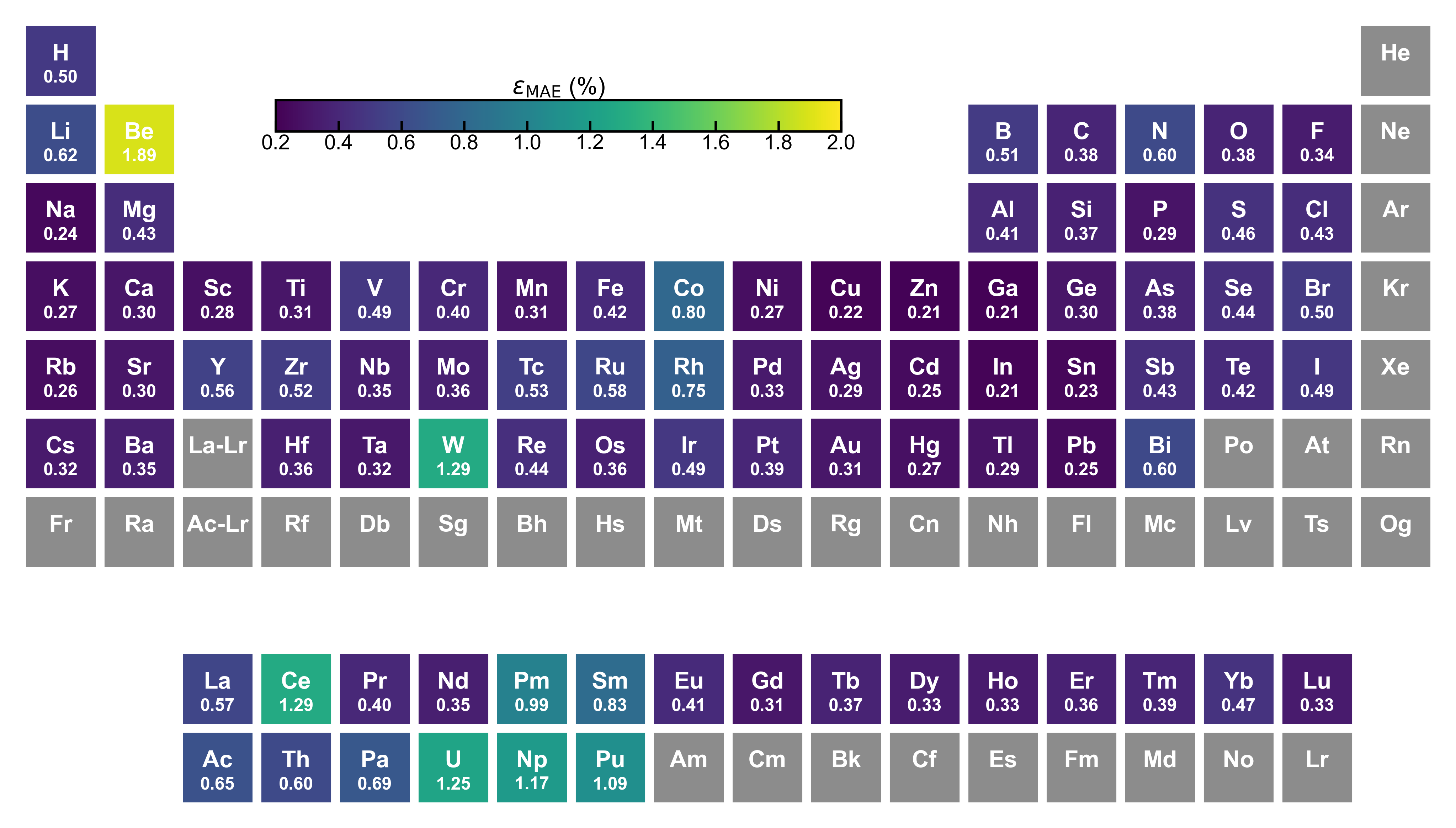}
    \caption{%
\textbf{Element-resolved density accuracy of a shared multielement model.}
Colors and numerical labels show normalized density errors (\%) for elements appearing in at least ten structures in the held-out Materials Project validation set. For each element, absolute density differences and reference density values are summed over validation probe points within 2~\AA\ of at least one atom of that species, and their ratio is reported. Within each structure, a point is counted once for that element even when multiple neighborhoods overlap. The resulting statistic characterizes density prediction in the sampled neighborhoods of that element. Gray cells indicate that no value is reported.
}
    \label{fig:generalization_table_acc}
\end{figure*}

Then, we evaluate the performance of SCALE-Charge in predicting densities for different chemical compositions. We train a single multielement model on the charge-density dataset of the Materials Project, which has been preprocessed in Ref.~\cite{2025-qin-eac}. The dataset contains 149,196 inorganic crystal structures, each with 5,000 randomly sampled real-space probe points. For training and validation we use a random $95\%:5\%$ structure level split. This test evaluates predictions on held-out structures within the chemical coverage of the dataset.

Fig.~\ref{fig:generalization_table_acc} reports element-resolved errors obtained by accumulating absolute density differences and reference densities over sampled points within 2~\AA\ of atoms of each species. These local, element-resolved errors differ from the complete-grid, structure-level errors used in the material-specific benchmarks. Most displayed elements have errors of approximately $0.2\%$--$0.8\%$, with a smaller group reaching roughly $1.2\%$--$1.9\%$. The larger values include Be, W, Ce, and U, while low errors occur across several chemical families. Elements appearing in fewer than ten validation structures are excluded from the element-resolved analysis to limit uncertainty from insufficient sampling.

\begin{figure*}[t]
    \centering
    \includegraphics[width=\textwidth]{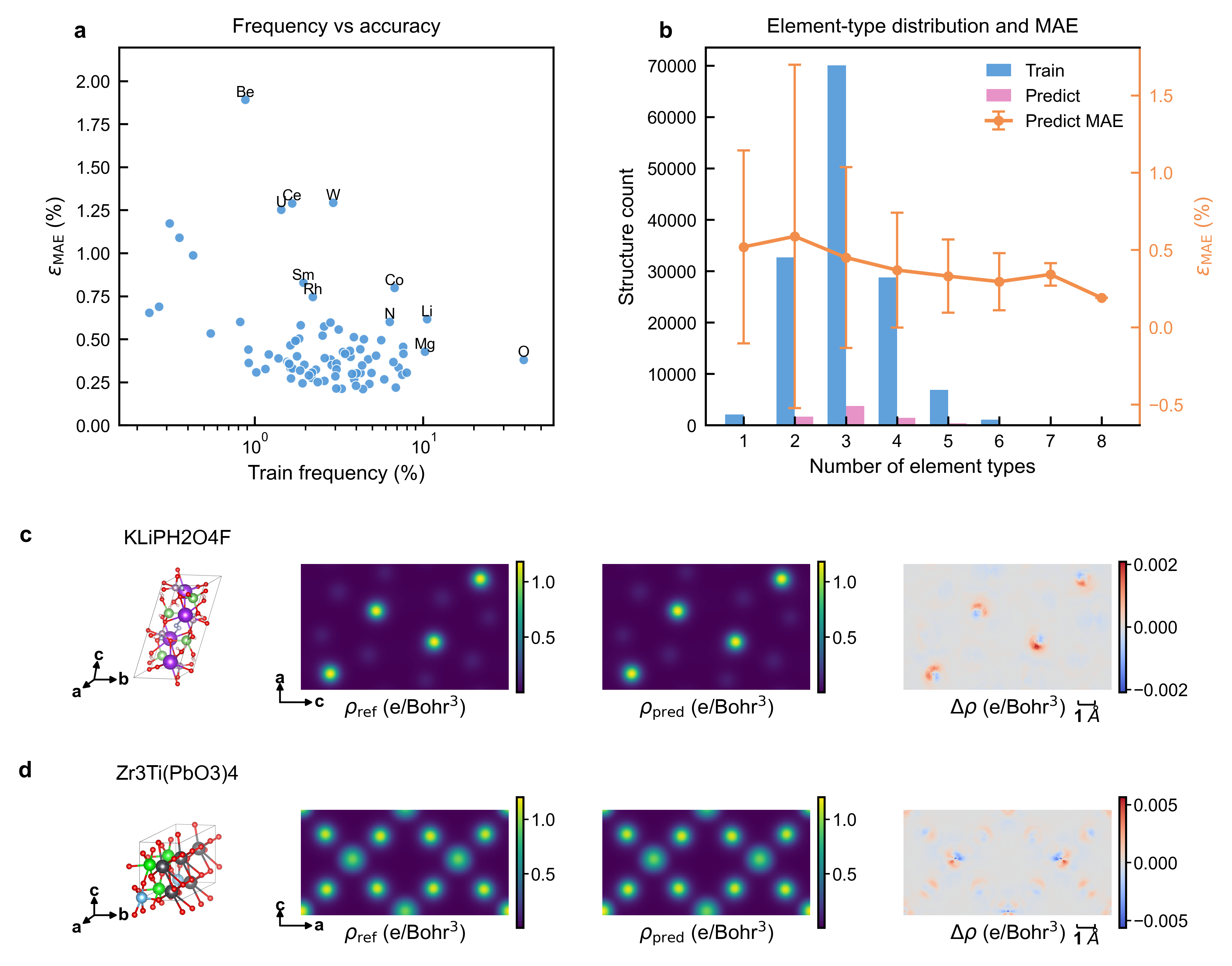}
    \caption{%
\textbf{Chemical coverage, compositional complexity, and real-space density reconstruction.}
\textbf{a}, Element-resolved normalized density error versus the fraction of training structures containing each element, irrespective of the number of atoms of that species in a structure. The same model and element set as in Fig.~\ref{fig:generalization_table_acc} are used.
\textbf{b}, Numbers of training and validation structures grouped by the number of distinct constituent elements, shown by blue and pink bars on the left axis. Both counts exclude structures containing any element omitted from the selected element set, namely elements appearing in fewer than 10 structures in the complete validation set. For each retained validation structure, the normalized density error is calculated over its 5,000 evaluation grid points as $\epsilon_{\mathrm{MAE}}=100\%\times\sum_p|\rho_p^{\mathrm{pred}}-\rho_p^{\mathrm{DFT}}|/\sum_p|\rho_p^{\mathrm{DFT}}|$. The orange curve shows the arithmetic mean of these structure-level errors within each group on the right axis, with error bars indicating one standard deviation of the structure-level errors within that group.
\textbf{c,d}, Representative predictions for $\mathrm{KLiPH_2O_4F}$ and $\mathrm{Zr_3Ti(PbO_3)_4}$, containing six and four elements, respectively. Each row shows the crystal structure, reference density, predicted density, and signed difference.
}
    \label{fig:generalization_analysis}
\end{figure*}

Fig.~\ref{fig:generalization_analysis}a relates these errors to the fraction of training structures containing each element. Larger errors occur mainly among less frequently represented elements, although some rare elements also have low errors and elements with comparable frequencies can differ substantially in accuracy. Occurrence frequency therefore provides only a partial measure of the training coverage relevant to prediction, which also depends on the diversity of local chemical environments.

Fig.~\ref{fig:generalization_analysis}b examines the dependence on compositional complexity. Ternary structures form the largest group, followed by substantial binary and quaternary populations, and the training and validation distributions are similar. The mean structure-level density MAE does not increase monotonically with the number of constituent elements. The groups with the most elements contain relatively few structures, limiting the statistical weight of their averages, but the displayed results show no systematic increase in error with compositional complexity over the sampled dataset.

Figs.~\ref{fig:generalization_analysis}c,d illustrate real-space predictions for $\mathrm{KLiPH_2O_4F}$ and $\mathrm{Zr_3Ti(PbO_3)_4}$, containing six and four elements, respectively. The shared model reproduces the principal atomic and bonding features and the lower-density regions of the reference fields. The displayed residual ranges are two to three orders of magnitude smaller than the corresponding density ranges. Together, these tests demonstrate the applicability of a shared density representation across the chemical environments covered by the dataset.

%% file: context/result_defect.tex
\begin{figure*}[t]
    \centering
    \includegraphics[width=\textwidth]{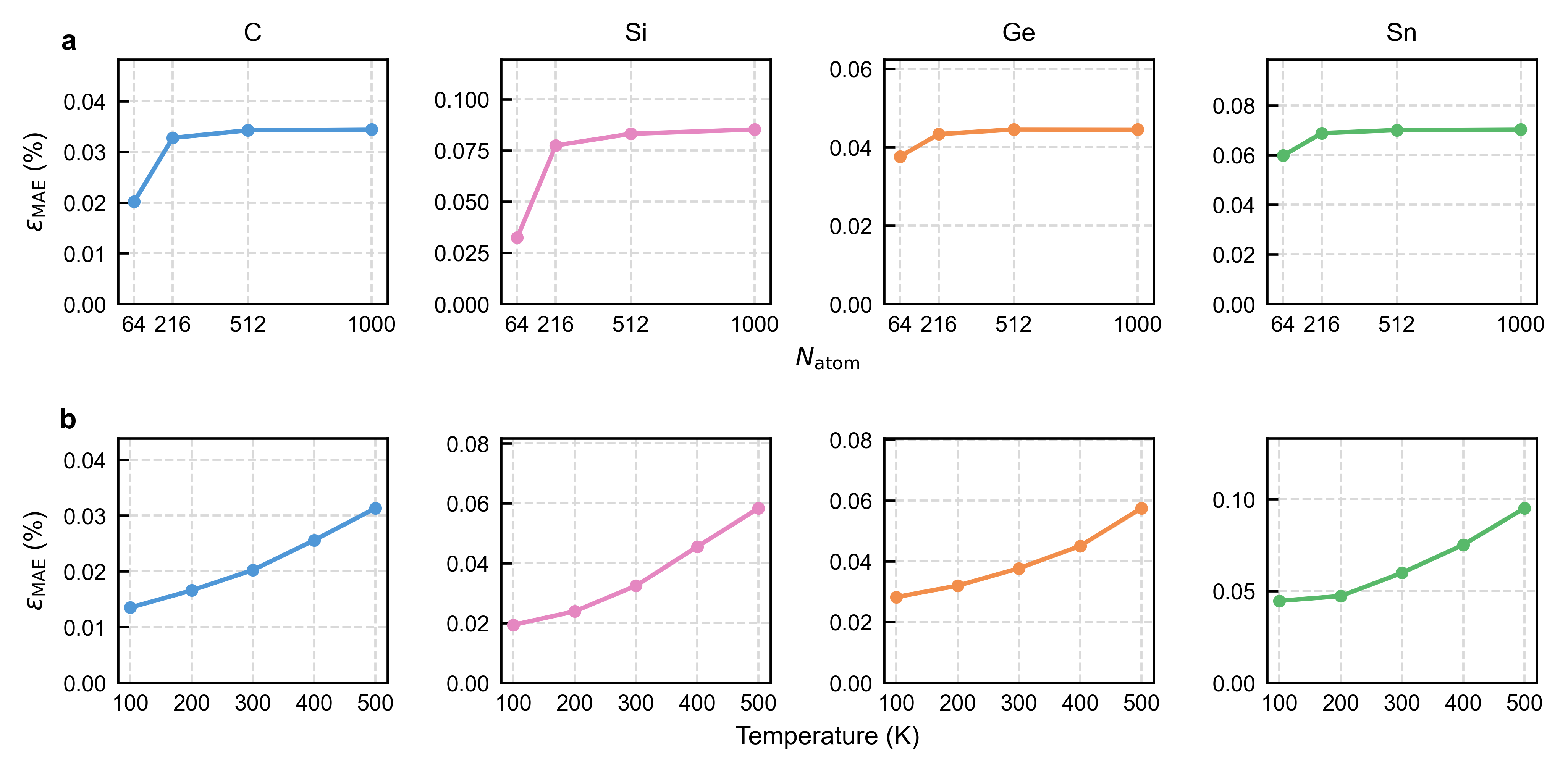}
    \caption{%
\textbf{Size and temperature transferability.}
\textbf{a}, Complete-grid prediction error versus atom number for C, Si, Ge, and Sn, shown from left to right. For each material, a model trained on 64-atom configurations at 300~K is evaluated on 100 separate 64-atom validation configurations. The same model is then applied without retraining to 216-, 512-, and 1000-atom supercells, with 10 validation configurations at each larger size.
\textbf{b}, Mean prediction errors for 64-atom supercells at 100, 200, 300, 400, and 500~K, evaluated using 100 validation configurations per material at each temperature. The same material-specific models trained at 300~K are used throughout, with the material ordering as in \textbf{a}.
Each points represents the arithmetic mean of the configuration-level electron-number-normalized absolute density errors $\epsilon_{\mathrm{MAE}}$.
}
    \label{fig:transferability_size_temp}
\end{figure*}
\begin{figure*}[t]
    \centering
    \includegraphics[width=\textwidth]{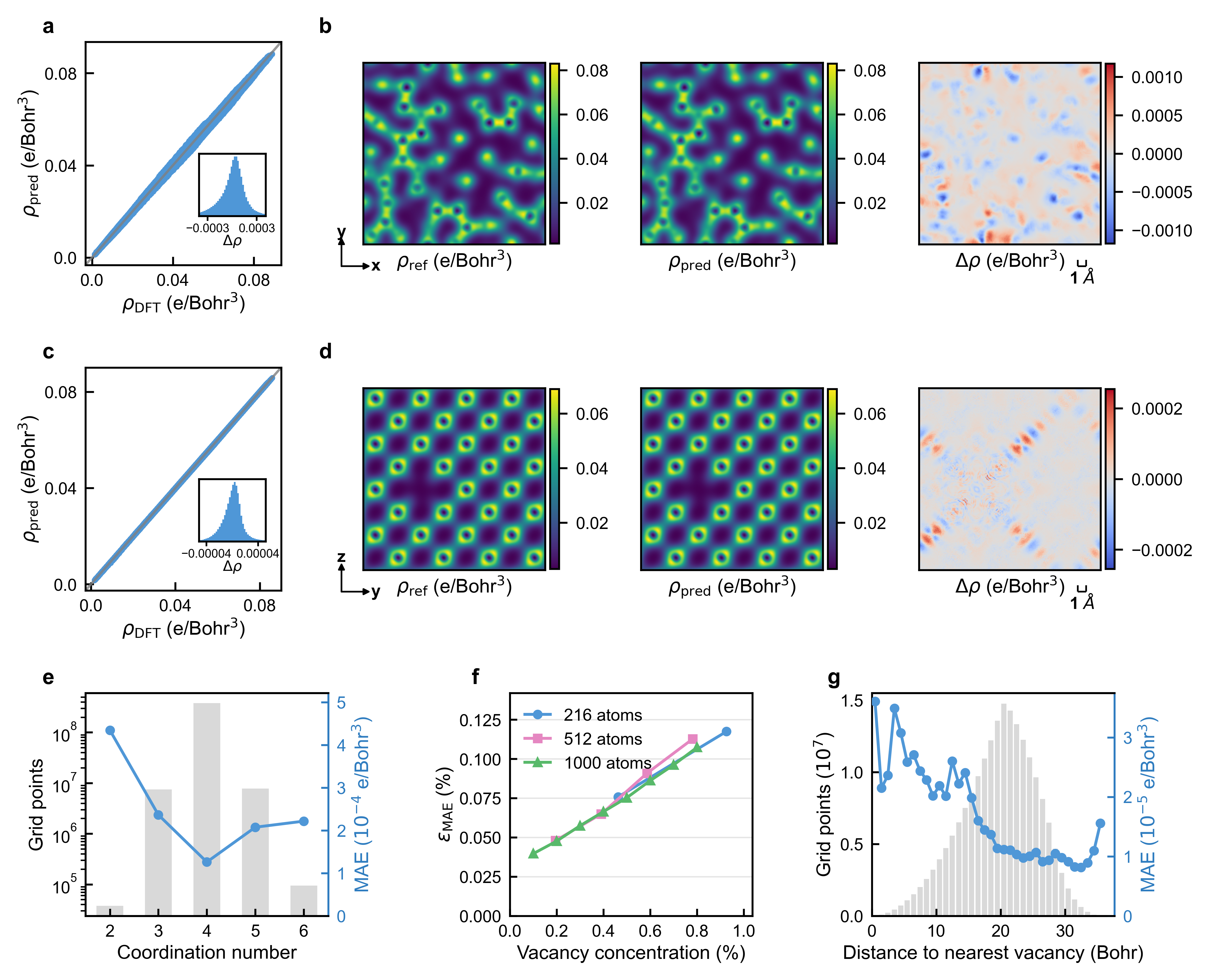}
    \caption{%
\textbf{Density prediction in amorphous and vacancy-containing Si.}
Separate models are trained for the two structural classes. The 512-site vacancy validation set contains 40 held-out relaxed structures, ten for each vacancy number from one to four.
\textbf{a,b}, Grid-point parity plot and density slices for a representative amorphous Si validation structure.
\textbf{c,d}, Corresponding results for a representative single-vacancy structure selected from the 512-site validation set. Insets in \textbf{a,c} show signed residuals $\Delta\rho=\rho^{\mathrm{pred}}-\rho^{\mathrm{DFT}}$. Slices in \textbf{b,d} show reference, predicted, and difference fields from left to right. Matched density fields share a color scale, and residual scales are centered at zero. Scale bars denote 1~\AA.
\textbf{e}, Mean absolute grid-point error grouped by coordination number across the 20 amorphous validation structures, with each grid point assigned to its nearest atom.
\textbf{f}, Normalized density error versus vacancy fraction for 216-, 512-, and 1000-site parent cells with 1--2, 1--4, and 1--8 vacancies, respectively. Each point represents the arithmetic mean over 10 validation configurations for the corresponding parent-cell size and vacancy count, giving 20, 40, and 80 evaluated configurations for the three sizes, respectively. Sizes refer to the parent crystals before atom removal, and vacancy fractions use the corresponding pristine-site counts.
\textbf{g}, Mean absolute grid-point error versus distance to the nearest vacancy center across ten single-vacancy validation structures derived from 512-site parent cells. Within each distance bin, absolute errors are averaged over all grid points pooled from these structures. In \textbf{e,g}, blue curves show the mean absolute grid-point error on the right axes, while gray bars show grid-point counts on the left axes. The count axis is logarithmic in \textbf{e} and scaled by \(10^7\) in \textbf{g}. Panel \textbf{f} reports $\epsilon_{\mathrm{MAE}}$ as a percentage. Densities and residuals are in $e\,\mathrm{Bohr}^{-3}$, while local errors in \textbf{e,g} carry factors of $10^{-4}$ and $10^{-5}$, respectively. Distances in \textbf{g} are in Bohr.
}
    \label{fig:transferability_defect}
\end{figure*}

After evaluating the shared model over chemical compositions, we then explore the size transferability of material-specific models and their performance in different structural environments. First, we apply the crystalline models of Sec.~\ref{sec:density_accuracy} to larger supercells and different temperatures without retraining (Fig.~\ref{fig:transferability_size_temp}). Separate models are then trained for amorphous and vacancy-containing Si to assess density prediction in disordered and defect-perturbed environments (Fig.~\ref{fig:transferability_defect}).

Models trained on 64-atom crystalline configurations retain low errors when applied directly to 216-, 512-, and 1000-atom supercells (Fig.~\ref{fig:transferability_size_temp}a). At 1000 atoms, more than fifteen times the training size, the mean normalized errors remain approximately $0.03\%$--$0.09\%$. An initial increase occurs between 64 and 216 atoms, particularly for Si, while further enlargement produces only weak changes. The results support the transfer of a learned local structure--density relation to larger crystalline configurations with similar local environments.

The same models, trained at 300~K, also retain low errors across 100--500~K (Fig.~\ref{fig:transferability_size_temp}b). At 500~K, the mean errors are approximately $0.03\%$, $0.06\%$, $0.06\%$, and $0.10\%$ for C, Si, Ge, and Sn, respectively, with smaller errors at lower temperatures. Their gradual increase is consistent with the broader distribution of local distortions at higher temperatures. Because temperature is not a model input, this test assesses the learned relation under changes in the atomic-configuration distribution.

For amorphous Si, a separate model is trained on 80 independently generated 512-atom structures and evaluated on 20 held-out structures prepared using the same melt--quench protocol. The mean normalized density error is approximately $0.45\%$, and the representative parity plot and slices reproduce the principal density features of the disordered network (Fig.~\ref{fig:transferability_defect}a,b). Coordination-resolved errors provide a more local assessment (Fig.~\ref{fig:transferability_defect}e). Four-coordinate environments account for most grid points and have the lowest mean absolute error, while the less common two-, three-, five-, and six-coordinate environments have higher errors. These differences identify local environments that are more challenging for the model and may reflect both their structural diversity and their coverage in the training data.

For vacancy-containing Si, a separate model is trained on relaxed and thermally perturbed structures derived from 512-site parent cells with one to four vacancies; data preparation and training details are given in Methods. The representative density comparison captures both the crystalline background and the variations around a relaxed vacancy, with larger residuals near the defect and neighboring bonds (Fig.~\ref{fig:transferability_defect}c,d). Across structures derived from 216-, 512-, and 1000-site parent cells, complete-grid errors follow similar trends at comparable vacancy fractions, increasing from approximately $0.04\%$ at $0.1\%$ vacancies to approximately $0.12\%$ near $0.93\%$ vacancies (Fig.~\ref{fig:transferability_defect}f). This correspondence supports transfer of the vacancy-trained model across the tested cell sizes and dilute vacancy concentrations.

Fig.~\ref{fig:transferability_defect}g complements the complete-grid averages with a spatially resolved assessment. Across ten single-vacancy validation structures derived from 512-site parent cells, bin-averaged absolute errors near vacancies reach approximately $(2$--$4)\times10^{-5}\,e\,\mathrm{Bohr}^{-3}$, while more distant regions typically have errors of order $10^{-5}\,e\,\mathrm{Bohr}^{-3}$. The higher errors are thus concentrated around defect-perturbed environments. Their spatial distribution suggests that the increase in error with vacancy concentration may reflect a larger proportion of grid points in these more challenging environments. Together, these material-specific tests establish density accuracy across the sampled crystalline, disordered, and defective structures.

%% file: context/result_properties.tex
\begin{figure*}[t]
    \centering
    \includegraphics[width=\textwidth]{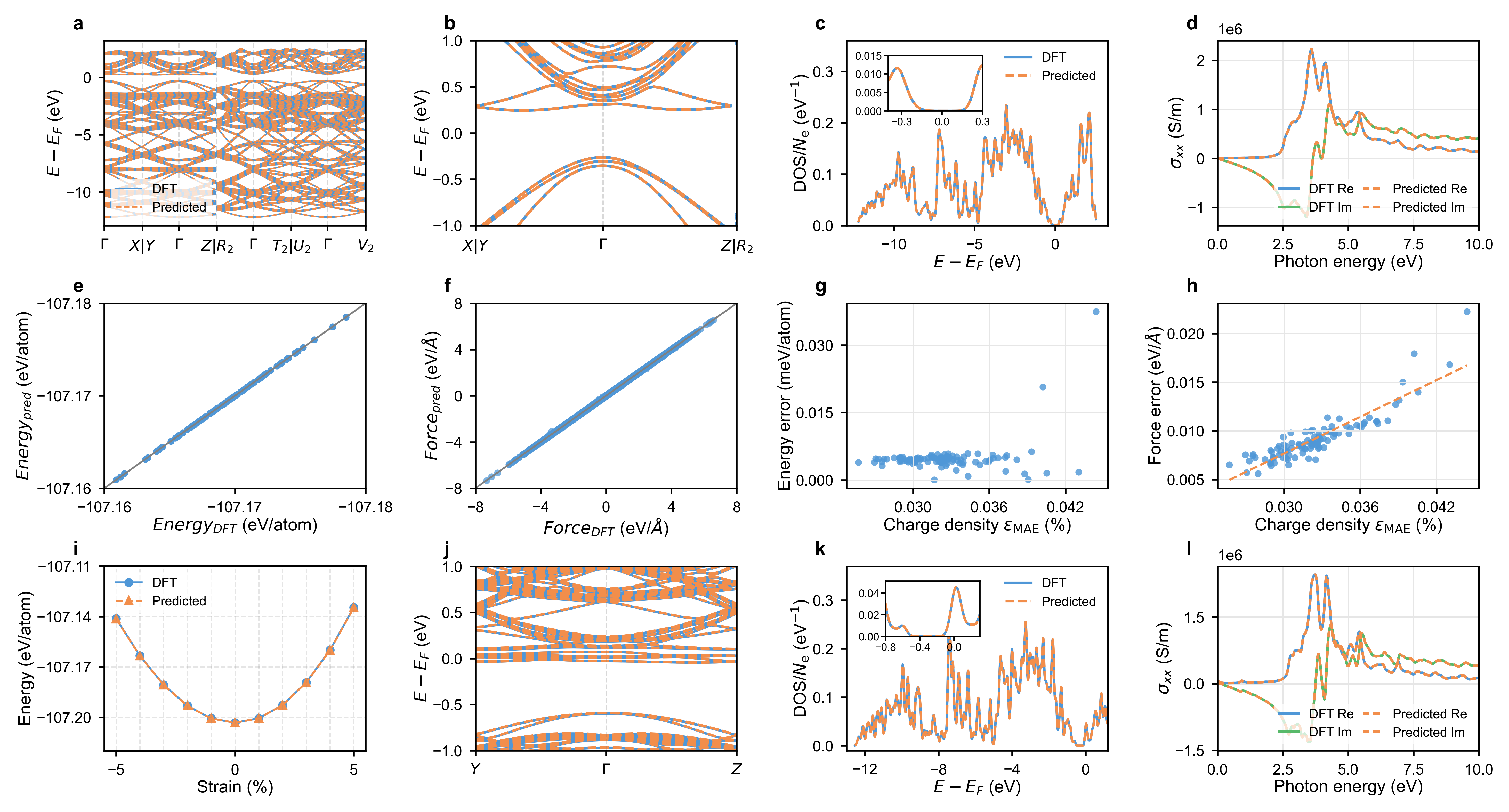}
    \caption{
    \textbf{Electronic-structure and optical properties from predicted densities for crystalline and vacancy-containing Si.}
    Property evaluations using predicted and self-consistent reference densities are compared under matched settings. No energy, force, or spectral labels are used to train the density models.
    \textbf{a,b}, Energy bands and a near-edge enlargement around $\Gamma$ for a 64-atom crystalline Si configuration at 300~K. Solid blue and dashed orange curves denote reference-density and predicted-density results.
    \textbf{c}, DOS averaged over 100 snapshots at 300~K. Before averaging, individual spectra are aligned to their corresponding Fermi energies. The inset enlarges the DOS near the Fermi level.
    \textbf{d}, Real and imaginary parts of the $xx$ optical conductivity for a representative crystalline Si configuration. Blue and green solid curves show the real and imaginary components obtained from the reference density, while orange dashed curves show the corresponding components obtained from the predicted density.
    \textbf{e,f}, Total energies per atom for 100 thermal snapshots at 300~K and Cartesian force components, respectively.
    \textbf{g,h}, Energy and force errors defined in Eq.~\eqref{eq:property_errors} versus normalized density error $\epsilon_{\mathrm{MAE}}$. Each point represents one configuration; the dashed line in \textbf{h} is a linear fit.
    \textbf{i}, Energy per atom under in-plane shear-like strain, imposed by changing the angle between the in-plane lattice vectors while keeping their lengths fixed. The strain is defined by a $-5\%$ to $5\%$ relative change in the [110] lattice-diagonal length.
    \textbf{j--l}, Corresponding spectral comparisons for a representative 510-atom Si structure obtained by removing two atoms from a 512-site parent supercell. 
    Panels \textbf{a--i} use the density model trained on 64-atom crystalline Si configurations at 300~K, whereas panels \textbf{j--l} use the density model trained on vacancy-containing Si configurations.
    }
    \label{fig:properties}
\end{figure*}

We next assess the accuracy of electronic-structure properties calculated from the predicted densities. None of these properties is used as a training label. For each crystalline Si configuration in Fig.~\ref{fig:properties}a--i, the electron-number-normalized prediction from the crystalline Si model of Sec.~\ref{sec:density_accuracy} is supplied to ABACUS for fixed-density property evaluation without further SCF refinement. Reference properties are evaluated from the converged self-consistent density using the same procedure. The ionic structure, exchange--correlation functional, pseudopotentials, orbital basis, and numerical settings are matched, isolating the effect of replacing the density input.

Figs.~\ref{fig:properties}a,b show that the predicted-density calculation closely reproduces the reference Kohn--Sham band dispersions for a representative configuration at 300~K, including the near-edge structure around $\Gamma$. Fig.~\ref{fig:properties}c compares DOS averaged over 100 thermal snapshots at 300~K using the model trained at the same temperature. Before averaging, individual spectra are aligned to their Fermi energies and divided by their electron numbers, giving an approximately unit integral below the Fermi level. The averaged curves retain the principal peak positions and relative intensities, including the low-DOS region near the Fermi level. Supplementary Fig.~S5\ignore{Fig.~\ref{supp-fig:si_dos_temp}} shows the corresponding comparisons at 100, 200, 400, and 500~K. Together, these results show that the predicted densities preserve the main spectral features both at the training temperature and across the other tested thermal ensembles.

Fig.~\ref{fig:properties}d extends the fixed-density comparison to the $xx$ optical conductivity for the representative Si configuration. Both the real and imaginary parts obtained from the predicted density closely follow the reference spectra over the plotted photon-energy range, reproducing the principal peak positions, amplitudes, and overall spectral shapes. This agreement shows that the predicted density also provides a suitable input for evaluating the optical response of this configuration.

Total energies per atom closely follow the reference values across 100 thermal snapshots at 300~K (Fig.~\ref{fig:properties}e), while Cartesian force components lie near the parity line over the sampled range of approximately $-8$ to $8$~eV/\AA\ (Fig.~\ref{fig:properties}f). To quantify these deviations and examine their relation to density accuracy, we define the configuration-level errors
\begin{equation}
\begin{aligned}
\delta E
&=
\frac{\left|E^{\mathrm{pred}}-E^{\mathrm{DFT}}\right|}{N_a},\\
\delta F
&=
\frac{1}{3N_a}
\sum_{i=1}^{N_a}\sum_{\alpha=x,y,z}
\left|F_{i\alpha}^{\mathrm{pred}}-F_{i\alpha}^{\mathrm{DFT}}\right|,
\end{aligned}
\label{eq:property_errors}
\end{equation}
where $E$ is the total energy and $N_a$ is the number of atoms. Figs.~\ref{fig:properties}g,h relate these errors to the normalized density error $\epsilon_{\mathrm{MAE}}$. For most configurations, $\delta E$ remains below approximately $0.01$~meV/atom, with a maximum of about $0.04$~meV/atom and only a weak dependence on density error over the sampled range. The force-component MAE shows a clearer positive correlation with $\epsilon_{\mathrm{MAE}}$, following an approximately linear trend with additional scatter, while remaining below approximately $0.023$~eV/\AA\ throughout the plotted set.

This contrast is qualitatively consistent with the stationary character of a variationally consistent DFT energy at self-consistency. At fixed atomic geometry and electron number, its first-order variation vanishes for small admissible electronic variations~\cite{2019-lin-numerical}. Atomic forces involve derivatives with respect to ionic positions and do not generally share this stationarity with respect to electronic variations. The observed trends are therefore consistent with greater sensitivity of force errors to the electronic input, although the density-error norm alone does not determine either property error.

Fig.~\ref{fig:properties}i examines an in-plane shear-like deformation parameterized by a $-5\%$ to $5\%$ change in the [110] lattice-diagonal length. The predicted-density energy curve follows the reference throughout this interval without retraining, reproducing the minimum near the undeformed structure and the energy increase on either side. Although no macroscopic strain was deliberately imposed when generating the training configurations, the model thus retains the energy response along this deformation path. Corresponding band, DOS, energy, and force comparisons for C, Ge, and Sn are provided in Supplementary Fig.~S6\ignore{Fig.~\ref{supp-fig:nscf_CGeSn}}.

Figs.~\ref{fig:properties}j--l extend these comparisons to a representative vacancy-containing Si configuration obtained by removing two atoms from a 512-site parent supercell. The density is predicted using the strictly local vacancy-trained model introduced in Sec.~\ref{sec:structural_transferability}, which is also used for the large-scale calculations below. After normalization to the target electron number, the predicted density is supplied to the same ABACUS-based property-evaluation workflow as the self-consistent reference density, without further SCF refinement. The atomic structure and electronic-structure settings are matched to isolate the effect of replacing the density input. The predicted-density calculation closely reproduces the near-edge band structure along $Y$--$\Gamma$--$Z$ (Fig.~\ref{fig:properties}j) and the principal DOS features, including the vacancy-related peak near the Fermi level (Fig.~\ref{fig:properties}k). Both the real and imaginary parts of the $xx$ optical conductivity agree closely with the reference results, including the dominant peaks and the weaker low-photon-energy features associated with the vacancies (Fig.~\ref{fig:properties}l).

Supplementary Fig.~S7\ignore{Fig.~\ref{supp-fig:vac_dos_abacus}} extends the DOS comparison to vacancy-containing structures derived from 216-, 512-, and 1000-site parent cells, covering several vacancy concentrations at each size. The predicted-density spectra reproduce the principal reference features across these tested structures. Supplementary Fig.~S8\ignore{Fig.~\ref{supp-fig:vac_ac_abacus}} further compares the optical conductivity and dielectric response for 215-, 510-, and 996-atom vacancy-containing structures, again showing close agreement between calculations using predicted and self-consistent densities under matched settings. These comparisons extend the spectral validation beyond the representative configuration in Fig.~\ref{fig:properties}j--l.

Together, the crystalline and vacancy benchmarks show that the predicted densities retain the electronic information needed to reproduce the DOS and optical responses examined here. These spectra are obtained without spectral training labels or further SCF refinement. In particular, the validation of the vacancy-trained model provides the basis for using the same model in the large-scale spectral calculations that follow.

%% file: context/result_ON.tex
\begin{figure*}[t]
    \centering
    \includegraphics[width=\textwidth]{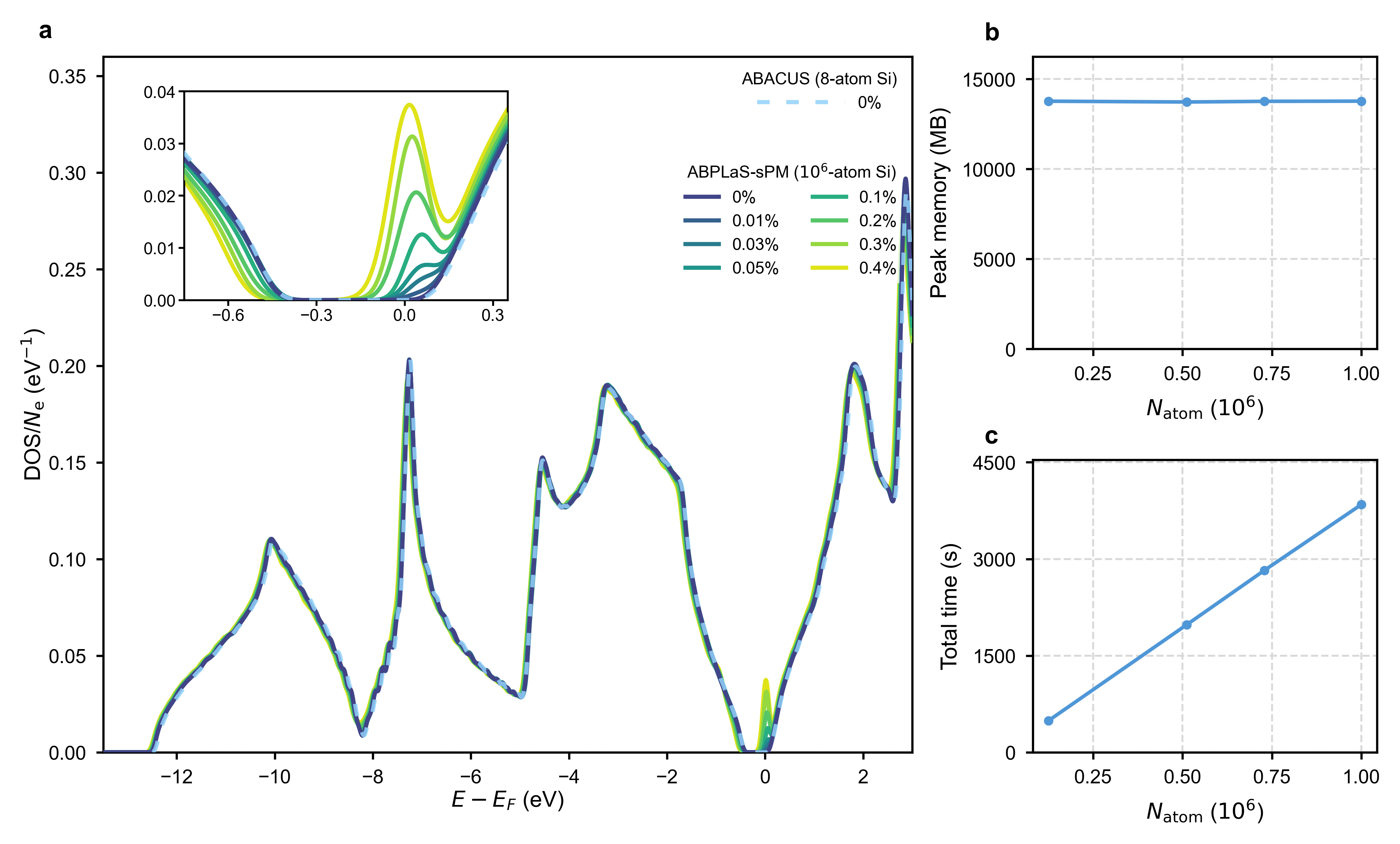}
    \caption{%
\textbf{Million-atom DOS calculations and density-inference resources.}
\textbf{a}, DOS for pristine and vacancy-containing Si supercells derived from a $10^6$-atom crystal, with vacancy concentrations from $0$ to $0.4\%$. Solid curves are obtained from fixed predicted-density Hamiltonians using ABPLaS--sPM, with one stochastic sample per structure. The dashed curve is the ABACUS reference for a pristine eight-atom Si cell using a $50\times50\times50$ $k$-point mesh, chosen to match the reciprocal-space sampling of the pristine million-atom supercell.
Each spectrum is divided by its corresponding electron number, giving an approximately unit integral below its own Fermi energy. All displayed spectra use a common Fermi-energy reference, and the inset enlarges the vacancy-related feature near this energy.
\textbf{b,c}, Peak recorded GPU memory across the atom-network and probe-decoding stages (\textbf{b}) and their summed execution time (\textbf{c}) for vacancy-free Si supercells.
These measurements exclude data preparation, Hamiltonian construction, and property evaluation.
}
    \label{fig:dos_large}
\end{figure*}

We now extend the calculations to system sizes beyond the practical reach of conventional self-consistent DFT followed by explicit diagonalization. The vacancy-trained, strictly local SCALE-Charge model validated in Sec.~\ref{sec:properties} supplies densities for larger Si supercells without retraining or target-system SCF iterations. These densities define fixed real-space Hamiltonians in ABPLaS, from which stochastic propagation evaluates spectral and transport properties without explicit diagonalization. Before presenting the large-scale results, we verify the consistency of the numerical representations and spectral solvers used in this extension.

Reference-data generation and large-scale propagation benefit from different numerical representations. The compact numerical atomic-orbital basis in ABACUS supports repeated reference DFT calculations~\cite{2026-zhou-review}, while the orthonormal real-space grid in ABPLaS avoids inverse-overlap operations during propagation~\cite{2023-zhou-time}. The learned real-space density connects these stages without transferring orbital coefficients or basis-dependent matrix elements. Independently converged ABACUS and ABPLaS calculations yield consistent reference DOS and optical-conductivity spectra for the same vacancy-containing Si structures (Supplementary Figs.~S9a--c and~S10a--c\ignore{Figs.~\ref{supp-fig:vac_dos_abplas}a--c and \ref{supp-fig:vac_ac_abplas}a--c}). Using the predicted density in ABPLaS also reproduces its self-consistent-density spectra under diagonalization (Supplementary Figs.~S9d--f and~S10d--f\ignore{Figs.~\ref{supp-fig:vac_dos_abplas}d--f and \ref{supp-fig:vac_ac_abplas}d--f}). This agreement demonstrates the flexibility that motivated our choice of the real-space density as an interface. A model trained on densities from atomic-orbital calculations can supply the electronic input for real-space-grid propagation, allowing reference-data generation and large-scale property evaluation to use numerical representations suited to their respective computational tasks.

For spectral evaluation, sPM estimates operator traces from propagated random states. Autocorrelations yield the DOS, while current correlations provide optical and transport responses without explicit construction of the eigenstates~\cite{2010-yuan-graphene,2023-tbplas,2026-fu-large}. For the same predicted-density ABPLaS Hamiltonian, sPM reproduces the diagonalization-based DOS and optical-conductivity spectra at the displayed resolution (Supplementary Figs.~S9g--i and~S10g--i\ignore{Figs.~\ref{supp-fig:vac_dos_abplas}g--i and \ref{supp-fig:vac_ac_abplas}g--i}). Together with the property benchmarks in Sec.~\ref{sec:properties}, these comparisons assess the predicted density, its use in the real-space representation, and the stochastic evaluation of the resulting spectra. Propagation acts on the Hamiltonian of the complete supercell, retaining electronic couplings across the density-inference blocks.

Fig.~\ref{fig:dos_large}a shows the DOS of pristine and vacancy-containing Si supercells derived from a $10^6$-site parent crystal, with vacancy fractions ranging from $0$ to $0.4\%$. For pristine Si, the million-atom predicted-density sPM spectrum agrees closely with the ABACUS reference obtained from an eight-atom cell using a $50\times50\times50$ $k$-point mesh. The principal peak positions and relative intensities are reproduced across the plotted energy range, providing a pristine spectral benchmark for the combined density-prediction and propagation workflow at this scale.

Introducing vacancies leaves the dominant host DOS peaks nearly coincident while producing a narrow feature near the Fermi level whose spectral weight increases with vacancy concentration. The changes are therefore concentrated in the low-DOS region, with comparatively small changes to the main host spectrum. Even the lowest nonzero concentration, $0.01\%$, corresponds to 100 vacancies in a single supercell. The large simulation volume accommodates many defect separations and local arrangements, allowing dilute, randomly distributed vacancies to be examined beyond the periodic repetition of a single defect in a small cell.

Generating the complete density field remains practical at these system sizes. Figs.~\ref{fig:dos_large}b,c report the GPU memory and network execution time for vacancy-free Si supercells under fixed model, grid, and blocking settings. From approximately $1.25\times10^5$ to $10^6$ atoms, the recorded peak GPU memory remains close to $1.4\times10^4$~MB, while network execution time increases approximately linearly and reaches about $3.8\times10^3$~s at the largest size. Increasing the system size requires more inference blocks while maintaining a bounded GPU working set.

\begin{figure*}[t]
    \centering
    \includegraphics[width=\textwidth]{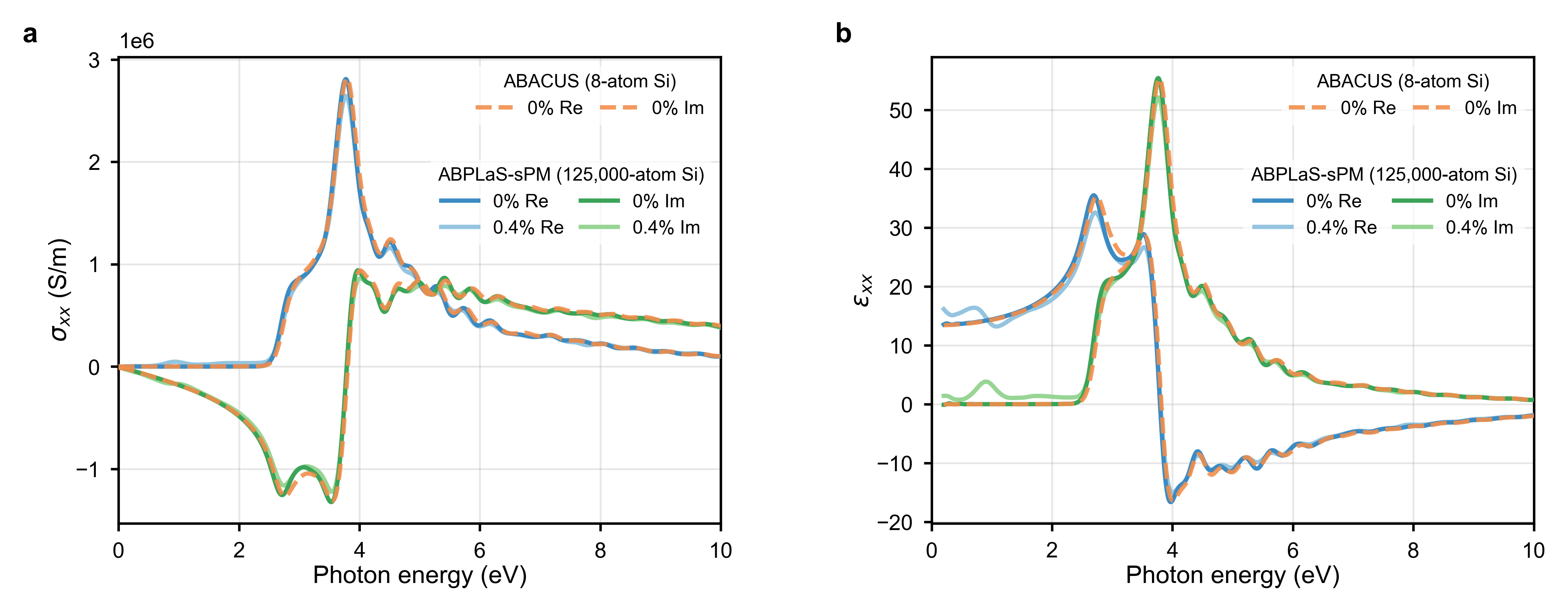}
    \caption{
    \textbf{Optical response of pristine and vacancy-containing Si.}
    \textbf{a}, Real and imaginary parts of the $xx$ optical conductivity $\sigma_{xx}(\omega)$ for pristine 125,000-atom Si and a 124,500-atom structure obtained by removing 500 atoms, corresponding to vacancy fractions of $0$ and $0.4\%$.
    \textbf{b}, Corresponding real and imaginary parts of the dielectric function $\varepsilon_{xx}(\omega)$.
    The solid curves are calculated with ABPLaS--sPM from fixed \ModelName{}-predicted densities. Blue and green denote the real and imaginary parts, respectively; darker shades indicate $0\%$ vacancies and lighter shades indicate $0.4\%$ vacancies. Orange dashed curves show the ABACUS–PYATB reference for pristine eight-atom Si using a $25\times25\times25$ $k$-point mesh, chosen to match the reciprocal-space sampling of the pristine 125,000-atom supercell. The ABACUS–PYATB reference spectra are convolved with a Gaussian kernel of standard deviation $0.1$~eV.
    }
    \label{fig:optical_12w}
\end{figure*}

To relate the spectral features to optical response, state character, and transport, we examine a common vacancy-containing structure obtained by removing 500 atoms from a 125,000-site Si supercell. The resulting 124,500-atom system has a vacancy fraction of $0.4\%$, and its pristine parent provides the baseline for identifying vacancy-induced changes in the optical response. All three analyses of the defective cell use the same predicted-density Hamiltonian.

Fig.~\ref{fig:optical_12w} compares the complex $xx$ optical conductivity and dielectric function of the pristine and defective supercells, together with an ABACUS–PYATB reference for pristine eight-atom Si sampled on a $25\times25\times25$ $k$-point mesh~\cite{2023-Jin-PYATB}. For the pristine system, the large-supercell calculation reproduces the principal resonances and overall spectral shapes of the small-cell reference in both quantities. This comparison provides a corresponding benchmark for the optical response at the hundred-thousand-atom scale.

Relative to the pristine supercell, vacancies modestly reduce the dominant peak intensity around $3$--$4$~eV and modify secondary features while introducing an additional low-energy response. This change is particularly visible as a weak absorptive feature near $1$~eV in the dielectric function and a modification of its low-energy real part. These changes are consistent with additional transitions involving defect-related states. Optical response depends on transition energies, occupations, and current matrix elements~\cite{2010-yuan-graphene,2023-tbplas}, providing information complementary to the DOS.

We examine the spatial character of the states using normalized quasi-eigenstates constructed by finite-time energy filtering of propagated random states~\cite{2023-tbplas,2026-fu-large}. 
The quasi-eigenstates are finite-energy-window superpositions of Hamiltonian eigenstates.
Three energies are selected from the defective-cell DOS, comprising a lower-energy reference $E_1=0.205$~Ha, the defect-peak energy $E_2=0.246$~Ha, and a higher-energy reference $E_3=0.271$~Ha. These values use the Hamiltonian energy reference also employed in Fig.~\ref{fig:vacancy_dc_diffusion}. Each filtered state is a superposition within a finite energy window.

\begin{figure*}[t]
    \centering
    \includegraphics[width=\textwidth]{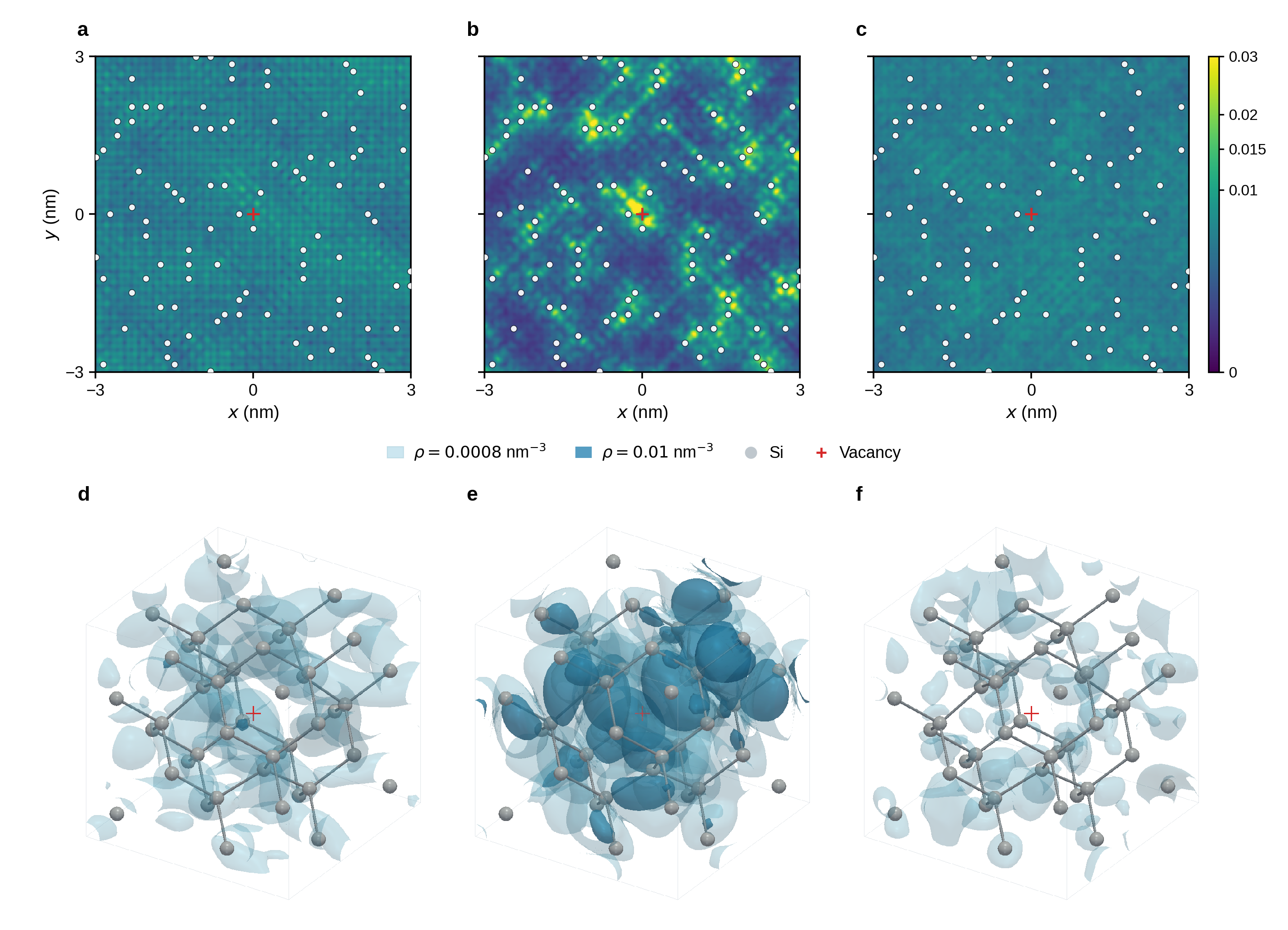}
    \caption{%
\textbf{Spatial character of energy-filtered states near the vacancy-induced spectral feature.}
All panels use the 124,500-atom Si supercell containing 500 vacancies. All quasi-eigenstates are normalized over the complete supercell.
\textbf{a--c}, Quasi-eigenstate probability densities projected onto the $xy$ plane by integration along $z$, shown within a common $6\times6\,\mathrm{nm}^2$ window.
\textbf{d--f}, Local probability-density isosurfaces within the same $1\times1\times1\,\mathrm{nm}^3$ region around a selected vacancy.
Columns show the lower-energy reference at $E_1=0.205$~Ha, the defect peak at $E_2=0.246$~Ha, and the higher-energy reference at $E_3=0.271$~Ha.
Energies use the Hamiltonian reference adopted in Fig.~\ref{fig:vacancy_dc_diffusion}.
The quasi-eigenstates are finite-energy-window superpositions of Hamiltonian eigenstates.
}
    \label{fig:vacancy_qe}
\end{figure*}

\begin{figure*}[t]
    \centering
    \includegraphics[width=\textwidth]{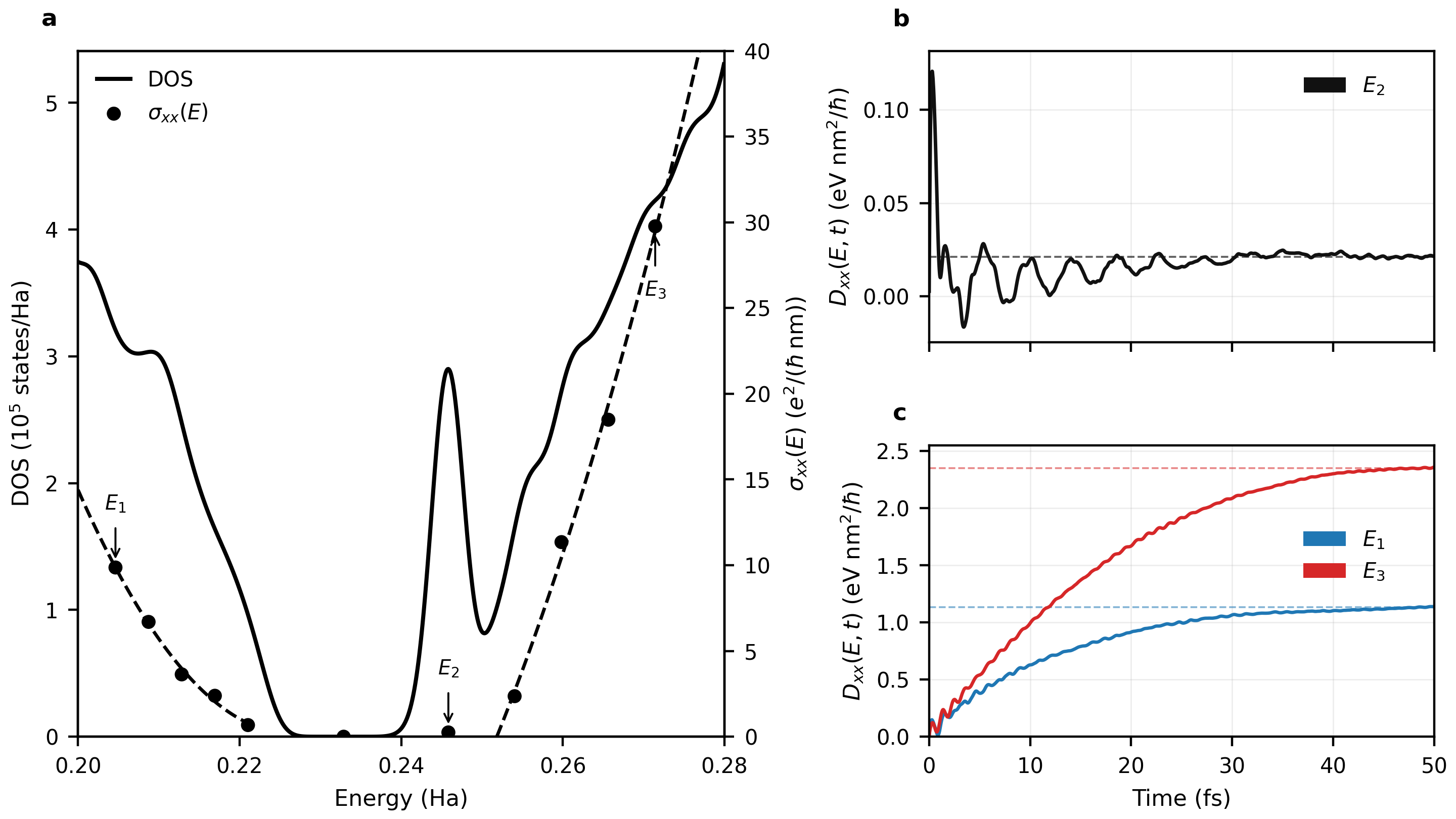}
    \caption{%
\textbf{Suppressed charge transport associated with vacancy-induced spectral weight.}
\textbf{a}, Total DOS (solid curve, left axis) and energy-resolved longitudinal conductivity $\sigma_{xx}(E)$ (points, right axis).
The marked energies $E_1=0.205$~Ha, $E_2=0.246$~Ha, and $E_3=0.271$~Ha denote the lower-energy reference, defect peak, and higher-energy reference, respectively.
\textbf{b,c}, Time-dependent diffusion coefficients at $E_2$ (\textbf{b}) and at $E_1$ and $E_3$ (\textbf{c}).
Horizontal dashed lines give averages over the same final-1000-step interval.
All panels refer to the 124,500-atom Si supercell used in Figs.~\ref{fig:optical_12w} and \ref{fig:vacancy_qe}.
}
    \label{fig:vacancy_dc_diffusion}
\end{figure*}

Fig.~\ref{fig:vacancy_qe}a--c compares the $xy$-projected quasi-eigenstate probability densities, obtained by integration along $z$, within the same enlarged region. With a common normalization and color scale, the distributions at $E_1$ and $E_3$ are comparatively uniform, whereas $E_2$ exhibits pronounced high-probability regions against a lower-probability background. Fig.~\ref{fig:vacancy_qe}d--f resolves the three-dimensional distribution within the same local region around a selected vacancy. At the same probability-density thresholds, the isosurfaces show stronger weight near this vacancy at $E_2$ than at the two reference energies. These spatial comparisons support the vacancy-related character of the spectral feature.

The energy-resolved longitudinal conductivity in Fig.~\ref{fig:vacancy_dc_diffusion}a shows that this enhanced spectral weight is accompanied by a weak transport contribution. Despite the pronounced DOS peak at $E_2$, $\sigma_{xx}(E_2)$ is only about $0.25\,e^2/(\hbar\,\mathrm{nm})$, while the conductivities at $E_1$ and $E_3$ are approximately 40 and 119 times the value at $E_2$, respectively.
In the propagation formulation, conductivity combines the DOS per unit volume with the energy-resolved diffusion coefficient~\cite{2023-tbplas,2026-fu-large}. The time dependence of the latter therefore provides a direct way to examine the suppressed transport.

At $E_2$, $D_{xx}(E_2,t)$ rises rapidly on a sub-femtosecond timescale, undergoes damped oscillations, and approaches a small late-time value (Fig.~\ref{fig:vacancy_dc_diffusion}b). At $E_1$ and $E_3$, diffusion coefficients continue to develop over tens of femtoseconds and approach substantially larger plateaus (Fig.~\ref{fig:vacancy_dc_diffusion}c). Averaging the final 1000 propagation steps gives approximately $2.1\times10^{-2}\,\mathrm{eV\,nm^2}/\hbar$ at $E_2$, about 53 and 111 times smaller than the values at $E_1$ and $E_3$, respectively.
The defect-related states thus support transient motion followed by much weaker diffusion over the simulated interval. This behavior is consistent with their spatially concentrated weight and accounts for the weak conductivity despite the enhanced DOS.

\begin{figure*}[t]
    \centering
    \includegraphics[width=\textwidth]{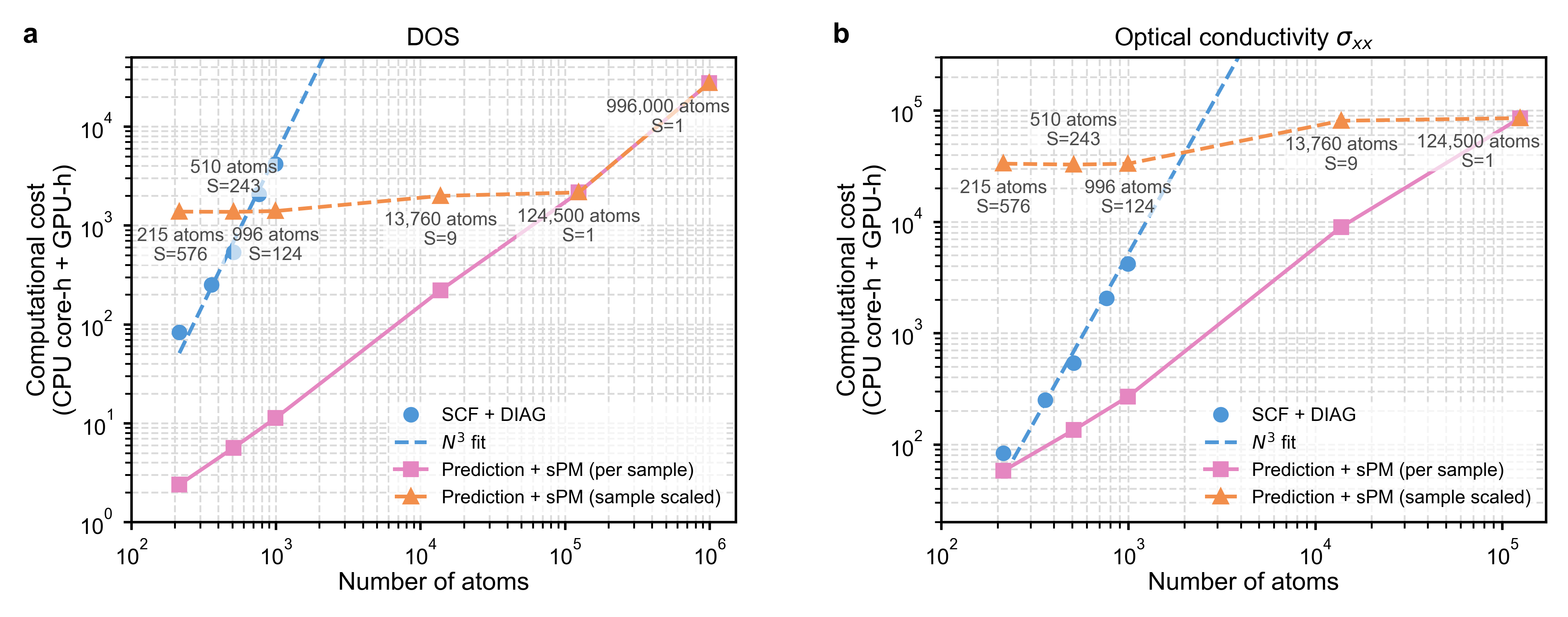}
    \caption{%
\textbf{Computational costs of conventional and predicted-density spectral calculations.}
\textbf{a,b}, Costs of the DOS (\textbf{a}) and $xx$ optical conductivity (\textbf{b}) for vacancy-containing Si at a single $k$ point.
Blue circles include SCF convergence, diagonalization, and spectral evaluation.
Blue dashed curves show fixed-$N_a^3$ fits and their extrapolation beyond the measured sizes.
Pink squares include one density prediction and the per-sample-equivalent sPM cost.
Orange triangles show sampling-adjusted estimates using the indicated random-state counts. For the prediction-based curves, GPU-hours for density inference are added numerically to sPM CPU core-hours under the resource-accounting convention detailed in Methods. Timing protocols and sample-count selection are described in Methods.
} \label{fig:scaling_comparison}
\end{figure*}

Fig.~\ref{fig:scaling_comparison} compares the computational cost of the predicted-density-plus-sPM workflow used above with that of conventional SCF-plus-diagonalization calculations for vacancy-containing Si. The propagation benchmarks use the numerical settings adopted for the large-scale spectra above, so the comparison assesses the cost of the production workflow. All timing measurements use a single $k$ point to compare solver costs at fixed Brillouin-zone sampling multiplicity.  At fixed discretization, spectral resolution, and random-state count, the leading cost of sPM propagation scales linearly with system size.

Alongside the per-sample-equivalent costs derived from measured runtimes, we report sampling-adjusted cost estimates calibrated to the DOS convergence test in Supplementary Fig.~S11\ignore{Fig.~\ref{supp-fig:abplas_dos_sample}}. For the 13,760-atom system, this test compares DOS averaged over groups of nine independent random states with the average over 36 states as an internal convergence reference. The nine-sample DOS averages have normalized errors of approximately $1\%$ relative to the 36-sample reference. We therefore adopt $S=9$ for the 13,760-atom system and use $(N_a^{\mathrm{ref}},S^{\mathrm{ref}})=(13760,9)$ as the baseline for adjusting the sampling effort with system size.

At fixed spectral resolution, the approximate self-averaging relation $\epsilon_{\mathrm{stoch}}\propto(N_aS)^{-1/2}$ allows larger systems to reach comparable statistical accuracy with fewer random states~\cite{2026-zhou-review}. Using this DOS sampling reference, the sample counts are chosen to keep $N_aS$ approximately constant, with integer rounding and a minimum of one sample. The resulting cost estimates therefore account for the decreasing sampling effort needed as the system grows, until the single-sample limit is reached. Density prediction is counted once per structure, and the measured per-sample-equivalent sPM cost is multiplied by the selected sample count.

Under the mixed CPU/GPU resource-accounting convention detailed in Methods, the conventional workflow has the lower reported cost for the smallest systems, whereas the sampling-adjusted DOS cost becomes lower between 510 and 996 atoms. For optical conductivity, extrapolation of the fixed-cubic fit to the conventional costs suggests a crossover around 2,000 atoms. These crossover estimates apply to the reported numerical settings, sampling prescription, and resource-accounting convention.

These calculations extend predicted-density electronic-structure evaluation to million-atom DOS and hundred-thousand-atom optical and transport properties. Within a common defective-cell Hamiltonian, the DOS, spatially resolved states, and diffusion reveal how a prominent vacancy-induced spectral feature can coexist with a weak transport contribution. The optical response provides complementary information on finite-frequency transitions. All these properties follow from the predicted density without property-specific retraining, while propagation over the complete supercell connects the locally learned electronic input to global spectral and dynamical behavior.

%% file: context/conclusion.tex
The density interface separates how electronic information is learned from how its physical consequences are calculated. SCALE-Charge learns a structure--density mapping from local atomic environments, while ABPLaS--sPM evaluates spectra and response from the resulting Hamiltonian of the complete supercell. Electronic propagation can therefore extend beyond the model's receptive field and density-inference blocks. This separation also permits different numerical representations for reference-data generation and property evaluation. For Si, compact numerical atomic orbitals in ABACUS are used to generate reference data, while an orthonormal real-space grid in ABPLaS supports large-scale propagation. The predicted density connects these stages without transferring orbital coefficients or overlap matrices. A common learned electronic input consequently supports multiple observables within an explicit quantum-mechanical framework.

The two stages address different sources of computational cost. Density inference replaces target-system SCF density generation, while hierarchical atom--probe blocking makes this inference practical on dense grids. Independent control of the atomic-context size and probe-batch size allows greater descriptor reuse together with smaller decoding batches. This reduces repeated atomic work and limits GPU working memory, improving the measured time--memory trade-off while retaining the required atomic environments. This organization suggests a broader inference strategy for models with reusable, query-independent atomic representations. The subsequent sparse propagation has linear leading cost per random sample at fixed resolution and reconstructs spectra from time correlations without explicit eigenstate construction. For suitably normalized global observables, self-averaging can additionally reduce the number of random states required to reach a given statistical precision as the system grows~\cite{2026-zhou-review}. Pointwise densities do not automatically benefit from the same volume averaging. Assigning density generation to learning and global spectral evaluation to stochastic propagation therefore exploits the complementary strengths of the two approaches.

The vacancy calculations illustrate the physical value of this access to several observables. Large supercells accommodate many randomly distributed vacancies even at dilute concentrations, allowing their spectral signatures to be examined beyond a small periodic repeat unit. The DOS develops a pronounced defect-related feature while retaining the principal host-spectrum structure. At a selected vacancy concentration, the quasi-eigenstate distributions and transport calculations associate this spectral weight with strong spatial inhomogeneity and substantially weaker diffusion and longitudinal conductivity than at the reference energies. The additional low-energy optical response is compatible with this suppressed transport, since finite-frequency transitions and sustained charge propagation probe different aspects of the electronic states. Examining these quantities within a common Hamiltonian distinguishes the spectral prominence of defect-related states from their ability to carry current, providing information that is not available from the DOS alone.

The present large-scale results are obtained from a fixed predicted-density Kohn--Sham Hamiltonian, so the quality of the electronic input and the numerical accuracy of the subsequent solution remain distinct. Increasing stochastic sampling or extending propagation improves the evaluation of the supplied Hamiltonian but cannot restore charge redistribution absent from the predicted density. The crystalline, disordered, and multielement benchmarks provide a numerical basis for using the learned representations within the tested structural and chemical environments, while extending their coverage calls for representative density references from additional environments.

In conclusion, we have established a framework that connects local charge-density learning to large-scale global electronic properties through memory-efficient inference and stochastic propagation. Hierarchical atom--probe blocking supports dense-grid prediction under GPU memory constraints, while the real-space solver provides access to spectra and electronic dynamics without complete diagonalization. Applied to Si, the framework enables million-atom-scale DOS calculations and combined optical, quasi-eigenstate, and transport analyses in systems containing over $10^5$ atoms. It provides a practical route from local structural information to the spectral and transport consequences of defects and disorder without property-specific retraining.

%% file: context/method_dataset.tex
\subsection{Atomic structures and datasets}
\label{sec:methods_structures}

The crystalline C, Si, Ge, and Sn datasets are constructed from diamond-type structures with space group $Fd\bar{3}m$ obtained from the Materials Project. Supercells containing 64, 216, 512, and 1000 atoms are used for the thermal and size-transfer tests. Atomic configurations are generated using LAMMPS with periodic boundary conditions in all three directions. Tersoff potentials are used for C, Si, and Ge, while Sn is described by a second-nearest-neighbor modified embedded-atom method potential~\cite{1989-sic-md, 1990-sic-md, 2017-ge-md, 2018-sn-md}.
At 300~K, simulations use the NPT ensemble at a target pressure of 0~bar, with Nos\'{e}--Hoover thermostat and barostat damping times of 0.1 and 1.0~ps, respectively. Each trajectory contains 550,000 steps of 1~fs. Configurations are saved every 500 steps, and the first 100 saved configurations are discarded as equilibration. For each element, 100 configurations are selected from the first 500 retained frames of the 64-atom trajectory for training and 100 from the last 500 for validation. Each larger-cell validation set contains ten configurations selected from the last 500 retained frames of its trajectory. Additional configurations at 100, 200, 400, and 500~K are used for temperature-transfer tests without retraining the 300~K models.

Amorphous Si is generated using the DPA-Semi interatomic potential~\cite{2024-liu-machine} and an NPT melt--quench protocol at 0~bar.
The system is heated from 10 to 2500~K over 2.5~ns, equilibrated at 2500~K for 1.0~ns, cooled to 1800~K over 0.5~ns, and equilibrated at 1800~K for 1.0~ns. It is then cooled to 300~K over 1.5~ns and equilibrated for another 1.0~ns. The time step and thermostat/barostat damping times are 1~fs, 0.1~ps, and 1.0~ps, respectively. The final configurations are minimized using conjugate gradients. Independent initial-velocity seeds produce 100 distinct 512-atom structures, of which 80 are used for training and 20 for validation.

The small-cell vacancy dataset is generated from 512-site Si parent cells by removing one to four atoms, with initial vacancy separations of at least 25~Bohr. Atomic coordinates and cell dimensions are relaxed using the Tersoff potential, with isotropic cell relaxation toward 0~bar. A 300~K trajectory is subsequently generated for each relaxed training structure, and its final thermal configuration is retained. The training set contains 82 configurations, comprising 41 relaxed structures and their 41 thermal counterparts. The validation set contains 40 separate relaxed structures, ten at each vacancy number. Additional validation structures derived from 216- and 1000-site parent cells contain one to two and one to eight vacancies, respectively. Vacancy fractions are defined relative to the pristine-site count before atom removal, whereas actual atom numbers are used when reporting the sizes of defective systems.

Large vacancy-containing Si structures are relaxed with alternating fixed-cell FIRE minimization and conjugate-gradient minimization with isotropic cell relaxation. The target pressure is 0~bar and the maximum fractional volume change per cell-relaxation iteration is 0.001. The requested force tolerances are progressively tightened from $10^{-4}$ to $10^{-6}$, $10^{-8}$, and $10^{-9}$~eV/\AA, with repeated minimization at the final stage. These are minimizer settings rather than a statement that every structure attains the nominal force tolerance. The million-site DOS systems contain randomly distributed vacancies at fractions of $0.01\%$--$0.4\%$. Optical, quasi-eigenstate, and transport calculations share the 124,500-atom structure obtained by removing 500 atoms from a 125,000-site crystal. The pristine parent is the optical reference. The performance comparison additionally includes the 13,760-atom structure obtained by removing 64 atoms from a 13,824-site parent.
Large-scale calculations use a separately trained strictly local Si model learned from vacancy-containing structures derived from 512-site cells and applied to larger structures without retraining. The model assignments and training procedures are described in Sec.~\ref{sec:methods_model}.

The multielement model uses the Materials Project charge-density dataset preprocessed in Ref.~\cite{2025-qin-eac}. It contains 149,196 inorganic crystal structures with 5000 stored probe points per structure. A random $95\%/5\%$ split is applied at the structure level. This imported dataset is separate from the ABACUS-generated reference datasets. The stored probe pool is distinct from the smaller set of points selected for each training evaluation.

%% file: context/method_DFT.tex
\subsection{Reference DFT calculations and fixed-density benchmarks}
\label{sec:methods_reference_dft}

Reference densities for the locally generated crystalline, amorphous, and vacancy datasets are obtained from self-consistent ABACUS~\cite{2016-li-large,2025-zhou-abacus} calculations using numerical atomic orbitals~\cite{2010-chen-systematically,2021-lin-strategy} and the PBE exchange--correlation functional~\cite{1996-perdew-generalized}.
The crystalline and amorphous datasets use SG15-V1.0 pseudopotentials with the corresponding double-zeta polarized orbitals and a 10~Bohr orbital cutoff. Vacancy-containing Si uses a PseudoDojo pseudopotential with the associated precision-series ABACUS-APNS-PPORBs-v1 orbitals. The reference Brillouin-zone meshes are $4\times4\times4$ for 64-atom crystalline cells, $3\times3\times3$ for 216-site cells, and $2\times2\times2$ for 512- and 1000-site crystalline or vacancy parent cells. Amorphous 512-atom Si uses a $4\times4\times4$ mesh. The single-$k$-point performance measurements are a separate benchmark described in Sec.~\ref{sec:methods_performance}.

For the fixed-density benchmarks in Fig.~\ref{fig:properties}, predicted and self-consistent densities are supplied to the same ABACUS evaluation procedure for each atomic configuration. The structure, orbital basis, exchange--correlation treatment, pseudopotentials, and numerical settings are matched. The predicted density is normalized to the target electron number before evaluation and is not iteratively refined. Band calculations use the reciprocal-space paths shown in the figure, while energies, forces, and DOS use the corresponding integration meshes. Energies and forces are outputs of the electronic-structure calculation and are not predicted by additional neural-network heads.

For the thermally averaged Si DOS in Fig.~\ref{fig:properties}c and Figs.~S5\ignore{Supplementary Fig.~\ref{supp-fig:si_dos_temp}}, 100 configurations are averaged at each temperature. Each spectrum is shifted by its corresponding Fermi energy, divided by its electron number, and interpolated onto a common grid with 0.01~eV spacing over the shared energy range before averaging. The 300~K crystalline Si model is used at all temperatures without retraining.

The energy--strain test applies an in-plane shear-like deformation, varying the length of the [110] lattice diagonal by $-5\%$ to $5\%$ in $1\%$ increments. Atomic fractional coordinates are held fixed, and the density model is not retrained.


%% file: context/model_detail.tex
\subsection{Density-model architecture and training}
\label{sec:methods_model}

The models used in Secs.~\ref{sec:density_accuracy}--\ref{sec:properties} employ recursive message passing, except for the vacancy-containing Si model introduced in Sec.~3.3, which uses the strictly local variant of SCALE-Charge\cite{2023-musaelian-learninglocaldynamics,2024-zhouyin-learninglocalqo}. This model is reused for the corresponding property benchmarks in Sec.~\ref{sec:properties} and the large-scale calculations in Sec.~\ref{sec:large_scale_properties}. Separate models are trained for each crystalline element, for amorphous Si, and for vacancy-containing Si, whereas the Materials Project calculations use one shared multielement model. The equations below specify the atom encoder and the atom--probe density readout.

For a directed atom--atom edge $j\to i$ and atom--probe edge $i\to p$, the relative displacements are $\mathbf{r}_{ij}=\mathbf{R}_i-\mathbf{R}_j$ and $\mathbf{r}_{pi}=\mathbf{r}_p-\mathbf{R}_i$, with distances $r_{ij}$ and $r_{pi}$ and unit directions $\hat{\mathbf{r}}_{ij}$ and $\hat{\mathbf{r}}_{pi}$. The neighborhoods are
\begin{equation}
\mathcal{N}_{\mathrm{aa}}(i)=\{j\ne i\mid r_{ij}<R_{\mathrm{aa}}\},
\qquad
\mathcal{N}_{\mathrm{pa}}(p)=\{i\mid r_{pi}<R_{\mathrm{pa}}\}.
\label{methods:neighborhoods}
\end{equation}
Periodic-image indices are omitted from the notation, but the neighbor lists and displacement vectors use the same images.

The atom encoder maintains invariant scalar edge features $\mathbf{x}_{ij}^{(\ell)}$, equivariant edge features $\mathbf{V}_{ij}^{(\ell)}$, and atom-centered features $\mathbf{n}_i^{(\ell)}$, where $\ell=0$ denotes initialization. Features are organized into irreducible representations up to $L_{\max}$. We use $\mathbf{Y}(\hat{\mathbf{r}})$ for real spherical harmonics, $\Vert$ for concatenation, $\otimes$ for an equivariant tensor product, and $\odot$ for scalar channel modulation with a common coefficient for all components of each irreducible channel. The operator $\Pi_0$ selects scalar channels, with even parity at the density output. The operators $\mathcal{L}_x$, $\mathcal{L}_V$, and $\mathcal{L}_n$ normalize the corresponding feature branches. Equivariant linear maps mix compatible representations only.

Atomic species enter through one-hot embeddings $\operatorname{emb}(Z_i)$, and distances are expanded in a radial basis $B(r)$. Initialization is
\begin{equation}
\begin{aligned}
\mathbf{x}_{ij}^{(0)}
&=\phi_x^{(0)}[\operatorname{emb}(Z_i)\Vert\operatorname{emb}(Z_j)\Vert B(r_{ij})]u_{\mathrm{aa}}(r_{ij}),\\
\mathbf{V}_{ij}^{(0)}
&=w_V^{(0)}[\mathcal{L}_x(\mathbf{x}_{ij}^{(0)})]\odot\mathbf{Y}(\hat{\mathbf{r}}_{ij}),\\
\mathbf{n}_i^{(0)}
&=W_n^{(0)}\left[
\frac{1}{\sqrt{N_{\mathrm{norm}}(i)}}
\sum_{j\in\mathcal{N}_{\mathrm{aa}}(i)}
w_n^{(0)}(\mathbf{x}_{ij}^{(0)})\odot\mathbf{Y}(\hat{\mathbf{r}}_{ij})
\right].
\end{aligned}
\label{methods:encoder_initialization}
\end{equation}
Here, $u_{\mathrm{aa}}$ is a smooth cutoff function and $\phi_x^{(0)}$ is an invariant multilayer perceptron. The aggregation factor $N_{\mathrm{norm}}(i)$ uses a species-conditioned mean coordination for the crystalline models and the incoming-edge count for the amorphous, multielement, and strictly local vacancy configurations listed in Supplementary Tables~S2--S4.

At layer $\ell$, the recursive message-passing context and intermediate edge feature are
\begin{equation}
\begin{aligned}
\mathbf{C}_{ij}^{(\ell)}
&=\mathcal{L}_n(\mathbf{n}_i^{(\ell)})\Vert
\mathcal{L}_n(\mathbf{n}_j^{(\ell)})\Vert
\mathcal{L}_V(\mathbf{V}_{ij}^{(\ell)}),\\
\widetilde{\mathbf{V}}_{ij}^{(\ell+1)}
&=\sigma\left\{
\mathbf{C}_{ij}^{(\ell)}\otimes
\left[w_T^{(\ell)}(\mathbf{x}_{ij}^{(\ell)})\odot\mathbf{Y}(\hat{\mathbf{r}}_{ij})\right]
\right\}.
\end{aligned}
\label{methods:recursive_context}
\end{equation}
The strictly local variant follows the same update form but omits the updated neighbor-atom features $\mathbf{n}_j^{(\ell)}$ from the context, while retaining neighboring species and relative geometry through the initialization.
The symbol $\sigma$ denotes a gated equivariant nonlinearity. Scalar modulation, invariant feedback, and atom-centered aggregation give
\begin{equation}
\begin{aligned}
\mathbf{V}_{ij}^{(\ell+1)}
&=a_V^{(\ell)}[\mathcal{L}_x(\mathbf{x}_{ij}^{(\ell)})]\odot
\left(W_V^{(\ell)}\widetilde{\mathbf{V}}_{ij}^{(\ell+1)}\right),\\
\mathbf{x}_{ij}^{(\ell+1)}
&=\phi_x^{(\ell)}[\mathbf{x}_{ij}^{(\ell)}\Vert
\Pi_0\widetilde{\mathbf{V}}_{ij}^{(\ell+1)}]u_{\mathrm{aa}}(r_{ij}),\\
\mathbf{n}_i^{(\ell+1)}
&=\frac{1}{\sqrt{N_{\mathrm{norm}}(i)}}
\sum_{j\in\mathcal{N}_{\mathrm{aa}}(i)}
a_n^{(\ell)}(\mathbf{x}_{ij}^{(\ell+1)})\odot\mathbf{V}_{ij}^{(\ell+1)}.
\end{aligned}
\label{methods:recursive_updates}
\end{equation}
The scalar feedback uses the intermediate feature $\widetilde{\mathbf{V}}$ before the subsequent channel modulation. Residual connections are applied between compatible feature branches according to the model configuration.

For spatial decoding, the final atom features after $T$ update layers are coupled to each probe direction and aggregated as
\begin{equation}
\begin{aligned}
\mathbf{x}_{pi}&=\phi_p[B(r_{pi})u_{\mathrm{pa}}(r_{pi})],\\
\mathbf{V}_{pi}&=W_p\sigma\left[\Pi_0\left(
\mathcal{L}_n(\mathbf{n}_i^{(T)})\otimes\mathbf{Y}(\hat{\mathbf{r}}_{pi})
\right)\right],\\
\hat{\rho}_p&=\phi_\rho\left\{W_\rho\left[
\sum_{i\in\mathcal{N}_{\mathrm{pa}}(p)}\mathbf{x}_{pi}\odot\mathbf{V}_{pi}
\right]\right\}.
\end{aligned}
\label{methods:probe_decoder}
\end{equation}
The nonlinear output map acts after aggregation at the probe. Directional tensor products use an $\mathrm{SO}(2)$-factorized implementation, rotating features into an edge-aligned frame, performing magnetic-index-resolved channel operations, and rotating back. Atomic descriptors depend on the atomic environment and can be reused across independently decoded probe batches. The raw output $\hat{\rho}_p$ is expressed in physical density units after reversing the training-time target scaling.

Training uses density labels only, with an $L_1$ loss on targets divided by a dataset-dependent scale factor. Energies, forces, eigenvalues, and response functions are not training labels. The staged schedules, cutoffs, channel multiplicities, probe counts, learning rates, and target scales are listed in Supplementary Tables~S2--S4. Unless otherwise stated, the models use $L_{\max}=4$; angular-order tests are reported separately in the Supplementary Information.

For configurations using mixed probe sampling, uniform draws are combined with density-dependent draws. The latter are generated from
\begin{equation}
\widetilde{p}_g^{\mathrm{G}}=
\exp\left[-\frac{1}{2\sigma_{\mathrm{samp}}^2}
\left(\frac{1}{\rho_g^{\mathrm{DFT}}+\epsilon_{\mathrm{samp}}}\right)^2\right],
\qquad
p_g^{\mathrm{G}}=
\frac{\widetilde{p}_g^{\mathrm{G}}}
{\sum_{g'\in\mathcal{G}}\widetilde{p}_{g'}^{\mathrm{G}}}.
\label{methods:probe_sampling}
\end{equation}
This weighting is Gaussian in inverse density. The uniform component has no density-dependent preference. Mixture fractions and probe counts depend on the dataset and training stage. For the Materials Project model, the 5000 stored probe points per structure define the available reference pool, whereas the training schedule uses 32 probes per structure in each sampled training example.

The vacancy training schedule reported in Supplementary Table~S3 begins with ten defect-free 512-atom configurations at 300~K before training on vacancy-containing structures. Its second stage uses vacancy-focused probe sampling. Si neighbors are identified under periodic boundary conditions with a 5.7-Bohr cutoff. Vacancy positions are estimated from undercoordinated atoms, along the missing-bond direction for threefold-coordinated Si. Grid points within 10.5~Bohr of these positions are vacancy-region candidates. For each training configuration, 40\% of the sampled points are drawn randomly from these regions and the remaining 60\% are drawn uniformly from previously unselected points across the cell.

\subsection{Hierarchical inference and density assembly}
\label{sec:methods_inference}

The complete probe grid is divided into outer blocks $\mathcal{P}_b$ of nominal edge length $B_A$ grid points. This size describes the target probes in the block, before adding the atomic context needed for inference. The atoms whose final descriptors are required by the decoder and the full atomic context are
\begin{equation}
\mathcal{C}_b=\bigcup_{p\in\mathcal{P}_b}\mathcal{N}_{\mathrm{pa}}(p),
\qquad
\mathcal{A}_b=\bigcup_{i\in\mathcal{C}_b}\mathcal{S}_i^{(T)},
\label{methods:block_context}
\end{equation}
where $\mathcal{S}_i^{(T)}$ is the atomic dependency domain of $\mathbf{n}_i^{(T)}$. For the recursive encoder in Sec.~\ref{sec:methods_model},
\begin{equation}
\mathcal{S}_i^{(0)}=\{i\}\cup\mathcal{N}_{\mathrm{aa}}(i),
\qquad
\mathcal{S}_i^{(\ell+1)}=
\mathcal{S}_i^{(\ell)}\cup
\bigcup_{j\in\mathcal{N}_{\mathrm{aa}}(i)}\mathcal{S}_j^{(\ell)}.
\label{methods:recursive_dependencies}
\end{equation}
The gathered context includes the initialization neighborhoods and subsequent message-passing dependencies. Atoms in $\mathcal{A}_b\setminus\mathcal{C}_b$ provide the additional context required to evaluate the decoded descriptors. For the strictly local model, $\mathcal{S}_i^{(T)}$ is determined by that encoder's fixed local dependency domain as specified in Sec.~\ref{sec:methods_model}.

The atom encoder is evaluated once per outer block, and its required final descriptors are retained as the cache $\{\mathbf{n}_i^{(T)}\mid i\in\mathcal{C}_b\}$. The target probes are partitioned into disjoint inner blocks $\mathcal{P}_{bq}$ of nominal edge length $B_P$ grid points. Each inner block is decoded using the same cache and its atom--probe connections. Probe intermediates are released between inner blocks, while the atomic cache is retained until the outer block is complete. The assembled outputs cover the complete target grid. Halo descriptors that are not used by the decoder need not equal full-system final descriptors, provided that all dependencies of the descriptors used for decoding are retained.

At fixed block sizes and bounded local atomic populations, the atomic features, cached descriptors, and active probe features form a bounded GPU working set. This statement concerns blockwise network evaluation and does not include storage of the complete output density or the downstream Hamiltonian.

After assembling the full raw density, a single global normalization enforces the target electron count,
\begin{equation}
\rho_g^{\mathrm{pred}}=
\frac{N_e\hat{\rho}_g}
{\sum_{g'\in\mathcal{G}}\hat{\rho}_{g'}\Delta v_{g'}},
\qquad
\sum_{g\in\mathcal{G}}\rho_g^{\mathrm{pred}}\Delta v_g=N_e.
\label{methods:density_normalization}
\end{equation}
Here, $\Delta v_g$ is the grid-volume element and $N_e$ is the sum of the valence-electron counts for the neutral pseudopotential systems. Normalization is applied once to the complete structure and is distinct from the target scaling used during training. Property calculations use this normalized density without further SCF updates.

\subsection{Error measures and statistical analysis}
\label{sec:methods_errors}

The complete-grid density error is the integrated absolute difference divided by the reference electron number, as defined in Eq.~\eqref{eq:density_mae}. Configuration-level errors are calculated before averaging over structures and are reported as percentages. The size- and temperature-transfer plots show the configuration mean. The predicted density is expressed in physical units after reversing the training-time target scaling.

Energy errors and mean absolute Cartesian force-component errors follow Eq.~\eqref{eq:property_errors}. The correlations in Fig.~\ref{fig:properties}g,h use one point per configuration, pairing each property error with the density error of that same configuration. No energy or force labels enter density-model training.

Local error statistics group grid points by atomic coordination or by distance to the nearest vacancy center. Nearest-atom assignments and distances are evaluated under periodic boundary conditions. For a group or spatial bin $\mathcal{B}$, the mean absolute pointwise error is
\begin{equation}
A_{\mathcal{B}}=
\frac{1}{|\mathcal{B}|}\sum_{g\in\mathcal{B}}
\left|\rho_g^{\mathrm{pred}}-\rho_g^{\mathrm{DFT}}\right|.
\label{methods:local_density_error}
\end{equation}
The accompanying grid-point distributions use the same grouping as the corresponding error statistics.

For coordination-resolved errors in amorphous Si, each grid point is assigned to its nearest atom and grouped by that atom's coordination number. Absolute errors are averaged within each coordination group over grid points pooled from the 20 validation structures. For the vacancy-distance profiles, absolute errors are averaged within each distance bin over grid points pooled from ten single-vacancy validation structures derived from 512-site parent cells. These local mean absolute errors have density units and are distinct from the electron-number-normalized complete-grid error.

For element-resolved errors in the Materials Project validation set, let $\mathcal{Q}_{k\alpha}$ be the union of stored probe points within 2~\AA\ of atoms of element $\alpha$ in structure $k$. A point is counted once for that element within a structure even when several neighborhoods overlap. The element-resolved error is
\begin{equation}
\epsilon_\alpha=
\frac{\sum_k\sum_{g\in\mathcal{Q}_{k\alpha}}
\left|\rho_{kg}^{\mathrm{pred}}-\rho_{kg}^{\mathrm{DFT}}\right|}
{\sum_k\sum_{g\in\mathcal{Q}_{k\alpha}}
\left|\rho_{kg}^{\mathrm{DFT}}\right|}.
\label{methods:element_density_error}
\end{equation}

The same set of elements, each appearing in at least ten validation structures, is used in Figs.~\ref{fig:generalization_table_acc} and \ref{fig:generalization_analysis}a, as well as Supplementary Fig.~S4. For each element, the normalized density error is obtained by accumulating absolute density errors and reference-density magnitudes before taking their ratio, which is reported as a percentage. Its training-set occurrence count is the number of training structures containing that element, with each structure counted once irrespective of the number of atoms of that species. Figure~\ref{fig:generalization_analysis}a relates the element-resolved error to this occurrence count expressed as a percentage of all training structures, whereas Supplementary Fig.~S4 reports the corresponding absolute structure counts.

For the compositional analysis in Fig.~\ref{fig:generalization_analysis}b, training and validation structures containing any element outside this selected set are excluded. The remaining structures are grouped by their number of distinct constituent elements. For each retained validation structure, the normalized density error $\epsilon_{\mathrm{MAE}}$ is calculated as the sum of absolute density differences divided by the sum of reference-density magnitudes. The plotted error is the arithmetic mean of these structure-level errors within each group, expressed as a percentage; error bars indicate one standard deviation within the group.

%% file: context/method_sPM.tex
\subsection{Real-space Hamiltonian and interface validation}
\label{sec:methods_hamiltonian}

The large-scale calculations use the real-space density predicted by the separately trained strictly local Si model as input to ABPLaS. The density is normalized to the target electron number as described in Sec.~\ref{sec:methods_inference} and remains fixed during spectral and transport evaluation. ABPLaS represents the Kohn--Sham Hamiltonian on a uniform real-space grid~\cite{2023-zhou-time}. In atomic units,
\begin{equation}
\hat{H}_{\mathrm{KS}}[\rho]
= -\frac{1}{2}\nabla^2
+V_{\mathrm{ion}}^{\mathrm{loc}}(\mathbf{r})
+\hat{V}_{\mathrm{ion}}^{\mathrm{nl}}
+V_{\mathrm{H}}[\rho](\mathbf{r})
+V_{\mathrm{xc}}[\rho](\mathbf{r}).
\label{methods:eq:hamiltonian}
\end{equation}
The kinetic operator uses a high-order finite-difference stencil~\cite{1994-chelikowsky-higher}, and the nonlocal ionic pseudopotential uses atom-centered projectors in the separable Kleinman--Bylander form~\cite{1982-kleinman-bylander}. In the orthonormal discrete grid basis, local potentials are diagonal, finite differences couple a bounded grid neighborhood, and the nonlocal projectors have finite spatial support. These properties permit sparse Hamiltonian--vector operations. The Hartree potential is obtained from the Poisson equation, and the exchange--correlation treatment is consistent with the corresponding reference calculation. The Hamiltonian acts over the complete periodic supercell, retaining couplings across density-inference block boundaries.

This representation connects the ABACUS reference calculations to large-scale propagation through the real-space density rather than orbital coefficients or basis-dependent Hamiltonian elements. In a nonorthogonal atomic-orbital basis with overlap matrix $S_{\mathrm{ov}}$, propagation formally involves the action of $S_{\mathrm{ov}}^{-1}H$. Although $H$ and $S_{\mathrm{ov}}$ can be sparse, the inverse overlap matrix is generally dense, so efficient propagation requires additional treatment of the overlap metric~\cite{2019-lin-numerical,2026-zhou-review}. The orthonormal grid representation avoids this inverse-overlap treatment.

The interface validation separates changes in density input, numerical representation, and spectral solver. In ABACUS, predicted and self-consistent reference densities are compared using the same diagonalization workflow, with matched structures, orbital basis, exchange--correlation functional, pseudopotentials, and spectral settings. Supplementary Fig.~S7\ignore{Fig.~\ref{supp-fig:vac_dos_abacus}} includes structures derived from 216-site cells with one or two vacancies, 512-site cells with one to four vacancies, and 1000-site cells with one to eight vacancies. The DOS is averaged over configurations at each size and vacancy number. Supplementary Fig.~S8\ignore{Fig.~\ref{supp-fig:vac_ac_abacus}} compares optical conductivity and dielectric response obtained with ABACUS and PYATB for the 215-, 510-, and 996-atom structures.

For these same three atom counts, independently converged self-consistent ground states in ABACUS and ABPLaS provide the reference spectra for the cross-representation comparison in Supplementary Figs.~S9a--c and~S10a--c\ignore{Figs.~\ref{supp-fig:vac_dos_abplas}a--c and \ref{supp-fig:vac_ac_abplas}a--c}. The exchange--correlation functional and pseudopotential settings are matched as closely as the two implementations permit. Within ABPLaS, diagonalization using the normalized predicted density is then compared with diagonalization using its self-consistent reference density under otherwise identical settings (panels d--f). Finally, DIAG and sPM evaluate the same fixed predicted-density ABPLaS Hamiltonian (panels g--i). The spectral normalization and broadening procedures are described in Sec.~\ref{sec:methods_spm}.

\subsection{Stochastic propagation and spectral response}
\label{sec:methods_spm}

For an orthonormal basis $\{|a\rangle\}_{a=1}^{N_b}$, stochastic trace estimation uses $S$ independent normalized states,
\begin{equation}
|\phi_s\rangle=\sum_{a=1}^{N_b}c_{as}|a\rangle,
\qquad
\sum_a|c_{as}|^2=1,
\qquad
\mathbb{E}[c_{as}^{*}c_{bs}]=\frac{\delta_{ab}}{N_b}.
\label{methods:eq:random_states}
\end{equation}
Consequently,
\begin{equation}
\operatorname{Tr}\hat{O}
\simeq\frac{N_b}{S}\sum_{s=1}^{S}
\langle\phi_s|\hat{O}|\phi_s\rangle.
\label{methods:eq:trace_estimator}
\end{equation}
Traces refer to the explicitly represented Hilbert space. Spin multiplicity not included in this space is applied when converting the traces to physical DOS and response functions. Time propagation uses $\hat{U}(t)=\exp(-i\hat{H}t/\hbar)$, where $\hat{H}$ denotes the fixed predicted-density Hamiltonian~\cite{2010-yuan-graphene,2023-tbplas,2026-fu-large}.

Before Chebyshev evaluation, the Hamiltonian is rescaled as $\widetilde{H}=(\hat{H}-E_c\hat{I})/\Delta E_H$, where $E_c=(E_{\max}+E_{\min})/2$ and $\Delta E_H=(E_{\max}-E_{\min})/2$, with $E_{\min}$ and $E_{\max}$ enclosing the represented spectrum. The propagator is approximated by
\begin{equation}
\hat{U}(t)\simeq e^{-iE_ct/\hbar}
\left[
J_0(\xi)\hat{I}
+2\sum_{n=1}^{N_C}(-i)^nJ_n(\xi)T_n(\widetilde{H})
\right],
\qquad
\xi=\frac{\Delta E_Ht}{\hbar}.
\label{methods:eq:chebyshev}
\end{equation}
Here, $J_n$ and $T_n$ are Bessel functions and Chebyshev polynomials. The polynomial sequence is generated through sparse Hamiltonian--vector products, and the truncation order is controlled by the neglected coefficient tail. This procedure avoids explicit eigenstate construction and orbital orthogonalization.

For the DOS, the autocorrelation is $C_s^{\mathrm{DOS}}(t)=\langle\phi_s|\hat{U}(t)|\phi_s\rangle$, and
\begin{equation}
D(E)\simeq\frac{N_b}{2\pi\hbar S}
\sum_{s=1}^{S}\int_{-t_{\max}}^{t_{\max}}
 w(t)e^{iEt/\hbar}C_s^{\mathrm{DOS}}(t)\,dt.
\label{methods:eq:dos}
\end{equation}
The window $w(t)$ controls finite-time ringing and broadening. This expression gives the total DOS in states per unit energy before any plotting normalization. Each autocorrelation yields the spectrum across the sampled energy interval without repeating propagation for every energy point.

Optical response uses the Fermi--Dirac operator $f(\hat{H})=[e^{\beta(\hat{H}-\mu)}+1]^{-1}$, with $\beta=(k_{\mathrm{B}}T_e)^{-1}$. The electronic temperature $T_e$ controls occupations and is distinct from the ionic temperature used for structural sampling. The one-particle current operator is
\begin{equation}
\hat{J}_{\alpha}=-e\hat{v}_{\alpha}
=-\frac{ie}{\hbar}[\hat{H},\hat{r}_{\alpha}],
\label{methods:eq:current}
\end{equation}
where $e>0$. The commutator includes the nonlocal pseudopotential contribution and is evaluated consistently with the periodic representation. With $\hat{J}_{\alpha}(t)=\hat{U}^{\dagger}(t)\hat{J}_{\alpha}\hat{U}(t)$, the optical correlation is
\begin{equation}
C_{\alpha\beta}^{\mathrm{OC}}(t)
=\operatorname{Tr}\!\left[
 f(\hat{H})\hat{J}_{\alpha}(t)
 [1-f(\hat{H})]\hat{J}_{\beta}
\right].
\label{methods:eq:optical_correlation}
\end{equation}
Its stochastic estimator uses $\langle\psi_{1,s}(t)|\hat{J}_{\alpha}|\psi_{2,\beta,s}(t)\rangle$ with the prefactor $N_b/S$, where $|\psi_{1,s}(t)\rangle=\hat{U}(t)f(\hat{H})|\phi_s\rangle$ and $|\psi_{2,\beta,s}(t)\rangle=\hat{U}(t)[1-f(\hat{H})]\hat{J}_{\beta}|\phi_s\rangle$~\cite{2023-tbplas,2026-fu-large}.

For $\omega>0$, the regular absorptive longitudinal conductivity is
\begin{equation}
\operatorname{Re}\sigma_{\alpha\alpha}^{\mathrm{reg}}(\omega)
=\frac{1-e^{-\beta\hbar\omega}}{2\hbar\omega\Omega}
\int_{-\infty}^{\infty}
 e^{i\omega t}C_{\alpha\alpha}^{\mathrm{OC}}(t)\,dt,
\label{methods:eq:optical_conductivity}
\end{equation}
where $\Omega$ is the supercell volume. Numerical evaluation uses the finite correlation interval and its window or damping. The zero-frequency Drude term is not included in this regular response. With the $e^{-i\omega t}$ convention for the applied field, the imaginary part follows from the Kramers--Kronig transform of the real part. The dielectric function in SI units is
\begin{equation}
\varepsilon_{\alpha\beta}(\omega)
=\delta_{\alpha\beta}+\frac{i\sigma_{\alpha\beta}(\omega)}{\varepsilon_0\omega}.
\label{methods:eq:dielectric}
\end{equation}
The displayed response is the $xx$ component, with photon energy $\hbar\omega$ in eV and conductivity in $\mathrm{S\,m^{-1}}$. The current correlations provide the optical spectrum across the sampled frequency interval in one propagation calculation.

The small-system DIAG--sPM validation spectra in Supplementary Figs.~S9g--i and~S10g--i\ignore{Figs.~\ref{supp-fig:vac_dos_abplas}g--i and \ref{supp-fig:vac_ac_abplas}g--i} use 256 and 128 random states, respectively. These counts are separate from the production sampling prescription in Sec.~\ref{sec:methods_performance}. The million-atom DOS spectra in Fig.~\ref{fig:dos_large}a each use one random state.

The validation spectra retain their specified normalization and broadening conventions. In Supplementary Fig.~S7\ignore{Fig.~\ref{supp-fig:vac_dos_abacus}}, each DOS is aligned to the corresponding structure's Fermi energy and divided by its electron number before configuration averaging. In Supplementary Fig.~S9\ignore{Fig.~\ref{supp-fig:vac_dos_abplas}}, each spectrum is divided by the atom number and its energy zero is the Fermi energy obtained by integrating the corresponding DOS to the target electron count. For the optical comparison in Supplementary Fig.~S10a--c\ignore{Fig.~\ref{supp-fig:vac_ac_abplas}a--c}, the ABACUS--PYATB spectra use Lorentzian broadening with $\eta=0.005$~eV followed by convolution with a Gaussian of standard deviation $0.10$~eV, truncated at five standard deviations. Convolution is performed on the full energy grid before selecting the plotted interval.

For the million-atom spectra in Fig.~\ref{fig:dos_large}a, $D(E)$ is divided by the corresponding electron number, so that its integral below that system's Fermi energy is approximately unity. The displayed spectra share a common Fermi-energy reference across vacancy concentrations.

\subsection{Quasi-eigenstates and energy-resolved transport}
\label{sec:methods_transport}

The quasi-eigenstate and transport calculations use the same fixed predicted-density Hamiltonian as the optical calculation for the 124,500-atom Si structure with 500 vacancies. Energy-filtered quasi-eigenstates are constructed from propagated random states through
\begin{equation}
\begin{aligned}
|\Psi_s(E)\rangle
&=\frac{1}{2\pi\hbar}
\int_{-t_{\mathrm{filt}}}^{t_{\mathrm{filt}}}
 w_E(t)e^{iEt/\hbar}\hat{U}(t)|\phi_s\rangle\,dt,\\
|\widetilde{\Psi}_s(E)\rangle
&=\frac{|\Psi_s(E)\rangle}
{\sqrt{\langle\Psi_s(E)|\Psi_s(E)\rangle}}.
\end{aligned}
\label{methods:eq:quasi_eigenstate}
\end{equation}
Here, $t_{\mathrm{filt}}$ is the filtering time and $w_E(t)$ is the window function, chosen here as a Hann window.
The finite interval selects a spectral window rather than a unique eigenstate~\cite{2023-tbplas,2026-fu-large}. Its duration is distinct from the subsequent dynamical time over which transport is evaluated. The selected energies are $E_1=0.205$~Ha, $E_2=0.246$~Ha, and $E_3=0.271$~Ha in the Hamiltonian reference used in Fig.~\ref{fig:vacancy_dc_diffusion}.

The probability density of each normalized filtered state is projected onto the $xy$ plane by integrating along $z$.
For orthonormal grid coefficients $u_g=\langle g|\widetilde{\Psi}_s(E)\rangle$, the corresponding real-space probability density is $\rho_E(\mathbf{r}_g)=|u_g|^2/\Delta v$ on a uniform grid with volume element $\Delta v$. 
For plotting, the projection along $z$ is $P_{xy}(x_i,y_j)=\sum_k|u_{ijk}|^2/(\Delta x\Delta y)$, in $\mathrm{nm}^{-2}$.
Panels a--c in Fig.~\ref{fig:vacancy_qe} show the same enlarged region and use a common color scale, with projected vacancy positions superimposed.
Panels d--f use a common $1\times1\times1\,\mathrm{nm}^3$ cube centered on the vacancy nearest the probability-density maximum of the defect-peak quasi-eigenstate.

Energy-resolved longitudinal transport uses the current-correlation formulation at a fixed electronic energy~\cite{2023-tbplas,2026-fu-large}. The correlation is estimated at finite energy resolution using energy-filtered stochastic states:
\begin{equation}
C_{\alpha}^{\mathrm{EC}}(t,E)
=
\frac{
\operatorname{Tr}\!\left[
\delta(E-\hat{H})\hat{J}_{\alpha}
e^{i\hat{H}t/\hbar}\hat{J}_{\alpha}
\right]
}{
\operatorname{Tr}\!\left[\delta(E-\hat{H})\right]
}.
\label{methods:eq:transport_correlation}
\end{equation}
The finite-time diffusion estimate and the associated conductivity are
\begin{equation}
\begin{aligned}
D_{\alpha\alpha}(E,\tau)
&=\frac{1}{e^2}\int_0^{\tau}
\operatorname{Re}\!\left[e^{-iEt/\hbar}C_{\alpha}^{\mathrm{EC}}(t,E)\right]dt,\\
\sigma_{\alpha\alpha}(E)
&=\frac{e^2D(E)}{\Omega}\overline{D}_{\alpha\alpha}(E).
\end{aligned}
\label{methods:eq:diffusion_conductivity}
\end{equation}
Here, $D(E)$ is the total DOS, including the same spin convention as the conductivity, and $\overline{D}_{\alpha\alpha}(E)$ is the late-time estimate from the diffusion time series. The energy argument, filter resolution, DOS normalization, and supercell volume are used consistently in combining these quantities. The reported diffusion averages use the final 1000 propagation steps over the same time interval at the three selected energies. The displayed units are $\mathrm{eV\,nm^2}/\hbar$ for diffusion and $e^2/(\hbar\,\mathrm{nm})$ for conductivity. This scan in electronic energy is distinct from the optical response as a function of photon energy.

\subsection{Computational performance and sampling-adjusted scaling}
\label{sec:methods_performance}

Density-inference performance is measured separately from Hamiltonian construction and spectral evaluation. In Figs.~\ref{fig:model_valid} and \ref{fig:dos_large}b,c, atom-encoder and probe-decoder execution times are accumulated over all calls covering the complete target grid. Their sum defines the network execution time. For each stage, peak memory is the maximum allocated GPU memory recorded over its calls using the PyTorch allocator. The measured allocations include model parameters and the graph and feature tensors present during that stage. Probe-stage memory also includes the retained atomic descriptors. The overall peak is the maximum of the stage peaks, rather than their sum. These measurements exclude data preparation and subsequent electronic-structure calculations.
The stage-resolved comparison in Fig.~\ref{fig:model_valid}c,d uses single-level blocking with $B=70$ and hierarchical blocking with $(B_A,B_P)=(200,50)$, with block edges measured in grid points.

For the halo statistic in Fig.~\ref{fig:model_valid}a, the geometric-core atom count is estimated as $N_{\mathrm{core}}=N_a\Omega_b/\Omega$, where $\Omega_b$ is the volume of the outer probe block and $\Omega$ is the supercell volume. The involved-atom count includes the complete atomic context required to construct the descriptors used by that block. Their ratio is $\eta_{\mathrm{halo}}=N_{\mathrm{involved}}/N_{\mathrm{core}}$. The system-size tests retain the same model, grid density, block sizes, numerical precision, and hardware, while varying the number of processed blocks.

The spectral-cost measurements in Fig.~\ref{fig:scaling_comparison} use the numerical settings employed for the large-scale spectral calculations described in Secs.~\ref{sec:methods_hamiltonian} and \ref{sec:methods_spm}. They therefore measure the size dependence of the production workflow. Both the conventional and prediction-based workflows use a single $k$ point, separating solver-dependent costs from changes in Brillouin-zone sampling multiplicity. The conventional measurements include systems with 215, 359, 510, 765, and 996 atoms. Their SCF calculations converge in 24, 27, 34, 35, and 39 iterations, respectively. The conventional cost includes all SCF iterations, the recorded diagonalization using the converged density, and the corresponding spectral evaluation. Prediction-based DOS timings use 215, 510, 996, 13,760, 124,500, and 996,000 atoms; optical timings use the first five sizes. Each prediction-based cost includes one density inference and the recorded sPM property calculation. Training-data generation and model training are excluded.

For CPU stage $q$, the allocated resource cost is
\begin{equation}
C_q^{\mathrm{CPU}}=\frac{p_qt_q}{3600},
\qquad
G_\rho=\frac{t_\rho}{3600},
\label{eq:methods_resource_time}
\end{equation}
where $p_q$ is the allocated CPU core count, $t_q$ is elapsed time in seconds, and $t_\rho$ is the single-GPU density-inference time in seconds. CPU costs are summed over the included stages, while $G_\rho$ is expressed in GPU-hours.

The sPM timing measurements in Fig.~\ref{fig:scaling_comparison} use a single random state per run. Sampling-adjusted sPM costs are estimated by multiplying the measured single-sample cost by the selected number of random states, rather than measured from separate multi-sample runs.

The DOS convergence analysis in Supplementary Fig.~S11\ignore{Fig.~\ref{supp-fig:abplas_dos_sample}} supplies the common sampling reference. For the 13,760-atom structure, 36 independent random states are divided into four groups of nine. Each group-averaged DOS is compared with the average over all 36 states as an internal convergence reference. Within each displayed comparison window, both spectra are normalized to unit integral before evaluating the reported $L_1$ and relative $L_2$ differences. Nine random states achieve the DOS statistical accuracy required here relative to this reference. We adopt this sampling reference for both DOS and optical-conductivity calculations.

At fixed spectral resolution, the approximate self-averaging relation $\epsilon_{\mathrm{stoch}}\propto(N_aS)^{-1/2}$ for suitably normalized global observables motivates adjusting the sample count with system size~\cite{2026-zhou-review}. Using $(N_a^{\mathrm{ref}},S^{\mathrm{ref}})=(13760,9)$, we assign
\begin{equation}
S_{\mathrm{adj}}(N_a)=
\max\!\left[
1,\operatorname{round}\!\left(
\frac{N_a^{\mathrm{ref}}S^{\mathrm{ref}}}{N_a}
\right)
\right].
\label{eq:methods_adjusted_samples}
\end{equation}
This prescription keeps $N_aS$ approximately constant until the single-sample limit is reached, with small deviations from integer rounding. The 996,000-atom system has already reached that limit and exceeds the reference product. The adjusted counts define workload estimates based on the measured per-sample-equivalent costs; they do not imply additional runs at every assigned count. The largest DOS and optical points use one sample and correspond to the production calculations presented in the Results.

Under the accounting convention used in Fig.~\ref{fig:scaling_comparison}, numerical values of GPU-hours for density inference are added to sPM CPU core-hours without a CPU--GPU conversion factor. The plotted indices are
\begin{equation}
C_1=G_\rho+C_{\mathrm{sPM}}^{(1)},
\qquad
C_{\mathrm{adj}}=G_\rho+S_{\mathrm{adj}}C_{\mathrm{sPM}}^{(1)}.
\label{eq:methods_mixed_cost}
\end{equation}
Density inference is counted once because the same density field is reused for all random states. These indices follow a mixed resource-time accounting convention and do not represent homogeneous CPU core-hours or complete-workflow elapsed times.

The blue dashed curves are fitted to the five measured conventional costs in each panel using $\log_{10}C=a+3\log_{10}N_a$, with only the intercept $a$ fitted. Portions beyond the measured sizes are extrapolations. The other connecting lines guide the eye and are not fitted scaling laws. The crossover estimates in the Results depend on the stated resource-accounting convention and sampling prescription; the optical crossover additionally uses the extrapolated conventional-cost fit.